\documentclass[preprint,12pt]{elsarticle}

\usepackage[english]{babel}
\usepackage[letterpaper,top=2cm,bottom=2cm,left=3cm,right=3cm,marginparwidth=1.75cm]{geometry}

\usepackage[title]{appendix}
\usepackage{graphicx}
 
\usepackage{float}
\usepackage{tikz}
\usepackage{amssymb}
 
\usepackage{listings}
 
\usepackage{subcaption}
\usepackage{rotating}

\usepackage{amsmath}

\biboptions{sort&compress}

\usepackage[colorlinks=true, allcolors=blue]{hyperref}
\usepackage{indentfirst}
\usepackage{color}
\usepackage{comment}

\usepackage{amsthm}
\graphicspath{{figs/}} 
\usepackage{booktabs}
\usepackage{multirow}
\usepackage[outline]{contour}
\usepackage{bm}

\makeatletter
\@addtoreset{table}{section}
\makeatother

\makeatletter
\@addtoreset{figure}{section}
\makeatletter

\begin{document}
\begin{frontmatter}
\title{Simulating  anisotropic diffusion processes  with smoothed particle hydrodynamics }	
\author{Xiaojing Tang}
\ead{xiaojing.tang@tum.de}
\author{Oskar Haidn}
\ead{oskar.haidn@tum.de}
\author{Xiangyu Hu\corref{mycorrespondingauthor}}
\cortext[mycorrespondingauthor]{Corresponding author.}
\ead{xiangyu.hu@tum.de}	
\address{TUM School of Engineering and Design,  Technical University of Munich, 85748 Garching, Germany}

\begin{abstract}
Diffusion problems with anisotropic features arise in the various areas of science and engineering fields.
However, the description of these processes has posed 
a challenge for many numerical methods because of the appearance of spurious oscillations and negative concentrations.
As a Lagrangian mesh-less method,  
SPH  has a special advantage in addressing the
diffusion problems due to the 
the benefit of dealing with the advection term.
But its application to solving anisotropic diffusion   is still limited since a robust and general 
SPH formulation is required to obtain 
accurate approximations of second derivatives.
In this paper, we modify a second derivatives model
based on the  SPH formulation to obtain a full version of Hessian matrix consisting of the Laplacian operator elements.
For solving diffusion with the diffusion coefficient being a second order tensor,
a coordinate transformation, 
is applied to conclude the anisotropic Laplacian operator. 
For solving diffusion in thin structure using the anisotropic kernel,
the anisotropic kernel and corresponding gradient functions are considered when employing the Taylor series expansion.
 
To verify the proposed SPH scheme,
firstly, the diffusion of a scalar which distributes following a pre-function within a thin structure is performed by using anisotropic resolution coupling anisotropic kernel. 
With various anisotropic ratios, 
excellent agreements with the theoretical solution are achieved. 
Then, the anisotropic diffusion of a contaminant in fluid is simulated.
The simulation results are very consistent with corresponding analytical solutions, 
showing that the present algorithm can obtain smooth solution without the spurious oscillations 
for contaminant transport problems with discontinuities, and achieve second-order accuracy.
Subsequently, we utilize this newly developed SPH formulation to tackle the problem of the fluid diffusion through a thin porous membrane and the anisotropic  transport of transmembrane potential within the left ventricle,
demonstrating the capabilities of the proposed SPH framework in solving the complex anisotropic problems.
\end{abstract}

\end{frontmatter}	

\section{introduction}   

Anisotropic diffusion problems,  
where the diffusion of some scalar quantity 
is direction dependent,
are  widely exists in the various areas of science
and engineering fields.
Examples include the  heat conduction \cite{van2014finite, biriukov2019stable}, transport in porous media  \cite{herrera2010multidimensional, herrera2009meshless, lian2021general}, 
plasma physics in fusion experiments and astrophysics \cite{ gunter2009mixed,sharma2007preserving, nishikawa2000plasma},    
petroleum engineering  \cite{aavatsmark1998discretization, crumpton1995discretisation,ertekin2001basic,mlacnik2006unstructured},
and image processing \cite{chan2001nontexture,
 	chan2003variational,karras2009new,weickert1998anisotropic}.
This directionally dependent or anisotropic diffusion
 may due to the heterogeneity of the porous media,
the turbulence in fluid, the property of materials and so on \cite{tran2016simulation}.
Addressing scenarios involve the anisotropic diffusion 
remains a significant challenge  since accurate expression 
of second derivative is required to achieve satisfying solutions.
 
As a mesh-free method,  
the smoothed particle hydrodynamics (SPH) method 
has gained increasing interest in solving various problems
\cite{liu2010smoothed, monaghan2012smoothed, 
Luo30Particle, Gotoh32On}.
Based on the principle of using particles to
discretize the computation domain, 
SPH has emerged as a versatile and 
robust method for simulating  fluid
dynamics \cite{Gotoh32On, monaghan1994simulating,	shao2006simulation,huang2019kernel}, 
solid mechanics \cite{libersky1991smooth, libersky1993high, wu2023sph,wang2021improved, lin2014efficient}, 
fluid-solid interaction \cite{antoci2007numerical, 
han2018sph, sun2019study}, 
and diffusion problems \cite{cleary1999conduction,
 monaghan2012smoothed, herrera2010multidimensional, 
herrera2009meshless, espanol2003smoothed} in recent years.

The application of SPH to diffusion problem to calculate 
the second derivative  is developed from Brookshaw  \cite{brookshaw1985method},  Cleary and Monaghan \cite{cleary1999conduction}, 
referring to isotropic heat conduction problems.
Espa$\tilde{n}$ol and Revenga  \cite{espanol2003smoothed}  
generalized the Brookshaw method to the anisotropic 
case considering the diffusivity coefficient as a tensor,
which has been the common choice in studies assessing 
anisotropic dispersion with SPH   \cite{herrera2013assessment,avesani2015smooth,alvarado2019anisotropic}.  
However, at least 50 neighbours in 2D is required to achieve 
the accuracy to within a few percent,
which is rather time consuming \cite{biriukov2019stable}.   
In recent decades, different SPH schemes for the anisotropic 
dispersive transport have been proposed \cite{herrera2013assessment, herrera2009meshless,  alvarado2019anisotropic,klapp2022approximately},
but independently of the schemes employed,
unphysical oscillations and negative values of the concentrations 
are exhibited in the numerical results when the off-diagonal 
terms of the tensorial dispersion coefficient are nonzero,  
even with large numbers of neighbors coupled to small kernel supports.
To minimize this effect, 
Avesani et al. \cite{avesani2015smooth} developed the
Moving-Least-Squares Weighted-Essentially Non-Oscillatory (MWSPH) method, 
which reduces the negative concentrations to limited values
 even with a high anisotropic dispersion tensor. 
However, the considerable gain of accuracy is 
at the expense of an increase of the computational 
cost due to the intermediate steps involved in 
the MLSWENO reconstruction and stencils construction.
Tran-Duc et al. \cite{tran2016simulation} proposed an anisotropic 
SPH approximation for anisotropic diffusion (ASPHAD), 
which first approximates the diffusion operator as 
an integral in a coordinate system where it is isotropic.
 An inverse transformation is then applied to 
 render the diffusion operator anisotropic. 
Although ASPHAD preserves the main diffusing directions,  
it is rather sensitive to particle disorder and 
reduces the degree of anisotropy in cases of anisotropic dispersion.
Another approach performs double SPH integration, 
known as two first derivative  method \cite{biriukov2019stable}.  
Even it is claimed that this TFD is mass-conservative and stable, 
non-physical oscillations may occur in the vicinity of sharp gradients \cite{perez2024smoothed}.  
Consequently, 
mathematically reliable and accurate approximations of 
second derivative  are necessary when solving the anisotropic diffusion process.

Our research introduces an innovative application of 
the second derivative models, including the cross-derivatives, from reference \cite{asai2023class}, 
to anisotropic diffusion within the SPH framework. 
This full expression of Hessian matrix allows  for 
the accurate incorporation of the anisotropic diffusion tensor,
ensuring second-order accuracy in Laplacian computations. 
By approximating the diffusion operator initially 
in a coordinate system where the 
diffusion operator is isotropic, 
the transformation tensor representing a coordinate rotation and a coordinate scaling,
which associated with the positive diffusion coefficient tensor,
is applied to achieve the final form of anisotropic SPH approximation.
Furthermore, anisotropic characteristics also exist in diffusion problems within thin structures modeling by utilizing anisotropic kernel within the SPH framework, since anisotropic resolution coupling anisotropic kernel can effectively reduce the computation time.
In addressing this kind of problem, 
the full representation of the second-derivative model
allows for a clear expression of the anisotropic kernel  and gradient function in the formalism,
providing an enhanced accuracy with efficiency guaranteed.  

In this paper, 
we detail the theoretical underpinnings of our method, 
its implementation within the SPH framework. 
The accuracy is examined in relation to the truncation error in pre-set function.
Through a series of benchmark tests and practical applications, 
we demonstrate the reliable application of this approach in simulating anisotropic diffusion process and diffusion in thin structure problems using anisotropic kernels.

\section{Kinematics and governing equations}  
\subsection{Fluid diffusion coupling elastic film deformation}
Considering  a fluid-structure interaction model accounting 
for the simultaneous diffusion 
of fluid through a thin porous solid,
resulting in heightened fluid pressure and 
deformation of the solid structure \cite{zhao2013modeling}. 
\begin{figure*}[htbp]
	\centering
	\includegraphics[trim = 40mm 40mm 60mm 60mm, clip,width=0.5\textwidth]{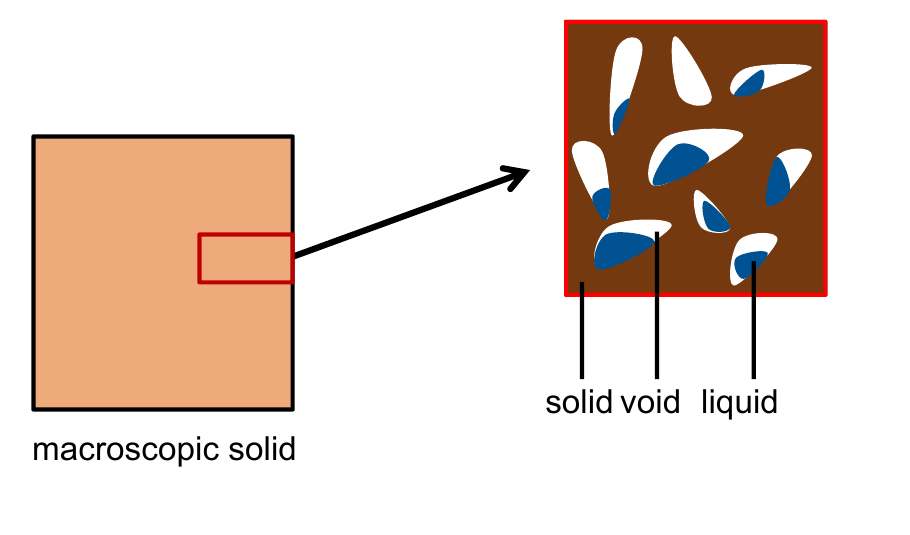}
	\caption{Partially saturated porous medium.}
	\label{continum-sketch}
\end{figure*}
In this model, the heterogeneous body
is considered as a continuous solid medium containing 
uniformly distributed small voids with a homogeneous porosity 
denoted by ${c}$.
Upon interaction with a fluid,
the fluid permeates into these small pores 
and diffuses within the medium, 
driven by the fluid concentration gradient,
resulting in the creation of a mixture 
consisting of solid and fluid components, as illustrated in Figure \ref{continum-sketch}. 

\subsubsection{Mass and momentum equations}
With a porosity $ {c}$ and fluid saturation level $ \widetilde{c}$,
as defined in reference  \cite{zhao2013modeling}, 
the locally effective fluid density $ \rho^l $  
can be expressed as 
\begin{equation}
	\label{fluid_density}
	\rho^l = \rho^L \widetilde{c},
\end{equation} 
where
$ \rho^L$ is the 
fluid density. 
For a porous solid partially-saturated by fluid,  
the total linear momentum $\mathbf{M}$ in 
the region $\mathcal{R}$ is the sum of fluid momentum 
and solid momentum
\begin{equation}
	\label{P_equation}
	\mathbf{M}= \rho \mathbf{v}=\rho^l \mathbf{v}^{l}+ \rho^s \mathbf{v}^{s},
\end{equation}
where  $ \rho $ and $\mathbf{v}$ are the total density and velocity respectively, 
$\mathbf{v}^l$ the velocity of fluid,
$ \rho^s $ and $\mathbf{v}^s$ the density and velocity of dry porous solid. 
Due to the difference between  
$\mathbf{v}^l$ and  $\mathbf{v}^s$, 
the fluid flux $\mathbf{q}$ on the element 
boundary $\partial V$,
considering a representative volume element $dV$, 
can then be expressed as
\begin{equation}
	\label{defination_q}
	\mathbf{q} = {\rho^l} (\mathbf{v}^{l}-\mathbf{v}^{s}).
\end{equation} 
Within an element $ dV$ of the mixture, 
the  momentum $\mathbf{M}$ 
is determined by both the stress exerted on the element
and  the fluid flux of linear momentum $\mathbf{v}^{l} \otimes \mathbf{q}$ 
on the boundary $ \partial V $,  
where the symbol $\otimes$ means the outer 
product of two vectors or tensors.  
It follows that 
the conservation of total linear 
momentum of the mixture can be expressed as
\begin{equation}
	\frac{d\mathbf{M}}{d t} = \nabla \cdot \boldsymbol{\sigma}- \nabla \cdot\left(\mathbf{v}^{l} \otimes \mathbf{q}\right),
	\label{totalmomentumupdate}
\end{equation}
where $\boldsymbol{\sigma}$ represents 
the Cauchy stress in the mixture
acting on the solid. 
$ \boldsymbol{\sigma} $ is determined by 
Cauchy stress $\boldsymbol{\sigma}^s$ due to deformation 
and the pressure stress $\boldsymbol{\sigma}^l$ 
resulting from the presence of the fluid phase, 
which is elaborated in Appendix \ref{appendixB3}.

\subsubsection{Fick's law} 
In a partially saturated  solid, 
variations in fluid saturation drive fluid movement 
regions with higher fluid fraction to those with lower fractions, 
and the resulting flux follows the Fick's law as
\begin{equation}
	\label{grad_q}
	\mathbf{q} = -D \rho ^l \nabla \widetilde{c},
\end{equation}
indicating that the fluid flux is proportional to the diffusivity $D$, 
the effective fluid density $\rho^{l} $, 
as well as  the gradient of the fluid saturation $ \widetilde{c} $.
Consequently,
the time derivative of fluid mass within an element $dV$ 
is attributed to the fluid flux $\mathbf{q}$ 
across the element boundary $\partial V$, considering Eq. 
\eqref{fluid_density}, written as 
\begin{equation}
	\frac{d \rho^{l}}{d t}= - \nabla \cdot \mathbf{q} =  D \rho ^L  \nabla  \cdot ( \widetilde{c} ~ \nabla \widetilde{c}) = \frac{1}{2} D \rho ^L  \nabla^2   \widetilde{c}^2.
	\label{fluid_mass_der}
\end{equation}
where the second-derivative operator $\nabla^2$ defines 
the time derivative of fluid density.

\subsection{Cardiac function} 
Following  Chi's algorithm \cite{zhang2021integrative},
a model for simulating the principle aspects of cardiac function,
including cardiac electrophysiology, passive mechanical response and electromechanical coupling is proposed.
In this scenario,  
the anisotropic transmembrane potential propagation behavior of the myocardium is
simulated by using our present anisotropic SPH diffusion discretization algorithm to achieve higher accuracy.
\subsubsection{Kinematics and governing equations} 
\begin{figure*}[htbp]
	\centering
	\includegraphics[trim = 0mm 0mm 0mm 0mm, clip,width=0.65\textwidth]{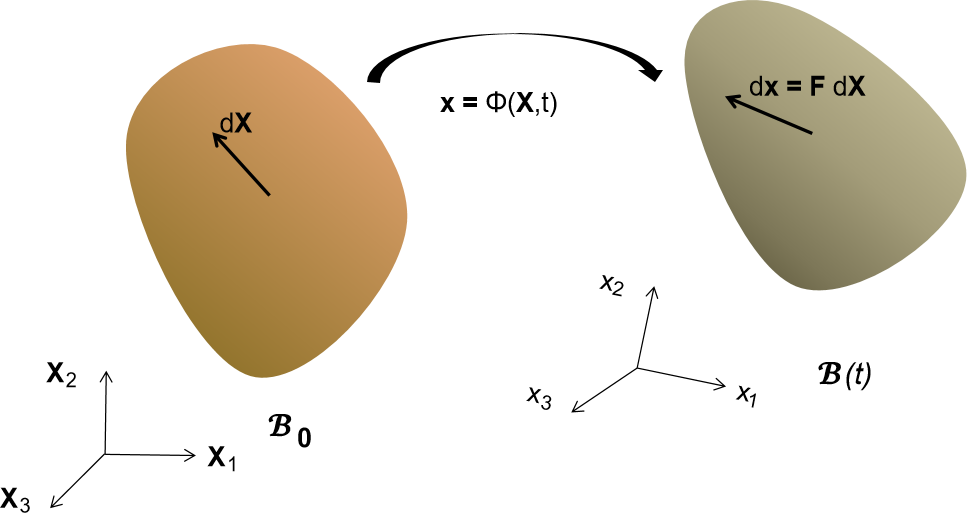}
\caption{Finite deformation process on a body $ \mathcal{B} $.}
\label{figure_solid_deformation}
\end{figure*}

In this section, we firstly concisely introduce
the fundamental physical concepts of solid deformation, 
along with the relevant notations and 
symbols within the total Lagrange framework. 
Our analysis considers a solid body $\mathcal{B}$ 
that occupies two regions: $\mathcal{R}_0$ and $\mathcal{R}$, 
representing the body configurations at
time $t_0$ (where $t=0$) and $t$, respectively.
In the initial configuration $\mathcal{R}_0$,
the position vector of a material point 
is denoted by $\mathbf{X}\in \mathcal{R}_0$,
while in the current configuration $\mathbf{x}\in \mathcal{R}$. 
The motion of the solid body, 
represented by the invertible mapping $\phi$, 
transforms a material point $\mathbf{X}$ 
to its corresponding vector 
$\mathbf{x}=\phi(\mathbf{X},t)$,
as  shown in Figure \ref{figure_solid_deformation}.
Accordingly, 
the Lagrangian velocity of a material point 
is given by $\mathbf{v}(\mathbf{X},t) = \frac{d\phi(\mathbf{X},t)}{dt}$.
The deformation gradient $	\mathbf{F}$, 
which characterizes the deviation of 
a material point from its initially 
undeformed position to its deformed position, 
can be calculated from the displacement 
vector $\mathbf{u} = \mathbf{x} - \mathbf{X}$
by the following equation: 
\begin{equation} \label{eq:deformationtensor-displacement}
	\mathbf{F} = \frac{d\mathbf{x}}{d\mathbf{X}} = \nabla^{0} {\mathbf{u}}  + \mathbf{I},
\end{equation}
where $\nabla^{0}$ denotes the gradient operator defined 
in the initial reference configuration, 
$\mathbf{I}$ the unit matrix.
 The corresponding  
Jacobian determinant term $ J = $ det($ \mathbf{F} $) 
indicates the local volume gain  $ J > 1 $ or loss $ J < 1 $.
The governing equations of solid deformation 
within the total Lagrange framework are derived as
\begin{equation}\label{eq:mechanical-mom}
\begin{cases}
\rho =  {\rho^0} \frac{1}{J}  \\
\rho^0 \frac{\text{d} \mathbf{v}}{\text{d} t}  =  \nabla^{0} \cdot \mathbf{P}^T
\end{cases},
\end{equation}
where $ \rho $ and $\rho_0$ are the densities 
in the  current configuration $ \mathcal{R} $
and the initial configuration $\mathcal{R}_0$, respectively,
$ \mathbf{v} $ the velocity
and  $\mathbf{P}$ the first 
Piola-Kirchhoff stress tensor.

Considering a coupled system of cardiac electrophysiology and microelectronics 
following the total Lagrangian formulations \cite{zhang2021integrative}. 
The cardiac electrophysiology runs the principle of the evolution of 
the transmembrane potential $ V_m $, expressed as
\begin{equation}
C_m \frac{d V_m}{d t}=  \nabla^0 \cdot (\mathbf{D} \nabla^0 V_m ) +  I_{ion},
\label{evolution-transmembrane}
\end{equation}
where $ C_m $ is the capacity of the membrane cell and
$ I_{ion} $ the ionic current. 
The conductivity tensor is defined by 
\begin{equation}
\mathbf{D}= d^{iso}\mathbf{I} +d^{ani} \mathbf{f^0} \otimes \mathbf{f^0}
\end{equation}
where $ d^{iso} $ represents the isotropic contribution and $ d^{ani} $ the anisotropic contribution to account for conductivity along the fiber direction $\mathbf{f^0} $.
To close equations, the constitutive laws for the ionic current  $I_{ion}$ and the first Piola–Kirchhoff stress $\mathbf{P}$ are required.
  
\subsubsection{Constitutive equations} 
Referring to the approaches described in Refs.\cite{ franzone2014mathematical, aliev1996simple},
a so-called reduced-ionic model is implemented 
for the ionic current. 
Specifically, we utilize the Aliev–Panfilov model, where
$I_{ion}(V_m, w)$ is considered as a function of the transmembrane potential $V_m$ and the gating variable $w$,
which indicates the percentage of open channels per unit area of the membrane. 
Particularly suitable for applications where 
electrical activity of the heart is the main interest, 
the Aliev–Panfilov model reads \cite{aliev1996simple}
\begin{equation}
\left\{\begin{array}{l}
	I_{i o n}\left(V_{m}, w\right)=-k V_{m}\left(V_{m}-a\right)\left(V_{m}-1\right)-w V_{m} \\
	\frac{\mathrm{d} w}{\mathrm{~d} t}=g\left(V_{m}, w\right)=\epsilon\left(V_{m}, w\right)\left(-k V_{m}\left(V_{m}-b-1\right)-w\right)
\end{array},\right.
	\end{equation}
where $\epsilon\left(V_{m}, w\right)=\epsilon_{0}+\mu_{1} w /\left(\mu_{2}+V_{m}\right) \text { and } k, a, b, \epsilon_{0}, \mu_{1} \text { and } \mu_{2}$ are constant parameters.
Note that dimensionless variables are employed here.
The actual transmembrane potential $\tilde V_m$ 
measured in millivolts (mV) and time measured in milliseconds (ms) 
can be derived through  scaling transformations,
expressed as follows
\begin{equation}
\tilde V_m = 100 V_m - 80.
\end{equation}

\subsubsection{Cardiac electromechanics} 
The first Piola–Kirchhoff stress $\mathbf{P}$ in Eq. \eqref{eq:mechanical-mom}  can be decomposed  
into passive $\mathbf{P}_{passive}$ and active response $\mathbf{P}_{active}$, 
expressed as \cite{nash2004electromechanical}
\begin{equation}
\mathbf{P} = \mathbf{P}_{passive} +\mathbf{P}_{active}.
\end{equation}
The passive first Piola–Kirchhoff stress $ \mathbf{P}_{passive} $,
characterizes the stress necessary to achieve a specified deformation of the passive myocardium,
computed as 
\begin{equation}
 \mathbf{P}_{passive}=  \mathbf{F}\mathbf{S},
\end{equation}
where by adopting the Holzapfel–Ogden model \cite{holzapfel2009constitutive}  and 
incorporating the anisotropic characteristics of the myocardium,
the second Piola–Kirchhoff stress $\mathbf{S}$ is defined as
\begin{equation}
	\begin{aligned}
		\mathbf{S} & =a \exp \left[b\left(I_{1}-3\right)\right] \mathbf{I}+\{\lambda \ln J-a\} \mathbf{C}^{-1} \\
		& +2 a_{f}\left(I_{f f}-1\right) \exp \left[b_{f}\left(I_{f f}-1\right)^{2}\right] \mathbf{f}^{0} \otimes \mathbf{f}^{0} \\
		& +2 a_{s}\left(I_{s s}-1\right) \exp \left[b_{s}\left(I_{s s}-1\right)^{2}\right] \mathbf{s}^{0} \otimes \mathbf{s}^{0} \\
		& +a_{f s} I_{f s} \exp \left[b_{f s}\left(I_{f s}\right)^{2}\right]\left(\mathbf{f}^{0} \otimes \mathbf{s}^{0}+\mathbf{s}^{0} \otimes \mathbf{f}^{0}\right)
	\end{aligned}
\end{equation}
where  a, b, $a_{f}$, $b_{f}, a_{s}, b_{s}, a_{f s} $ and  $b_{f s}$  are eight positive material constants, 
with the $ a $ parameters having dimension of stress and $  b $  parameters being dimensionless. Here,  $\lambda$  is Lamé parameter and $\mathbf{C}$ the
left Cauchy–Green deformation tensor, 
expressed as
\begin{equation}
\mathbf{C} =  \mathbf{F}^T \cdot \mathbf{F}
\end{equation}
which  possesses principal invariants, given by 
\begin{equation}
I_{1}=\operatorname{tr} (\mathbf{C}), \quad I_{2}=\frac{1}{2}\left[I_{1}^{2}-\operatorname{tr}\left(\mathbf{C}^{2}\right)\right], \quad I_{3}=\operatorname{det}(\mathbf{C})=J^{2}.
\end{equation}
Additionally,
there are three other independent invariants 
that account for directional preferences:
\begin{equation}
I_{ff}=\mathbf{C}: \mathbf{f}^{0} \otimes \mathbf{f}^{0}, \quad I_{s s}=\mathbf{C}: \mathbf{s}^{0} \otimes \mathbf{s}^{0}, \quad I_{f s}=\mathbf{C}: \mathbf{f}^{0} \otimes \mathbf{s}^{0},
\end{equation}
where  $ \mathbf{f}^{0} $ and  $ \mathbf{s}^{0} $
represent the undeformed myocardial fiber and sheet unit directions, respectively.
The structure-based invariants $ I_{ff}$ and 
$ I_{s s}$ correspond to the isochoric fiber 
and sheet stretch squared, 
representing the squared lengths of the deformed fiber and sheet vectors,
i.e., $ \mathbf{f}=\mathbf{F} \mathbf{f}^{0} $  and  $ \mathbf{s}=\mathbf{F} \mathbf{s}^{0} $,
and $ I_{f s}$ signifies the fiber-sheet shear \cite{holzapfel2009constitutive}.

The active component $\mathbf{P}_{active}$, 
represents the tension produced by the depolarization
associated with the propagating transmembrane potential, 
which is  aligned with the fiber direction and is represented mathematically as \cite{nash2004electromechanical} 
 \begin{equation}
 \mathbf{P}_{active} ={T}_{a} \mathbf{F}   \mathbf{f}^{0} \otimes \mathbf{f}^{0} , 
 \end{equation}
with $ {T}_{a} $ being the active cardiomyocite contraction stress,
which evolves according to an ordinary differential equation (ODE) given by 
 \begin{equation}
\frac{\mathrm{d} T_{a}}{\mathrm{~d} t}=\epsilon\left(V_{m}\right)\left[k_{a}\left(V_{m}-V_{r}\right)-T_{a}\right],
 \end{equation}
where parameters $k_{a}$  and $V_{r}$ 
control the maximum active force, the resting transmembrane potential and the activation function \cite{wong2011computational}
\begin{equation}
\epsilon\left(V_{m}\right)=\epsilon_{0}+\left(\epsilon_{\infty}-\epsilon_{-\infty}\right) \exp \left\{-\exp \left[-\xi\left(V_{m}-\bar{V}_{m}\right)\right]\right\},
\end{equation}
where $\epsilon_{-\infty}$ and  $ \epsilon_{\infty} $ are the limiting values  as $ V_{m} \rightarrow-\infty $
 and $  V_{m} \rightarrow \infty $.
The parameters $ \bar{V}_{m} $  and $\xi$ represent the phase shift and the transition slope, respectively, 
ensuring a smooth activation of the muscle contraction.

\section{Theory of ASPH}  
Derived from smoothed particle hydrodynamics (SPH) method,
the adaptive smoothed particle hydrodynamics (ASPH) method 
is predicated on an integral formulation,  
wherein pertinent physical quantities are approximated 
via the integration of neighboring particles, 
but the kernel function being evolved is adaptive.   
Spherical kernels of radius given by a scalar smoothing length in SPH
is replaced by an anisotropic smoothing 
involving ellipsoidal kernels in ASPH. 
The real position vector is generalized to a normalized vector.  
This can be portrayed as a localized, linear shift of coordinates, 
which is described by a transformation tensor $ \mathbf{G} $. 
This shift transforms the coordinates to a system where 
the anisotropic volume changes appear uniform in all directions. 
 
\subsection{Fundamental and theory of ASPH} 
Following Ref. \cite{owen1998adaptive},   
the real position vector $ \boldsymbol{r}$,
is generalized to a normalized form 
$\boldsymbol \eta$  in ASPH through a linear 
coordinate transformation tensor  $\mathbf{G}$.  
This transformation is expressed as $\boldsymbol{\eta} = \mathbf{G}\boldsymbol{r}$,
resulting in the representation of the kernel function
$W(\boldsymbol{\eta}) = W(\mathbf{G}\boldsymbol{r})$. 
In contrast to the isotropic kernel, 
the normalization undergoes a change:
SPH:  $\boldsymbol{\eta} = \boldsymbol{r} / h \rightarrow $  
ASPH:  $ \boldsymbol{\eta} =\mathbf{G} \boldsymbol{r}$.
It is evident that SPH can be regarded as a 
degenerate case of ASPH, where the tensor 
$\mathbf{G}$ becomes diagonal with a constant component of $1/h$.

Under such notations,  the gradient of the kernel function 
$ \nabla W (\boldsymbol{\eta})$ can be expressed as \cite{owen1998adaptive}
\begin{equation}
\label{kernel-gradient}
\nabla W (\boldsymbol{\eta}) = \nabla W^{\prime}(\mathbf{G} \boldsymbol{r}) =
\dfrac{\partial W(\mathbf{G} \boldsymbol{r})}{\partial \boldsymbol{r}} = 
\dfrac{\partial \boldsymbol{\eta}}{\partial \boldsymbol{r}} \dfrac{\partial W}{\partial \boldsymbol{\eta}}=
\mathbf{G}
\frac{\boldsymbol{\eta}}{\eta} \dfrac{\partial W}{\partial \boldsymbol{\eta}},
\end{equation}

Incorporating the particle summation,
one has the approximation of derivative of 
a variable field $f$ at particle $i$ in a weak form by
\begin{equation}
\label{eq:gradsph-weak}
\nabla f_i =  \nabla f_i -f_{i}\nabla 1 \approx  -2 \sum_{j} ||\mathbf{G}||  V_j \widetilde{f}_{ij}  \nabla_i W_{ij},  
\end{equation}
and a strong form as
\begin{equation}
	\label{eq:gradsph-strong}
	\nabla f_i = f_{i}\nabla 1 + \nabla f_i \approx  \sum_{j} ||\mathbf{G}||  V_j {f}_{ij}    \nabla_i W_{ij}, 
\end{equation}
where $ V_j $ is the volume of the $j$-th neighboring particle,  $\widetilde{f}_{ij} = (f_{i} + f_{j} ) /2$ is the inter-particle average value,
 $f_{ij} = f_{i} - f_{j}$ is the inter-particle difference value. 

Considering a spatially varying smoothing tensor,   
the kernel function needs to be symmetrized to
ensure the conservation of quantities such as linear momentum.
The symmetrization of kernel function $ W_{ij} $   
and the gradient of which between two  particles $ i $ and $ j $ 
can be implemented as the averaged formalism as
\begin{equation}	 
W_{ij}=\frac{1}{2} (W(\boldsymbol\eta_{i,ij}) + W(\boldsymbol\eta_{j,ij}) ),  \quad 
\nabla W_{ij}=\frac{1}{2} (\nabla W(\boldsymbol\eta_{i,ij}) + 	\nabla W(\boldsymbol\eta_{j,ij})) ,
\end{equation}
where 
\begin{equation}	 
\boldsymbol\eta_{i,ij} = \mathbf{G}_i \boldsymbol{r}_{ij}, \quad \boldsymbol\eta_{j,ij} =  \mathbf{G}_j \boldsymbol{r}_{ij}.
\end{equation}
 
\subsection{Kernel function with nonisotropic smoothing}
 In the following cases, we use the Wenland kernel function and its first derivative which can be further written to match the nonisotropic ellipsoidal smoothing kernel as
 \begin{equation}
 	\label{kernel-asph}
 	W^{v-D}(\eta)=A^{v-D}\left\{\begin{array}{ll}
 		(1- \frac{\eta}{2}) ^{4}(1+2\eta), & 0 \leq \eta \leq 2  \\
 		0, & \eta >2
 	\end{array}\right.
 \end{equation}
 
 \begin{equation}
 	\nabla W^{v-D}(\eta)=A^{v-D} \mathbf{G}  \frac{\boldsymbol{\eta}}{\eta}\left\{\begin{array}{ll}
 		-5 \eta (1- \frac{\eta}{2}) ^{3}, & 0 \leq \eta\leq 2 \\
 		0, & \eta>2
 	\end{array}\right.
 \end{equation}
 where $ v $ means the dimension and 
 \begin{equation}
 	\label{kernel-gradient-asph}
 	A^{1-D}=\frac{3}{4}||\mathbf{G}||, \quad A^{2-D}=\frac{7}{4 \pi}||\mathbf{G}||, \quad A^{3-D}=\frac{21}{16\pi}||\mathbf{G}||.
 \end{equation}
 Benefiting from  the tensor $ \mathbf{G}  $, 
 the displacement between two particles 
 is mapped to the generalized position vector $ \boldsymbol{\eta} $, 
 the norm of which is compared with the cutoff radius to 
 calculate the kernel function and kernel gradient value.
 Using normalized position vector $ \boldsymbol{\eta} $
 rather than $\mathbf{r}/h $ in the discretization of quantities, 
 the expression of dynamic equations in SPH and ASPH are identical.
 
\subsubsection{Transformation tensor $\mathbf{G}$}
Defined as a linear transformation that maps 
from real position space ($ \mathbf{r} $) to 
normalized position space ($ \boldsymbol\eta$),
$\mathbf{G}$ is determined by the coupling of the geometrically scaling transformation 
and the rotational transformation, 
involving the smoothing lengths in different directions 
and the rotation angle of the axes deviated from the real frame.
Detailed information can be found in \ref{appendix-G}   and the reference paper \cite{owen1998adaptive}. 

\section{First and second derivatives in ASPH}
\subsection{Correction of the first derivatives}
To discretize the solid mechanics, we employ the initial 
undeformed configuration as the reference.  
First, aiming to restore the 1st-order consistency, 
the configuration of particle $ i $ is corrected with a 
tensor $ \mathbf{B} $ in the total Lagrangian formalism, expressed as \cite{vignjevic2006sph, randles1996smoothed} 
\begin{equation}
\label{F-consitency}
\mathbf{F}^0_i  = \left( \sum_j V_j \left( \mathbf{r}^0_j - \mathbf{r}^0_i \right) \otimes \nabla^0_i W_{ij}  \right) \mathbf{B}^0_i = 	\mathbf{I},
\end{equation}
where
$\boldsymbol{r}^0_i$ and $\boldsymbol{r}^0_j$ denote the positions 
of particles $i$ and $j$ in the reference configuration.
Equivalently, 
\begin{equation}
\label{B-consitency}
\mathbf{I} =   \sum_j V_j \left( \mathbf{r}^0_j - \mathbf{r}^0_i \right) \otimes (\nabla^{0^T}_iW_{ij} \mathbf{B}^0_i) =  \sum_j V_j \left( \mathbf{r}^0_j - \mathbf{r}^0_i \right) \otimes (\mathbf{B}^{0^T}_i \nabla^{0}_iW_{ij} )^T.
\end{equation}
The gradient correction $ \mathbf{B} $ is operated over $\nabla^{0} $ to correct the kernel. 
We define 
\begin{equation}
	\label{kernel_gradient_correction}	 
\tilde{\nabla}^{0} = \mathbf{B}^{0^T}_i \nabla^{0}  ,
\end{equation}
where the symbol  $ \tilde{\nabla}^{0}_i $ represents the corrected approximation of the differential operator with respect to the initial material coordinates.
From Eq. \ref{F-consitency}, involving the tensor $\mathbf{G} $,
the correction matrix $\mathbf{B}^0$  of particle $ i $ in ASPH
can be consequently calculated as:
\begin{equation} \label{eq:sph-correctmatrix}
\mathbf{B}^0_i = \left( \sum_j V_j \left( \boldsymbol{r}^0_j - \boldsymbol{r}^0_i \right) \otimes \nabla^0_i W_{ij} \right) ^{-1}
= \left( \sum_j V_j \left( \boldsymbol{r}^0_j - \boldsymbol{r}^0_i \right) \otimes ( \mathbf{G}_{i}	\frac{\boldsymbol{\eta_{ij}}}{\eta_{ij}}   \frac{\partial W} {\partial \eta _{ij} } )\right) ^{-1}.
\end{equation}
In total Lagrangian formulation,
the neighborhood of particle $i$ is 
defined in the initial configuration,  
and this set of neighboring particles remains 
fixed throughout the entire simulation.
Therefore, $\mathbf{B}^0_i $ is 
computed only once under the initial reference configuration.
 
\subsection{Second derivatives coupling with nonisotropic resolution}

\begin{figure*}[htbp]
	\centering
	\begin{subfigure}[b]{0.30\textwidth}
		\includegraphics[trim =0mm 0mm 1mm 0mm, clip,width=0.95\textwidth]{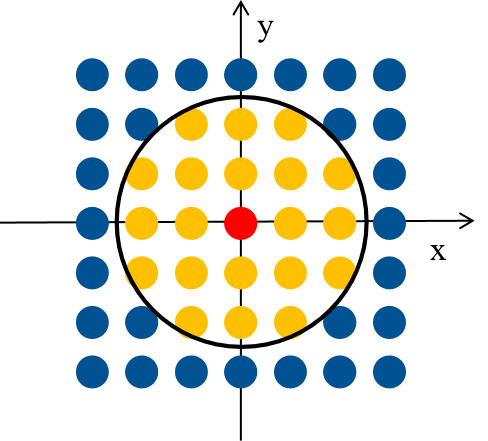}
		\caption {}
	\label{ken-iso}	
\end{subfigure}
\begin{subfigure}[b]{0.60\textwidth}
		\includegraphics[trim =0mm 0mm 1mm 0mm, clip,width=0.95\textwidth]{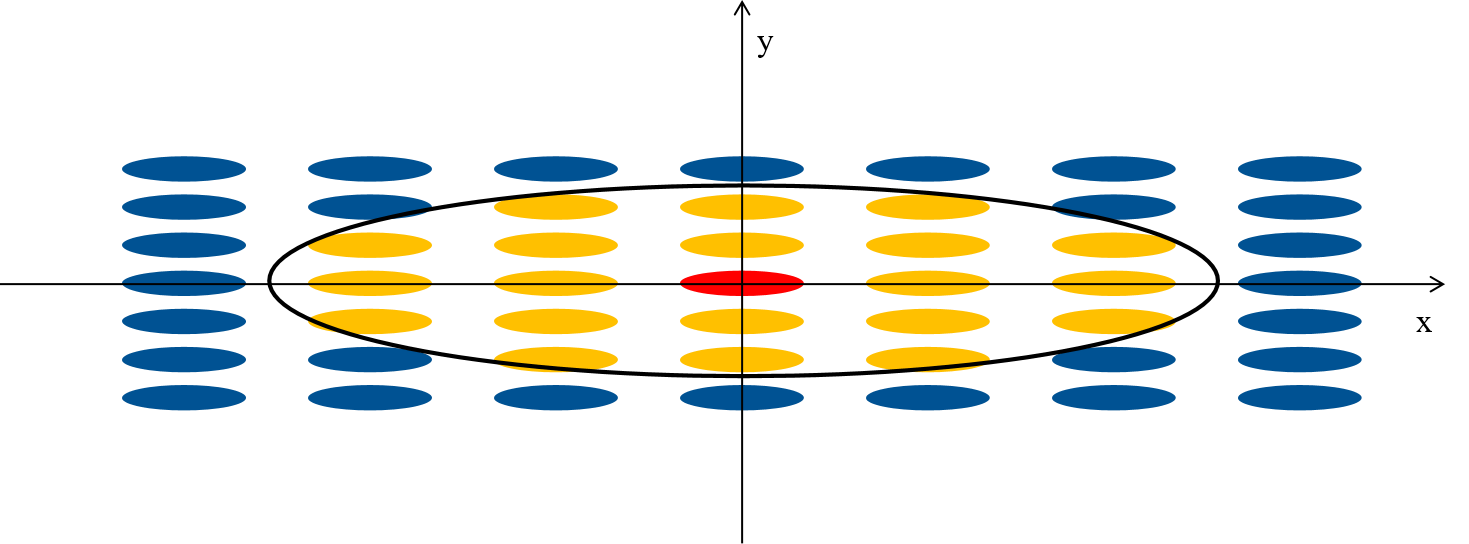}
		\caption {}
		\label{ken-aniso-par-aniso}
\end{subfigure}
	\caption{(a) Isotropic with spherical smoothing kernel and (b) anisotropic resolutions with  elliptical smoothing kernel.}
	\label{ASPHken-aniso-par-aniso}
\end{figure*}

Considering the diffusion in a
thin structure, 
with large aspect ratio,
it is favorable to discretize 
the computation domain with nonisotropic resolution to reduce total particle number,
as shown in Figure. \ref{ASPHken-aniso-par-aniso}.
To cope with this nonisotropic resolution,
we apply a nonisotropic kernel, 
where an ellipsoid smoothing 
replaces the normal sphere one,
as shown in Figure. \ref{ken-aniso-par-aniso}.

With the nonisotropic resolution, presenting the Taylor series expansion of function 
at $x_i$ in $ \boldsymbol{x}$-coordinate which is nonisotropic,
there is 
\begin{equation}
\phi_{i} = \phi_j + \frac{\partial \phi }{\partial \boldsymbol{x} } \cdot \boldsymbol{r}_{ij} + \frac{1}{2}\boldsymbol{r}_{ij} \cdot \frac{\partial^2 \phi_i }{\partial \boldsymbol{x}^2 } \boldsymbol{r}_{ij}
+R_{(3)}
= \phi_j + \frac{\partial \phi }{\partial \boldsymbol{x} } \cdot \boldsymbol{r}_{ij}+R_{(2)},
\end{equation}
where $ \phi_j $ means the value of $\phi $ at particle $ j $ and $ \boldsymbol{r}_{ij} $ means the particle distance vector
between particle $ i $ and particle $ j $,
the  $R_{(2)}$ and $R_{(3)}$ represent 
the truncation error of first
and second order, respectively.
Consequently,
\begin{equation}
\phi_{ij} := \phi_i - \phi_j
 =\nabla \phi_i \cdot \boldsymbol{r}_{ij} + R_{(2)},
 \label{phi_ij}
\end{equation}
where 
\begin{equation}
\begin{aligned}
R_{(2)} & \approx \frac{1}{2} r_{i j} \cdot \frac{\partial^{2} \phi_{i}}{\partial x^{2}} r_{i j} \\
& =\frac{1}{2} \sum_{ {\mathrm{i}}=1}^{\text {ndim }} \sum_{ {\mathrm{j}}=1}^{\text {ndim}} r^{\mathrm{i}}_{i j} r^{\mathrm{j}}_{i j}\frac{\partial^{2} \phi_{i}}{\partial x^{\mathrm{i}} \partial x^{\mathrm{j}}}.
\label{second-order-error}
\end{aligned}  
\end{equation}
where the term with superscript $  r^{\mathrm{i}}_{i j} $ and $  r^{\mathrm{j}}_{i j} $  mean the components of the vector  $ \boldsymbol{r}_{ij} $.

Following Asai Mitsuteru \cite{asai2023class},
in total Lagrangian formulation,
the term $\nabla \phi$ in the right hand of Eq. \eqref{phi_ij} at particle $ i $ can be approximated in SPH
 as
\begin{equation}
\begin{aligned}
\langle\nabla \phi_{i}\rangle_{(2)} & =\bigcup_{j \in \mathbb{S}^{i}} \frac{m_{j}}{\rho_{j}} \tilde{\nabla}^0_{i(\boldsymbol{x})} W_{i j}\left(\phi_{i}-\phi_{j}-  R_{(2)}  \right) \\
& \approx \bigcup_{j \in \mathbb{S}^{i}} \frac{m_{j}}{\rho_{j}} \tilde{\nabla}^0_{i(\boldsymbol{x})} W_{i j}\left(\phi_{i}-\phi_{j}-\frac{1}{2} \sum_{ {\mathrm{i}}=1}^{\text {ndim }} \sum_{ {\mathrm{j}}=1}^{\text {ndim}}  r^{\mathrm{i}}_{i j} r^{\mathrm{j}}_{i j} \frac{\partial^{2} \phi_{i}}{\partial x^{\mathrm{i}} \partial x^{\mathrm{j}}}\right),
\label{gradient-firstcase}
\end{aligned}
\end{equation}
satisfying the second order accuracy.
Note that $\tilde{\nabla}^0_{i(\boldsymbol{x})} W_{i j}   $ is the gradient kernel approximated in the nonisotropic ${\boldsymbol{x}}$-coordinate 
with kernel correction  matrix $\mathbf{B}^0$ applied, written as
$\tilde{\nabla}^0_{i(\boldsymbol{x})} W_{i j} = \mathbf{B}^{0^T} \mathbf{G}_{i}	\frac{\boldsymbol{\eta_{ij}}}{\eta_{ij}}   \frac{\partial W} {\partial \eta _{ij} }$. 
Consequently, by applying Eq.  \eqref{gradient-firstcase} and
Eq.  \eqref{second-order-error} to  Eq. \eqref{phi_ij},
\begin{equation}
\begin{aligned}
 \phi_i - \phi_j
 &\approx \boldsymbol{r}_{ij} \cdot \langle\nabla \phi_{i}\rangle_{(2)}  +
\frac{1}{2} \sum_{ {\mathrm{i}}=1}^{\text {ndim }} \sum_{ {\mathrm{j}}=1}^{\text {ndim}} r^{\mathrm{i}}_{i j} r^{\mathrm{j}}_{i j}\frac{\partial^{2} \phi_{i}}{\partial x^{\mathrm{i}} \partial x^{\mathrm{j}}}\\
& \approx \boldsymbol{r}_{ij} \cdot
 \left \{\bigcup_{k \in \mathbb{S}^{i}} \frac{m_{k}}{\rho_{k}} \tilde{\nabla}^0_{i(\boldsymbol{x})} W_{i k}\left(\phi_{i}-\phi_{k}-\frac{1}{2} \sum_{ {\mathrm{i}}=1}^{\text {ndim }} \sum_{ {\mathrm{j}}=1}^{\text {ndim}}  r^{\mathrm{i}}_{ik} r^{\mathrm{j}}_{i k} \frac{\partial^{2} \phi_{i}}{\partial x^{\mathrm{i}} \partial x^{\mathrm{j}}}\right) \right\}\\
 & \ \qquad \qquad \qquad \qquad +
  \frac{1}{2} \sum_{ {\mathrm{i}}=1}^{\text {ndim }} \sum_{ {\mathrm{j}}=1}^{\text {ndim}} r^{\mathrm{i}}_{i j} r^{\mathrm{j}}_{i j}\frac{\partial^{2} \phi_{i}}{\partial x^{\mathrm{i}} \partial x^{\mathrm{j}}}.
  \end{aligned}
\end{equation}
It can be transformed to

\begin{equation}
\left[\begin{array}{l}
\left\langle\frac{\partial^{2} \phi_i}{\left(\partial x^{1}\right)^{2}}\right\rangle_{(2)}  \\
\left\langle\frac{\partial^{2} \phi_i}{\left(\partial x^{2}\right)^{2}}\right\rangle_{(2)}  \\
\left\langle\frac{\partial^{2} \phi_i}{\left(\partial x^{3}\right)^{2}}\right\rangle_{(2)}  \\
2\left\langle\frac{\partial^{2} \phi_i}{\partial x^{1} \partial x^{2}}\right\rangle_{(2)}  \\
2\left\langle\frac{\partial^{2} \phi_i}{\partial x^{2} \partial x^{3}}\right\rangle_{(2)}  \\
2\left\langle\frac{\partial^{2} \phi_i}{\partial x^{3} \partial x^{1}}\right\rangle_{(2)} 
\end{array}\right]
=  2 \mathbf{M}^{-1} \ \bigcup_{j \in \mathbb{S}^{i}}\left[\frac{m_{j}}{\rho_{j}} \frac{\boldsymbol{r}_{i j} \cdot \tilde{\nabla}^0_{\boldsymbol{x}} W_{i j}}{\left|\boldsymbol{r}_{i j}\right|^{4}}\left\{\phi_{i j}-\boldsymbol{r}_{i j} \cdot\left(\bigcup_{k \in \mathbb{S}^{i}} \frac{m_{k}}{\rho_{k}} \phi_{i k} \tilde{\nabla}^0_{\boldsymbol{x}} W_{i k}\right)\right\} \mathbf{s}_{i j}\right] ,
\label{2d-universal_formation}
\end{equation}
where $\mathbf{s}_{i j} = [({r^{1}_{ij}})^2, \  ({r^{2}_{ij}})^2, \ ({r^{3}_{ij}})^2, \ 
 r^{1}_{ij}r^{2}_{ij}, \ r^{2}_{ij}r^{3}_{ij}, \  
 r^{3}_{ij}r^{1}_{ij} ]^T $, 
 and 
 \begin{equation}
\mathbf{M}_i^\mathrm{(i,j,k,l)} = 
 \bigcup_{j \in \mathbb{S}^{i}} \frac{m_{j}}{\rho_{j}} \frac{\boldsymbol{r}_{i j} \cdot \tilde{\nabla}^0_{\boldsymbol{x}}  W_{i j}}{\left|\boldsymbol{r}_{i j}\right|^{4}} \left\{(r_{ij}^\mathrm{i}\  r_{ij}^\mathrm{j}\  r_{ij}^\mathrm{k} \ r_{ij}^\mathrm{l}) - (r_{ij}^\mathrm{i}\  r_{ij}^\mathrm{j})\boldsymbol{r}_{i j} \cdot \bigcup_{k \in \mathbb{S}^{i}} \frac{m_{k}}{\rho_{k}}(r_{ik}^\mathrm{k}\  r_{ik}^\mathrm{l}) \tilde{\nabla}^0_{\boldsymbol{x}} W_{i k}\right\} 
\end{equation} 
Accordingly, 
the diffusion rate of $ \phi $ at particle $ i $ is obtained by
\begin{equation}
\langle\frac{\partial \phi_i }{\partial t } \rangle_{(2)} = \left\langle\frac{\partial^{2} \phi_i}{\left(\partial x^{1}\right)^{2}}\right\rangle_{(2)} +
\left\langle\frac{\partial^{2} \phi_i}{\left(\partial x^{2}\right)^{2}}\right\rangle_{(2)}  +
\left\langle\frac{\partial^{2} \phi_i}{\left(\partial x^{3}\right)^{2}}\right\rangle_{(2)} ,
\end{equation}
which has the second order accuracy,
while in two dimensional case, it is simplified to 
\begin{equation}
\langle\frac{\partial \phi_i }{\partial t } \rangle_{(2)}= \left\langle\frac{\partial^{2} \phi_i}{\left(\partial x^{1}\right)^{2}}\right\rangle_{(2)} +
\left\langle\frac{\partial^{2} \phi_i}{\left(\partial x^{2}\right)^{2}}\right\rangle_{(2)}.
\end{equation}

\subsection{Nonisotropic diffusion with nonisotropic diffusion coefficients}
\label{anisotropic-diffusion}
Considering the diffusion equation  
 \begin{equation}
  \frac{\partial c}{\partial t} =\nabla \cdot(\mathbf{D} \nabla c),
\label{diffusion-equation}
 \end{equation}
where $c$ is the concentration function and $\mathbf{D} $
is the diffusion coefficient tensor, which is commonly 
influenced by the material properties 
or the geometry of the computational domain and so on,
behaving nonisotropic features, 
expressed by a symmetric positive-definite tensor. 
In the case,
the  $\mathbf{D}$ is written as
\begin{equation}
\mathbf{D} = \begin{bmatrix}
 D_{11} &D_{12} &D_{13}  \\ 
 D_{21} &D_{22} &D_{23} \\
 D_{31} &D_{32} &D_{33}   
\end{bmatrix}.
\end{equation}
 Usually, in  an  isotropic diffusion case,  
 $\mathbf{D}$ is considered to be a scalar diffusion coeffici
 ent as \cite{cleary1999conduction}
 \begin{equation}
(\nabla \cdot(\mathbf{D} \nabla c))_{i}=2 \sum_{j} V_{j}  {D}  \frac{c_{i j}}{r_{i j}} \frac{\partial W_{i j}}{\partial r_{i j}},
 \end{equation}
 where $c_{ij} = c_i - c_j$.
In anisotropic cases,  $\mathbf{D}$ is considered to be a symmetric positive-definite matrix and can be decomposed by Cholesky decomposition as \cite{tran2016simulation}, 
 \begin{equation}
\mathbf{D} = \mathbf{L} \mathbf{L}^\mathbf{T}.
\label{D_decomposition}
\end{equation}
By changing the coordinate 
 \begin{equation}
\mathbf{X} =  \mathbf{L}^{-1} \mathbf{x},
\label{coordinate_change}
\end{equation} 
the original 
nonisotropic diffusion operator 
is transformed to be isotropic in $\mathbf{X}$-coordinate system as 
 \begin{equation}
 \nabla \cdot(\mathbf{D} \nabla c)  =
 \nabla_\mathbf{X} \cdot(  \nabla_\mathbf{X} c) =  (\nabla \cdot \nabla )_\mathbf{X}c.
\end{equation} 
Since there is a Hessian matrix formulation as 
\begin{equation}
\mathbf{H}_X = \begin{bmatrix} 
\vspace{1.3ex} \frac{\partial^{2} \phi_{i}}{\left(\partial X^{1}\right)^{2}} & \frac{\partial^{2} \phi_{i}}{\partial X^{1}\partial X^{2}}& 
\frac{\partial^{2} \phi_{i}}{\partial X^{1}\partial X^{3}}  \\ 
\vspace{1.3ex} \frac{\partial^{2} \phi_{i}}{\partial X^{2}\partial X^{1}}&
\frac{\partial^{2} \phi_{i}}{\left(\partial X^{2}\right)^{2}}  &  	\frac{\partial^{2} \phi_{i}}{\partial X^{2}\partial X^{3}} 
\\ 
\vspace{1.3ex} \frac{\partial^{2} \phi_{i}}{\partial X^{3}\partial X^{1}} & 
\frac{\partial^{2} \phi_{i}}{\partial X^{3}\partial X^{2}}  &	\frac{\partial^{2} \phi_{i}}{\partial X^{3}\partial X^{2}}
\end{bmatrix}, ~~
\mathbf{H}_x = \begin{bmatrix} 
\vspace{1.3ex} \frac{\partial^{2} \phi_{i}}{\left(\partial x^{1}\right)^{2}} & \frac{\partial^{2} \phi_{i}}{\partial x^{1}\partial x^{2}}& 
\frac{\partial^{2} \phi_{i}}{\partial x^{1}\partial x^{3}}  \\ 
\vspace{1.3ex} \frac{\partial^{2} \phi_{i}}{\partial x^{2}\partial x^{1}}&
\frac{\partial^{2} \phi_{i}}{\left(\partial x^{2}\right)^{2}}  &  	\frac{\partial^{2} \phi_{i}}{\partial x^{2}\partial x^{3}} 
\\ 
\vspace{1.3ex} \frac{\partial^{2} \phi_{i}}{\partial x^{3}\partial x^{1}} & 
\frac{\partial^{2} \phi_{i}}{\partial x^{3}\partial x^{2}}  &	\frac{\partial^{2} \phi_{i}}{\partial x^{3}\partial x^{2}}
	\end{bmatrix},
\end{equation}
while obviously $\mathbf{H}_X = \mathbf{L} ~\mathbf{H}_x ~ \mathbf{L}^{T}$. 
According to Eq. \eqref{laplacian_3d} in \ref{full-expression}, 
the elements of Hessian matrix $\mathbf{H}_x $ 
can be concluded from
the full expression of the second derivative model.
Therefore, applying the relation of $\mathbf{H}_X = \mathbf{L} ~\mathbf{H}_x ~ \mathbf{L}^{T}$, 
the Laplacian with nonisotropic diffusion tensor coefficient can be obtained by
\begin{equation}
\nabla \cdot(\mathbf{D} \nabla c)  = (\nabla \cdot \nabla c)_\mathbf{X}=  \frac{\partial^{2} \phi_{i}}{\left(\partial X^{1}\right)^{2}} +  \frac{\partial^{2} \phi_{i}}{\left(\partial X^{2}\right)^{2}}+ \frac{\partial^{2} \phi_{i}}{\left(\partial X^{3}\right)^{2}}.
	\end{equation}

\section{ ASPH discretization}
\subsection{ASPH discretization for fluid-structure interaction}
In the fluid-structure interaction model discretization, 
each particle carries the location 
$\mathbf{x}_n = \phi (\mathbf{X}, t_n)$  
at time $t_n$, along with an initial 
representative volume $V^0$ that partitions 
the initial domain of the macroscopic solid. 
The deformation gradient $\mathbf{F}_n$ of the solid phase 
is stored to update the solid current 
volume $V_n $ and density $\rho^s_n$. 
Additionally, the fluid mass $m_n^l$, 
saturation $\widetilde{a}_n$, and  density-weighted 
velocity of the fluid relative to 
solid $\mathbf{q}_n$ are stored. 
The fluid mass equation 
Eq. \eqref{fluid_mass_der} of particle $ i $ 
is discretized as
\begin{equation} 
	\label{fluid_mass_sph}
	\frac{\text{d} m_i^l}{\text{d} t} =  2 V_i \sum_j\frac{m_j}{\rho_j}(\mathbf{q}_{i} -\mathbf{q}_{j}) \tilde{\nabla_{i}} W_{i j}  .
\end{equation}
Note that with the Eq \eqref{eq:deformationtensor-displacement},
we have the relation of gradient kernel function 
in the total Lagrangian and updated Lagrangian  $ \tilde{\nabla_{i}} W_{i j} =\mathbf{B}^{0^T}_i \mathbf{F}^{-1}   \nabla_{i}^0 W_{i j} $.
Once fluid mass is updated, the locally effective
fluid density $ \rho^l $ is obtained subsequently.
According to Eq. \eqref{fluid_density} and  Eq. \eqref{grad_q}, 
we update the fluid  
saturation $ \widetilde{a} $ and the fluid flux $ \mathbf{q} $ in 
the particle form 
\begin{equation} 
	\label{ralative_velocity_sph}
	\mathbf{q}_i =  -K\rho^l V_i \sum_j\frac{m_j}{\rho_j}(\widetilde{c}_{i} -\widetilde{c}_{j})\tilde{\nabla_{i}} W_{i j}.
\end{equation}
With the fluid flux and the stress in hand, we obtain discrete formulations for the momentum 
balance Eq. \eqref{totalmomentumupdate} as
\begin{equation}
	\label{totalmomentumupdatesph}
	\frac{d\mathbf{M}_i}{d t} = 2 \sum_j V_j ( \boldsymbol{\sigma}_i +  \boldsymbol{\sigma}_j) \tilde{\nabla_{i}}  W_{ij}  - 2 \sum_j V_j ( {\mathbf{v}_i^{l} \otimes \mathbf{q}_i} + {\mathbf{v}_j^{l} \otimes \mathbf{q}_j}) \tilde{\nabla_{i}}  W_{ij},
\end{equation}
where $ \boldsymbol{\sigma}_i$ and $ \boldsymbol{\sigma}_j$ are the stress 
tensors of particles $i$ and $j$.
We then compute the updated solid 
velocity $\mathbf{v}^s$ using the 
total momentum definition 
Eq. \eqref{P_equation}, where the total density 
of the mixture is the sum of the solid and 
fluid densities $ \rho = \rho^s+\rho^l $, written as  
\begin{equation}
	\label{solid_velocity_update}
	\mathbf{v}^{s} =\frac{\mathbf{M} -\mathbf{q}}{\rho} = \frac{\mathbf{M}-\mathbf{q}}{\rho^s+\rho^l }.
\end{equation}
Subsequently, the fluid velocity $\mathbf{v}^l$ is calculated  
using Eq. \eqref{defination_q} as 
\begin{equation}
	\label{fluid_velocity_update}
	\mathbf{v}^l =  \mathbf{v}^s-  \frac{\mathbf{q}}{\rho^l}.
\end{equation}  

\subsection{ASPH discretization of cardiac function}
The anisotropic ASPH discretization of diffusion is given in section \ref{anisotropic-diffusion}. 
For the discretization of momentum equation, 
considering the kernel correction,  
Eq. \eqref{eq:mechanical-mom} can be approximated
in the weak form  as 
\begin{equation}\label{eq:sph-mechanical-mom}
\frac{ d \mathbf{v}_i}{ {d}t} = \frac{2}{\rho_i} \sum_j V_j \tilde{\mathbf{P}}_{ij} {\nabla}^0_i W_{ij} , 
\end{equation} 
where $\rho_i$ represents the density of particle $i$,
$\tilde{\mathbf{P}}_{ij}$ is the averaged first Piola-Kirchhoff stress of the particle pair $(i,j)$,
and to keep the conservative of particles, 
the correction matrix $\mathbf{B}^{0^T}$ is performed on each particle, 
thus $\tilde{\mathbf{P}}_{ij}$  is stated as
\begin{equation}\label{P-corrected}
\tilde{\mathbf{P}}_{ij} = \frac{1}{2} \left( \mathbf{P}_i \mathbf{B}^{0^T}_i + \mathbf{P}_j \mathbf{B}^{0^T}_j  \right). 
\end{equation}  
The first Piola-Kirchhoff stress tensor is dependent on the deformation tensor $\mathbf{F}$, referring to 	Eq. \eqref{F-consitency},
the time derivative of which is computed from 
\begin{equation}
\label{rate_F}
\frac{d\mathbf{F}_i }{dt} = \left( \sum_j V_j \left( \mathbf{v}_j - \mathbf{v}_i \right) \otimes \nabla^0_i W_{ij}  \right) \mathbf{B}^0_i.
\end{equation}

\section{Numerical examples}
\subsection{Diffusion with nonisotropic resolution}
In this section, we discretize the computation with 
different anisotropic ratios
to form nonisotropic resolutions.
Firstly, the second-order accuracy of the algorithm is verified
by comparing the result generated by the current method 
with the analytical solution.
Subsequently, a block diffusion coupling with 
periodic boundary conditions is tested.
Finally, the fluid-solid interaction model is simulated to 
show the applicability of this algorithm.

\subsubsection{Verification of second-derivative operator in a square patch}
Considering a square patch composed of a number of SPH particles,
and according to its position,
each particle receives the values of the following function
\begin{equation}
\phi (x,y) = x^2 + y^2,  ~~0 \leq  x, ~y \leq 1
\label{test-1}
\end{equation}
and the analytical solution of this function is 
\begin{equation}
\frac{d\phi (x,y)}{dt} = \frac{\partial^2 \phi }{\partial  x^2} + \frac{\partial^2 \phi }{\partial  y^2}  = 4.
\label{test-1-solution}
\end{equation}
The numerical value  $ \frac{d\phi }{dt}$
is evaluated on the particles that compose the square patch.
Figure. \ref{} shows the contour
obtained from the analytical solution and
the present ASPH method with different ratios $ r $ = 1.0, 2.0, 4.0, 8.0. 
Regardless of the anisotropic ratios, 
all results precisely correspond to the analytical solution,
demonstrating the algorithm's accuracy.

\begin{figure*}[htbp]
\centering
\begin{subfigure}[b]{0.45\textwidth}
	\includegraphics[trim =2mm 2mm 2mm 2mm, clip,width=0.95\textwidth]{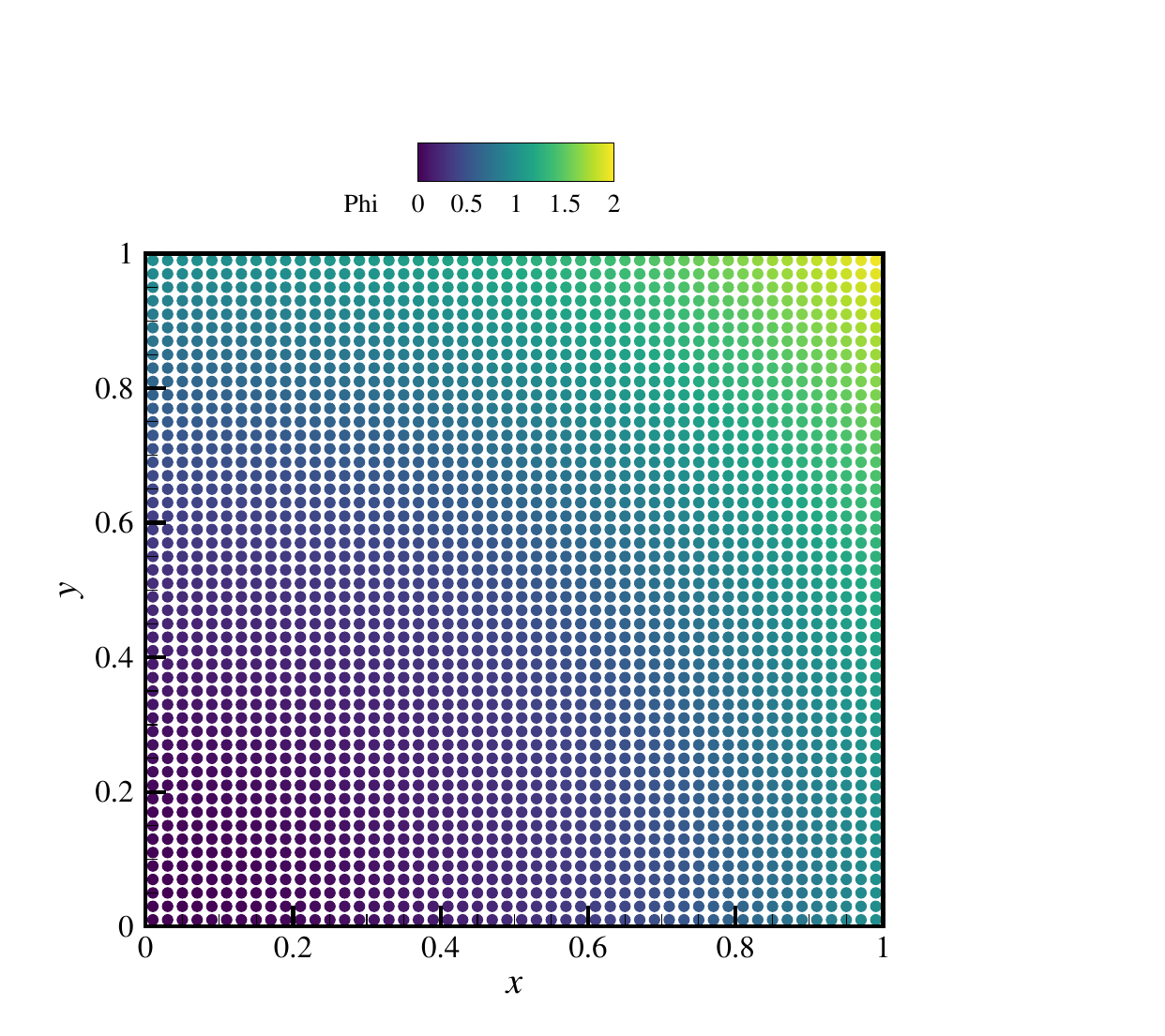}
	\caption {$\phi$ distribution.}
	\label{case1-r1}
\end{subfigure}
\begin{subfigure}[b]{0.45\textwidth}
\includegraphics[trim =2mm 2mm 2mm 2mm, clip,width=0.95\textwidth]{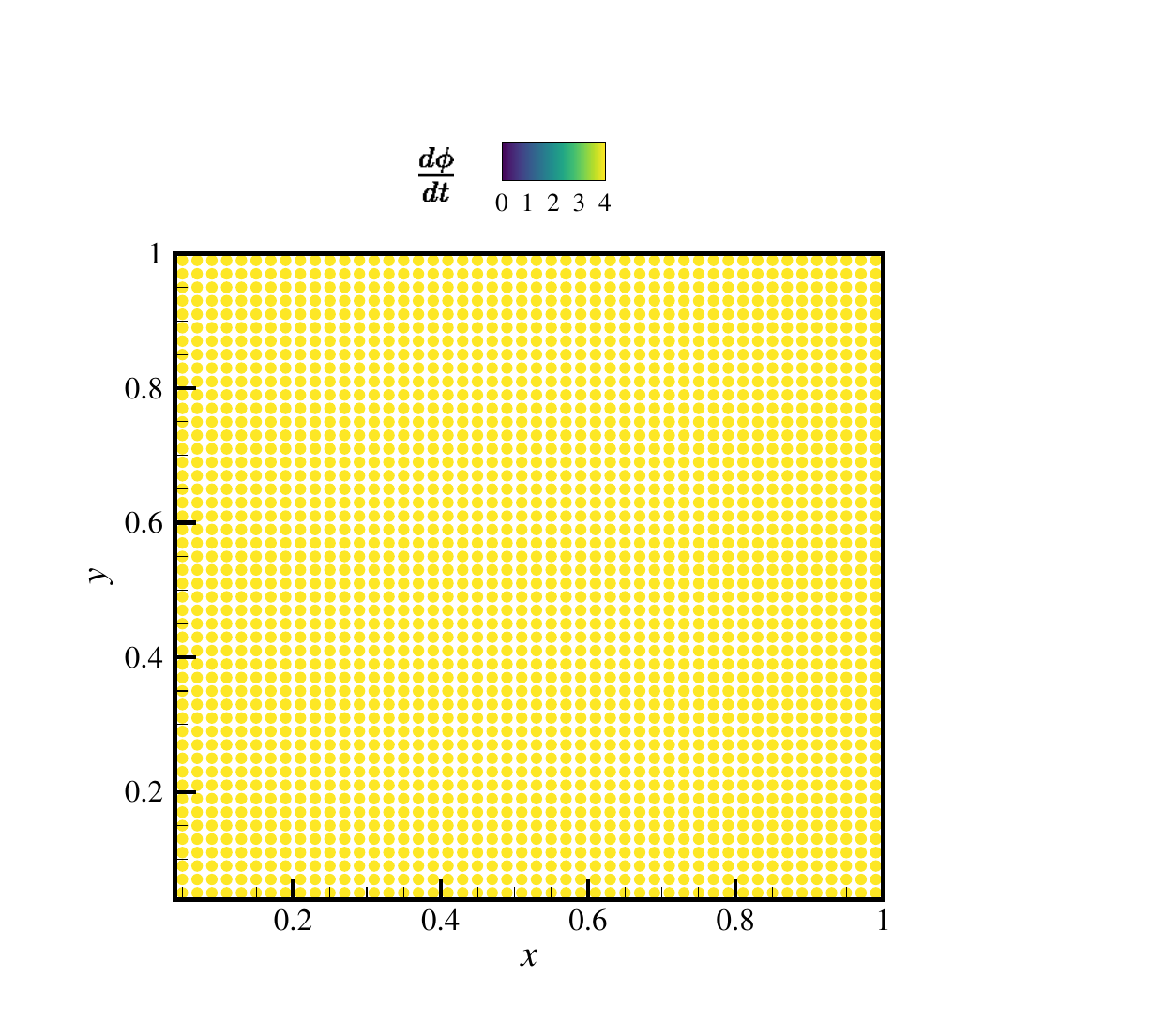}
\caption {The analytical solution of $ \frac{d\phi}{dt} $ .}
\label{case1-analytical-solution}
\end{subfigure}
\begin{subfigure}[b]{0.45\textwidth}
\includegraphics[trim =2mm 2mm 2mm 2mm, clip,width=0.95\textwidth]{function-1-laplacian-eps-converted-to.pdf}
\caption  {The ASPH method with r = 1.0.}
\label{case1-r1-function}
\end{subfigure}
\begin{subfigure}[b]{0.45\textwidth}
	\includegraphics[trim =2mm 2mm 2mm 2mm, clip,width=0.95\textwidth]{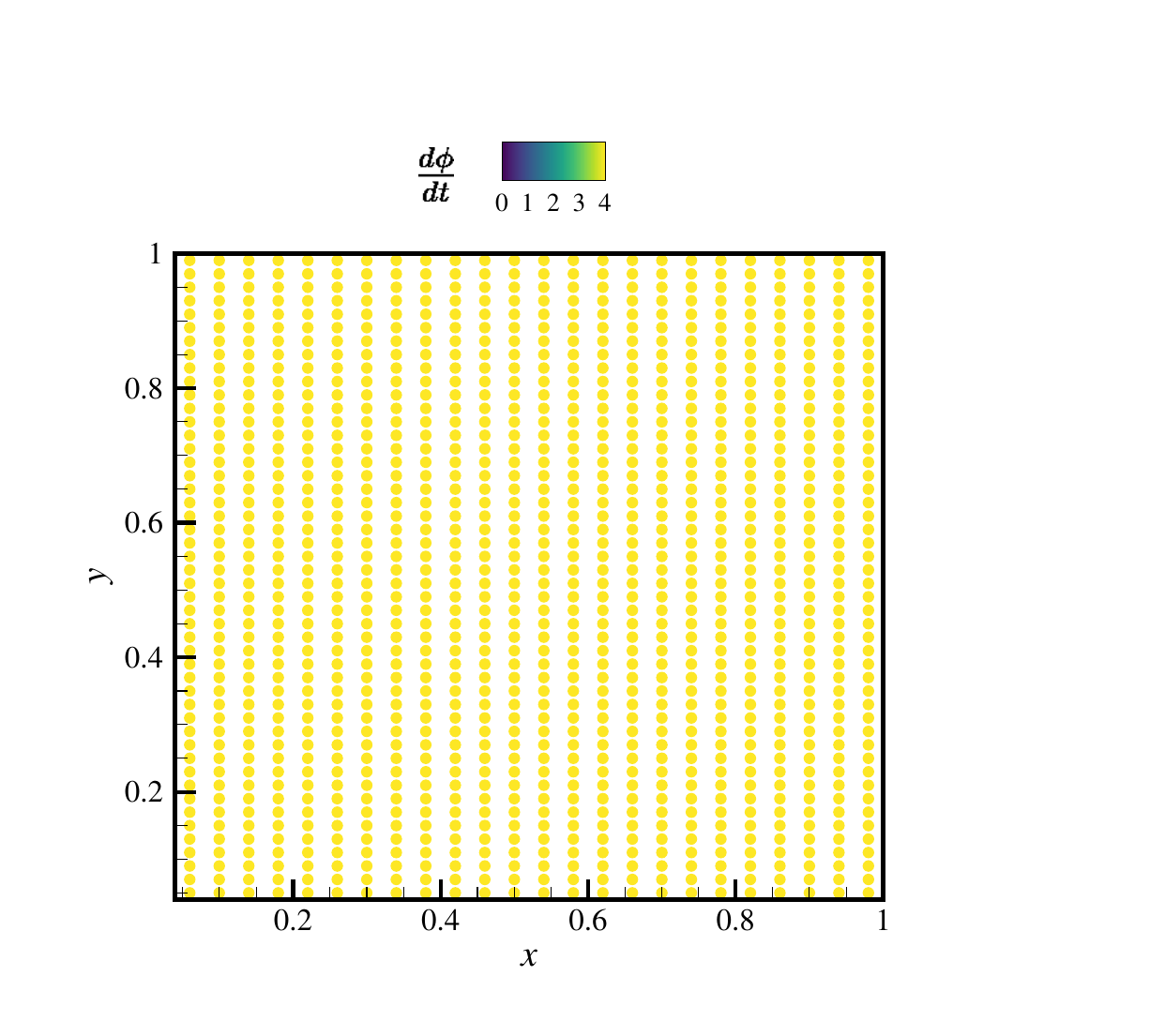}
	\caption {The ASPH method with r = 2.0.}
	\label{case1-r2}
\end{subfigure}
\begin{subfigure}[b]{0.45\textwidth}
	\includegraphics[trim =2mm 2mm 2mm 2mm, clip,width=0.95\textwidth]{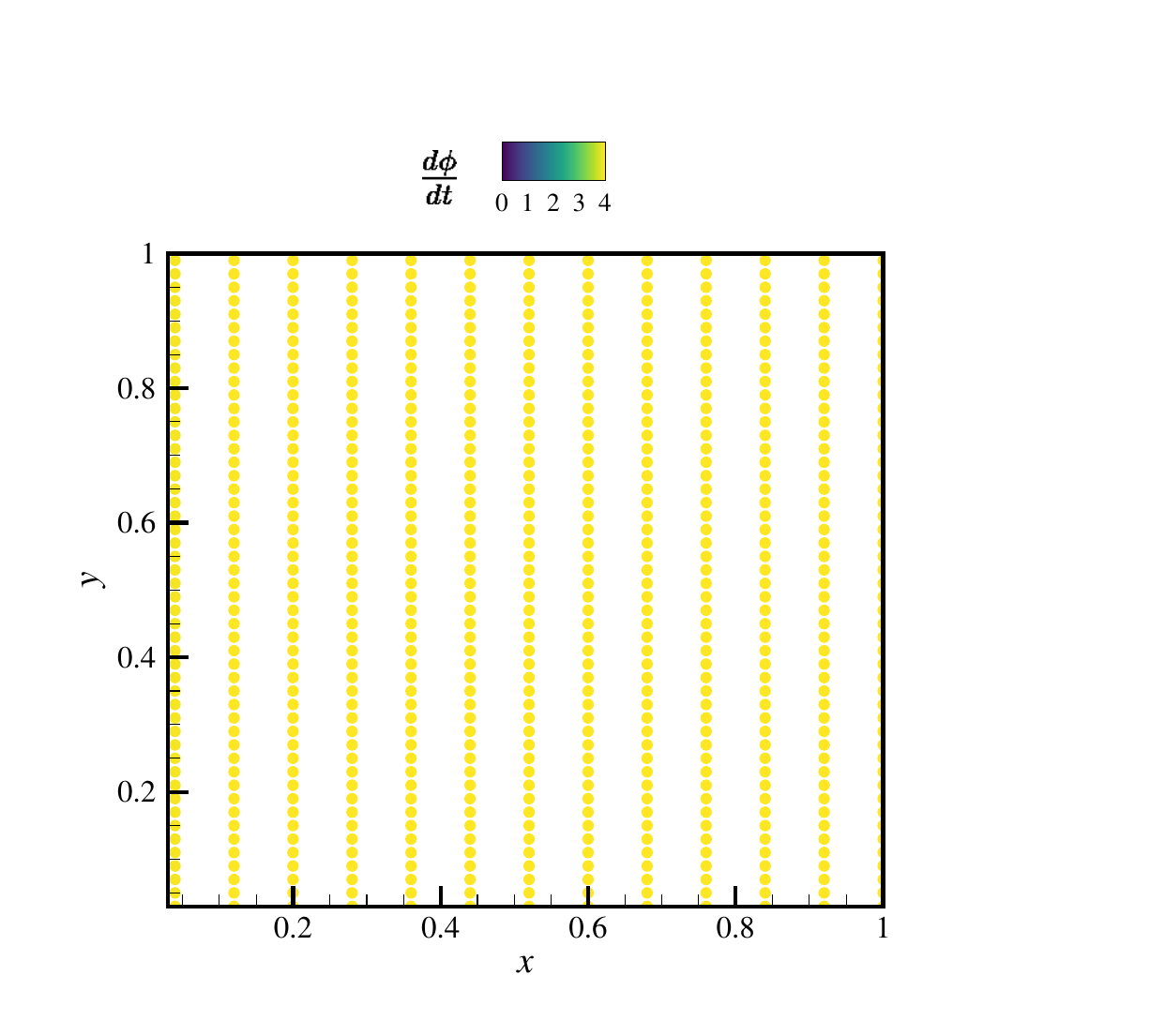}
	\caption {The ASPH method with r = 4.0.}
	\label{case1-r4}
\end{subfigure}
\begin{subfigure}[b]{0.45\textwidth}
	\includegraphics[trim =2mm 2mm 2mm 2mm, clip,width=0.95\textwidth]{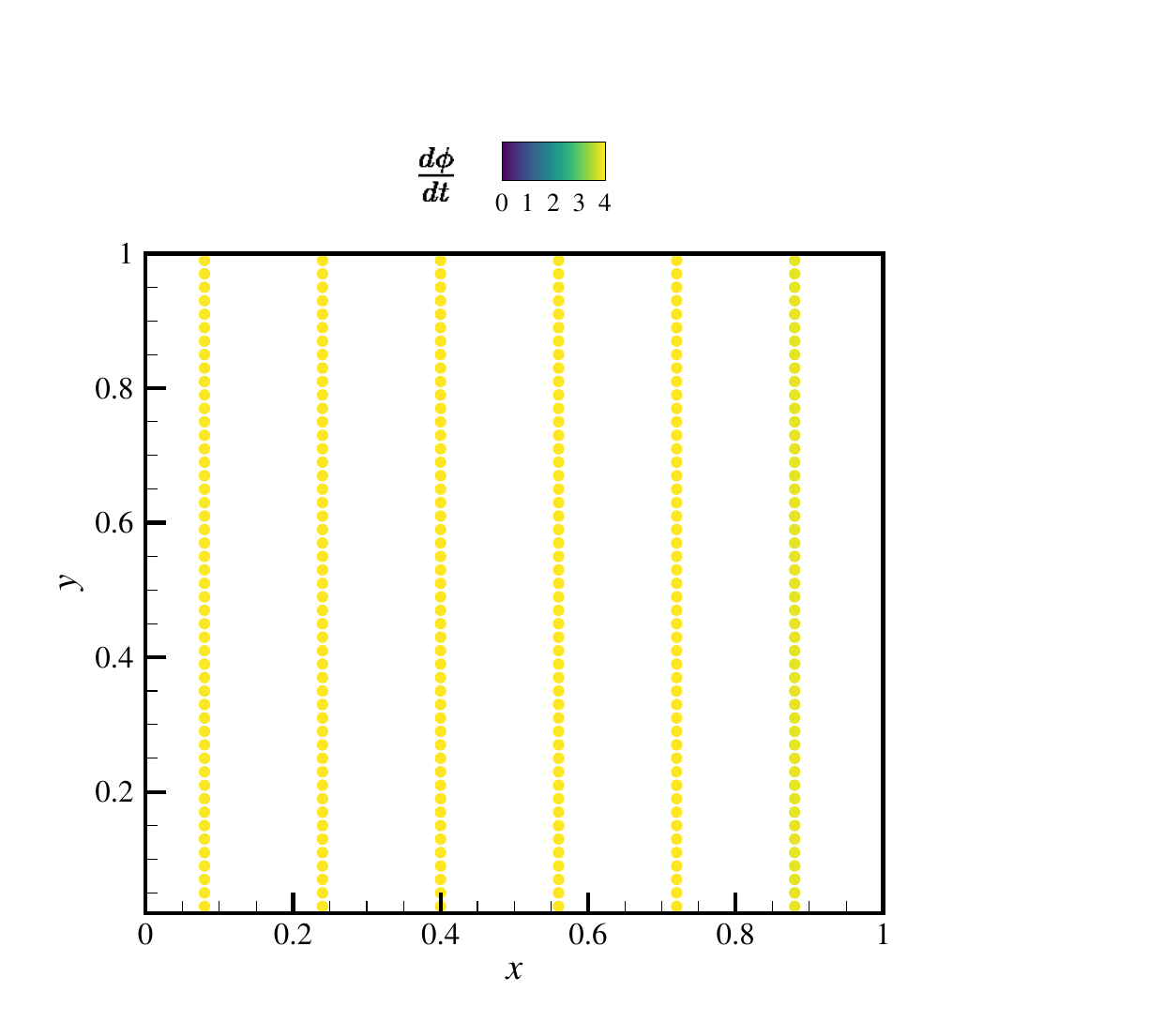}
	\caption  {The ASPH method with r = 8.0.}
	\label{case1-r8}
\end{subfigure}
\caption{2D nonoisotropic diffusion: comparisons of the $ \frac{d\phi}{dt} $ contour obtained from (a) the analytical solution and (b)-(f) the present ASPH method with different anisotropic ratios.}
\label{case1}
\end{figure*}

\newpage

\subsubsection{Diffusion in a rectangle}
For a more complex initial condition,
Considering the a two-dimensional rectangle with a dimensionless length of 1 and width of 0.1, 
with a initial physical parameter $\phi$ setup as  
\begin{equation}
\phi= \{ \begin{array}{ll}
1.0, & 0.4\leq x\leq 0.6 \\
0,   & \rm else
\end{array} 
\end{equation}
The dimensionless diffusion coefficient $D$ is set as 1.0. 
Neumann boundary conditions are applied.
The diffusion is simulated by the present ASPH algorithm 
and various anisotropic ratios are settled to compare the efficiency.

\begin{figure*}[htbp]
	\centering
	\begin{subfigure}[b]{0.8\textwidth}
\includegraphics[trim =20mm 10mm 40mm 130mm, clip,width=0.95\textwidth]{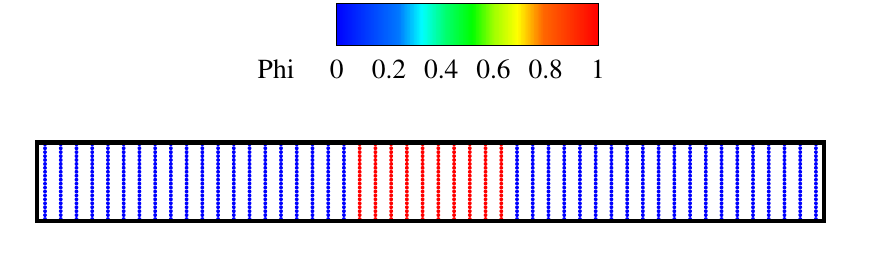}
		\caption {t = 0.}
		\label{case1.3-t0}
	\end{subfigure}
	\begin{subfigure}[b]{0.8\textwidth}
		\includegraphics[trim =20mm 10mm 40mm 145mm, clip,width=0.95\textwidth]{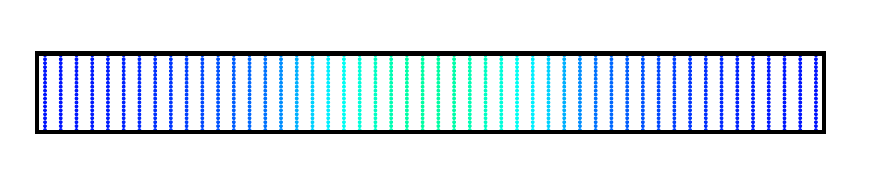}
		\caption {t = 0.02.}
		\label{case1.3-t0.02}
	\end{subfigure}
	\begin{subfigure}[b]{0.8\textwidth}
		\includegraphics[trim =20mm 10mm 40mm 145mm, clip,width=0.95\textwidth]{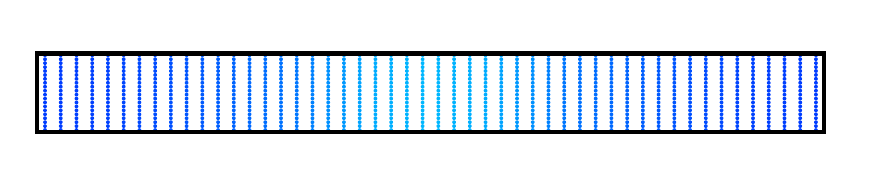}
		\caption{t = 0.04.}
		\label{case1.3-t0.04}
	\end{subfigure}
	\begin{subfigure}[b]{0.8\textwidth}	\includegraphics[trim =20mm 10mm 40mm 145mm, clip,width=0.95\textwidth]{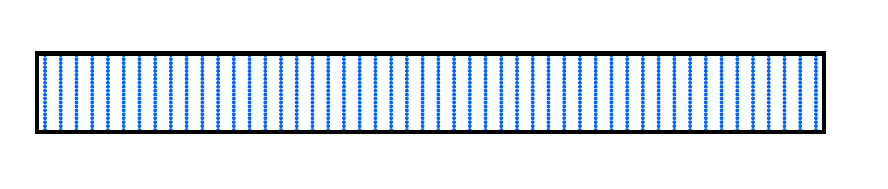}
		\caption {t = 0.2.}
		\label{case1.3-t0.2}
	\end{subfigure}
\caption{Diffusion in a rectangle: the time evolution of   $ \phi$ contour at different time instants  
using ASPH with anisotropic ratio r = 4.0, $ N_y $ = 20.
}
\label{case1.3}
\end{figure*}

\begin{figure*}[htbp]
 	\centering
 	\includegraphics[trim = 2mm 2mm 2mm 7mm, clip,width=0.65\textwidth]{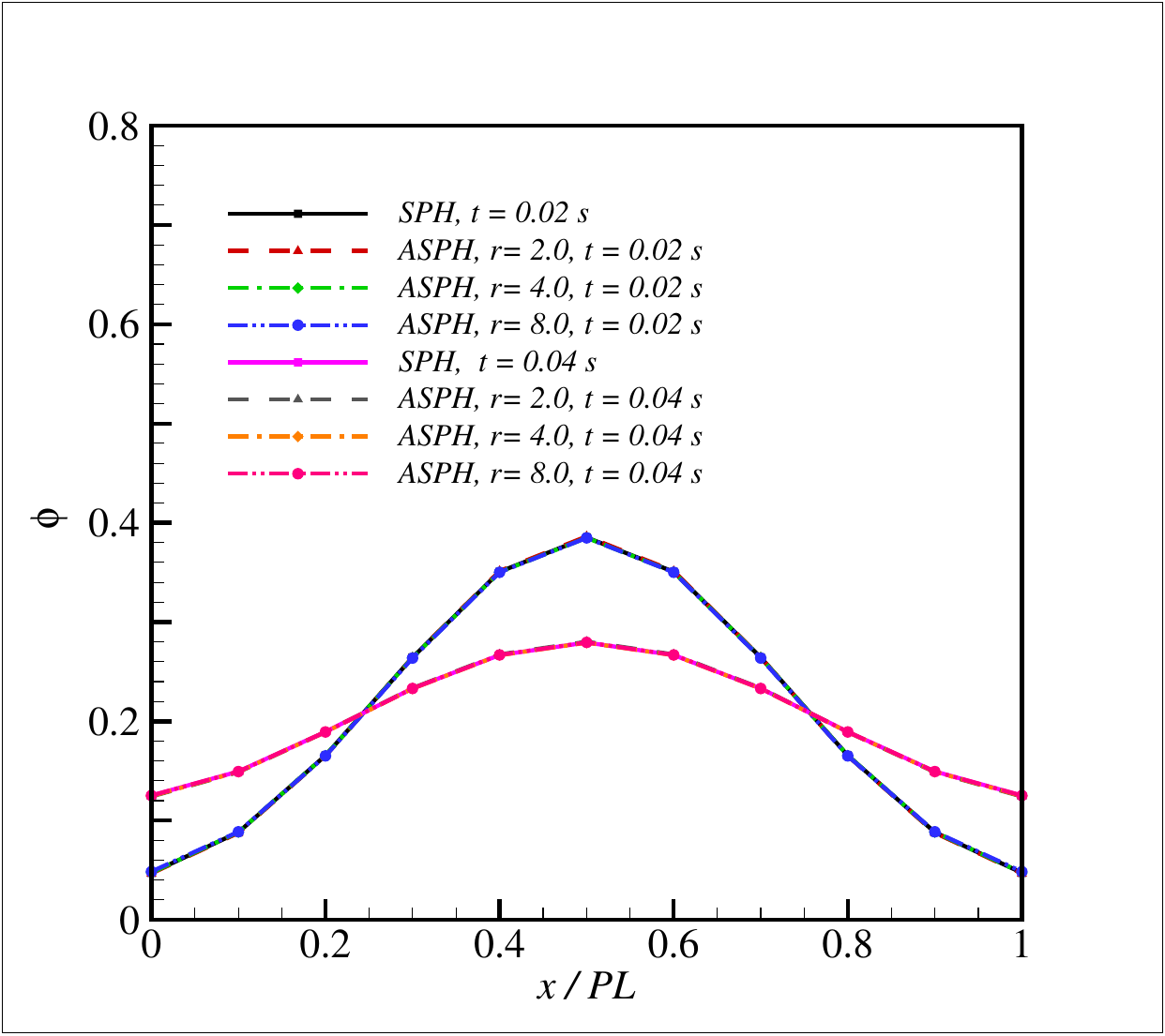}
 	\caption {Diffusion in a rectangle: the $ \phi $ variation on the horizontal line of the  }
 	\label{diffusion-phi-variation-line}
 \end{figure*}

\begin{figure*}[htbp]
	\centering
	\includegraphics[trim = 2mm 2mm 2mm 7mm, clip,width=0.65\textwidth]{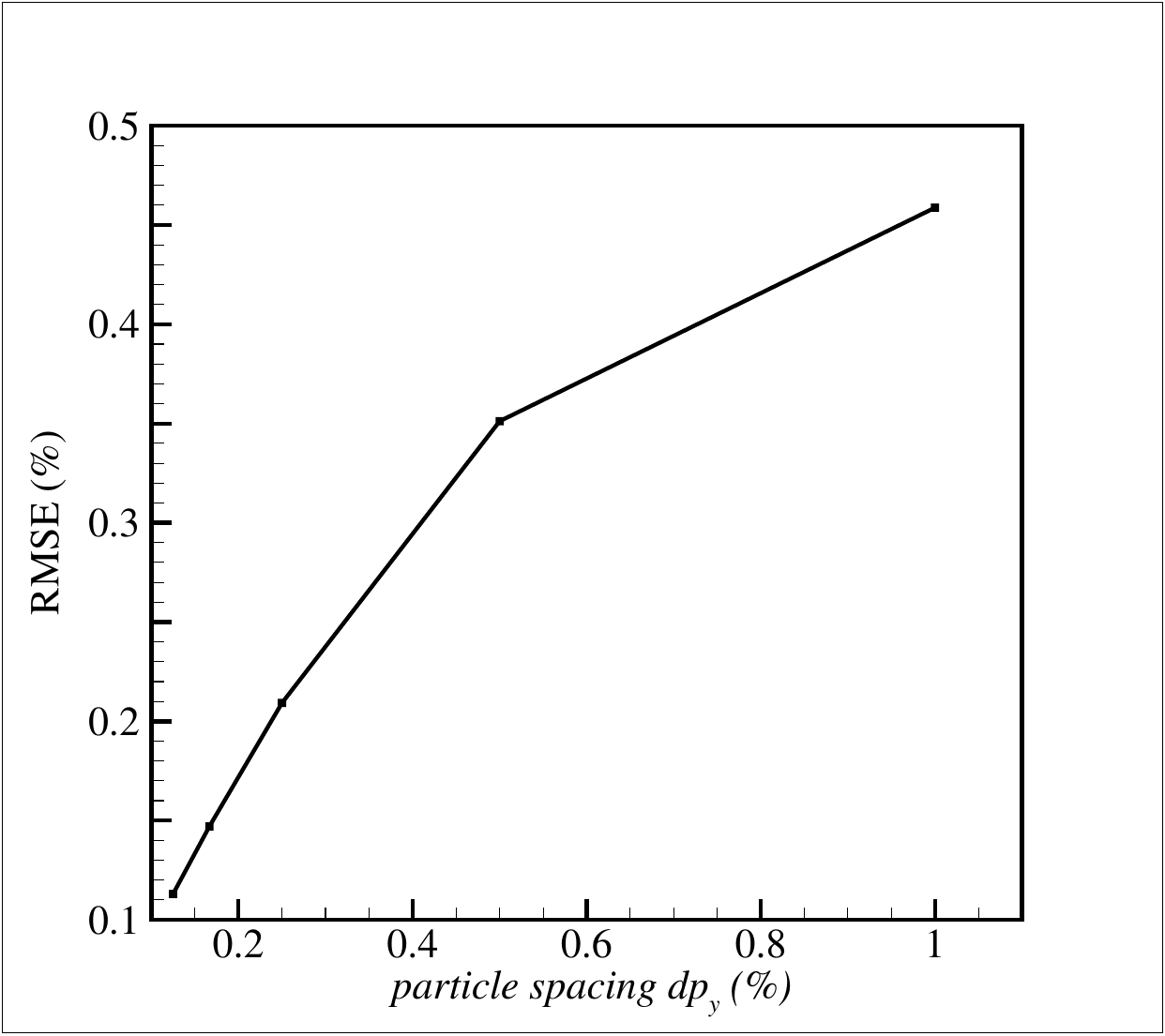}
	\caption {Diffusion in a rectangle: the RMSE error under different resolution with anisotropic ratio = 4.0. }
	\label{diffusion-error-line}
\end{figure*}

Using the ASPH method with anisotropic ratio $r$ = 4.0,
particle number in $ y $ direction (the vertical direction) $ N_y $ = 20, 
Figure \ref{case1.3} illustrates snapshots of the diffusion at different time instances. 
We record the $\phi$ value across the horizontal line of $y$ = 0.05,
and diffusion process at different time instances
yielded by the  ASPH algorithm with 
different $r$ are depicted in Figure. \ref{diffusion-phi-variation-line}.
With the analytical solution being $\phi $ = 0.1 throughout the domain at the final stable state, 
very close results are obtained by using SPH and ASPH algorithm  with various anisotropic ratios.
The convergence study is performed, 
taking r = 4.0 as example,
as particle number increases, 
the final stable  result of $\phi$ are converged to 
the analytical solution being  0.1 throughout the domain.
To quantitatively verify the accuracy,
the normalized root mean square error (RMSE) is evaluated  
between the analytical solution and the numerical values 
computed by the second-derivative operator,
defined as
\begin{equation}
\rm{RMSE}~ ( \left<\phi\right>^i) = \frac {\sqrt{ \sum_i (\phi^i - \left<\phi\right>^i )^2}}{\sqrt{ \sum_i (\phi^i )^2}},
\label{case-1.2-error}
\end{equation}
where $ \phi^i$ indicates the analytical solution of these test functions, 
and  $ \left< \phi \right>^i$  is the numerical value derived from 
ASPH method.
The normalized RMSE error comparison is shown in  
Figure. \ref{diffusion-error-line}. 

\subsubsection{Two dimensional fluid diffusion coupling solid deformation}
 
\begin{figure*}[htbp]
	\centering
	\includegraphics[trim = 20mm 80mm 20mm 70mm, clip,width=0.65\textwidth]{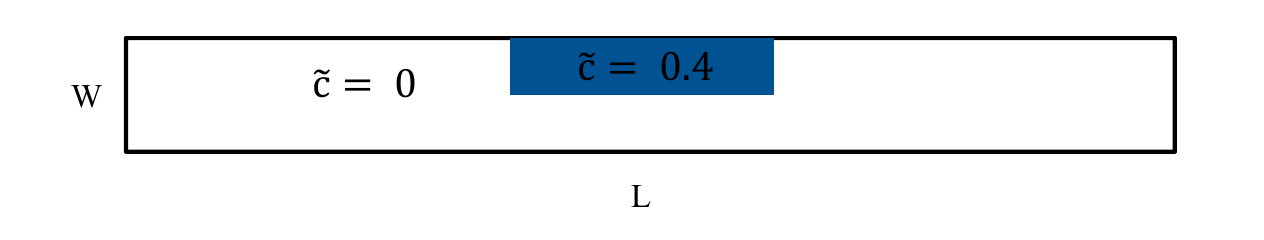}
	\caption {2D fluid-structure interaction: physical configuration of the thin porous beam.}
	\label{configuration}
\end{figure*}

In this section, we perform a two-dimensional simulation of fluid diffusion 
coupling with porous solid deformation 
to validate the efficiency of the presented method. 
As illustrated in Figure. \ref{configuration},
a thin porous beam with a length of $ L $ = 5.0 mm and
width of $ W $ = 0.125 mm is considered. 
The left and right sides are constrained to prevent any curling or movement. 
The simulation starts with a fluid droplet
contacting the center part of the beam,
extending to a length of $0.3L$.  
The total physical time is  set to 500 seconds.
Given the slender nature of the beam, 
we assume that initially,
all pores in the upper half part 
are filled with fluid.
As mentioned earlier,
the relationship between fluid
saturation $\widetilde{c}$
and solid porosity ${c}$ is
$0 \leq \widetilde{c} \leq {c} < 1$.
For this 2D and 3D cases discussed later, 
we assume a solid porosity of ${c} = 0.4$,
implying that the fluid saturation $ \widetilde{c} $ 
in the central part($0.5W \times 0.3L$) 
is constrained to $ \widetilde{c} = {c} = 0.4$ initially,
while in other regions $ \widetilde{c}_0 = 0.0$.

In alignment with the experimental setup,  
the solid material is modeled as 
a porous and elastic Nafion membrane, 
with water serving as the fluid.
The physical properties and  material parameters 
of this membrane  
are listed in Table \ref{parameter-table}.
The pressure coefficient C has been calibrated to fit the experimentally measured flexure curves,
while other parameters are obtained from previous research papers \cite{motupally2000diffusion,goswami2008wetting}.
\begin{table}[htb!]
	\centering
	\caption{Fluid-structure interaction: physical material parameters value of Nafion film. 
		Data estimated from Motupally  and Goswami \cite{motupally2000diffusion,goswami2008wetting}.}
	
	\resizebox{\textwidth}{!}{
		\renewcommand{\arraystretch}{1.1} 
		\begin{tabular}{cccccc} 
			\hline
			\centering
			Parameters & $\rho$  $ {\rm(kg/m^3)}$ & $ D $ $ {\rm(m^2/s)} $ & Pressure coefficient C  $ {\rm(MPa)}$& Young modulus $ {\rm(MPa)}$ & Poisson ratio \\ 
			\hline
			\centering
			Value & 2000  & $ 1.0\times 10^{-10} $	& $ 3.0  $  & $ 8.242  $& 0.2631\\
			\hline	
	\end{tabular}}
	\label{parameter-table}
\end{table}
In the simulation, 
considering the thin nature, 
the anisotropic resolution is applied when 
discretizing the computation domain.
Firstly, 
four particles are placed in the vertical direction, 
with a vertical particle spacing of $dp_y = W/6= 2.08 \times 10^{-2}$ mm. 
Different anisotropic ratios $ r= 2.0, ~4.0, ~6.0$ are applied, 
resulting in the horizontal particle spacing as  $dp_x = 2.0~dp_y, ~4.0~dp_y, ~6.0~dp_y$.

\begin{figure*}[htbp]
	\centering
	\includegraphics[trim = 1mm 40mm 2mm 10mm, clip,width=0.85\textwidth]{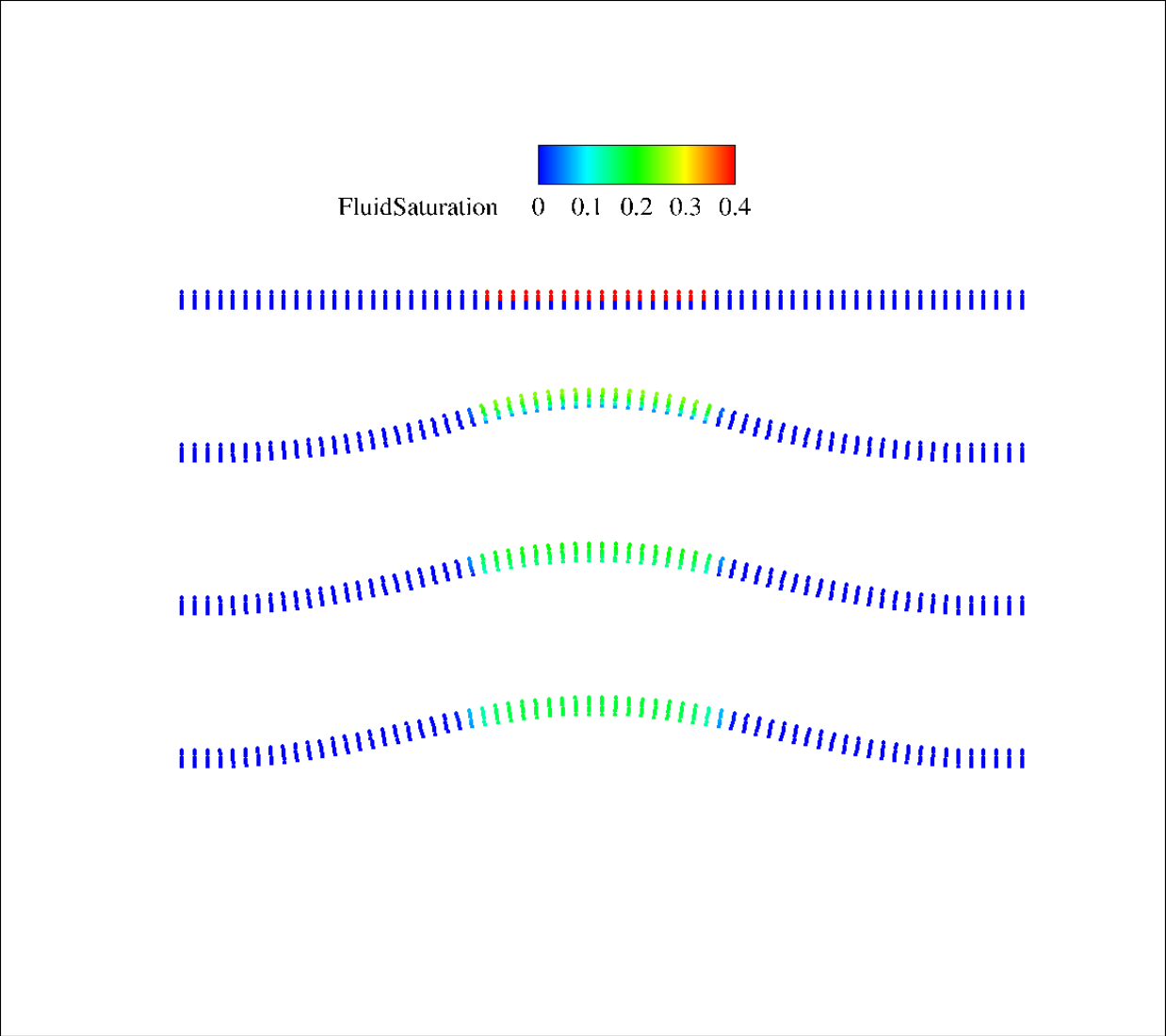}
	\caption {2D fluid-structure interaction: the physical configuration with anisotropic ratio = 4.0 colored by the fluid saturation at different time instants, from up to down, $ t = 0,~100 ~200,~500$ s.}
	\label{2dmembrane-diffusion-coutour}
\end{figure*}
 
 \begin{figure*}[htbp]
 	\centering
 	\includegraphics[trim = 2mm 2mm 2mm 10mm, clip,width=0.65\textwidth]{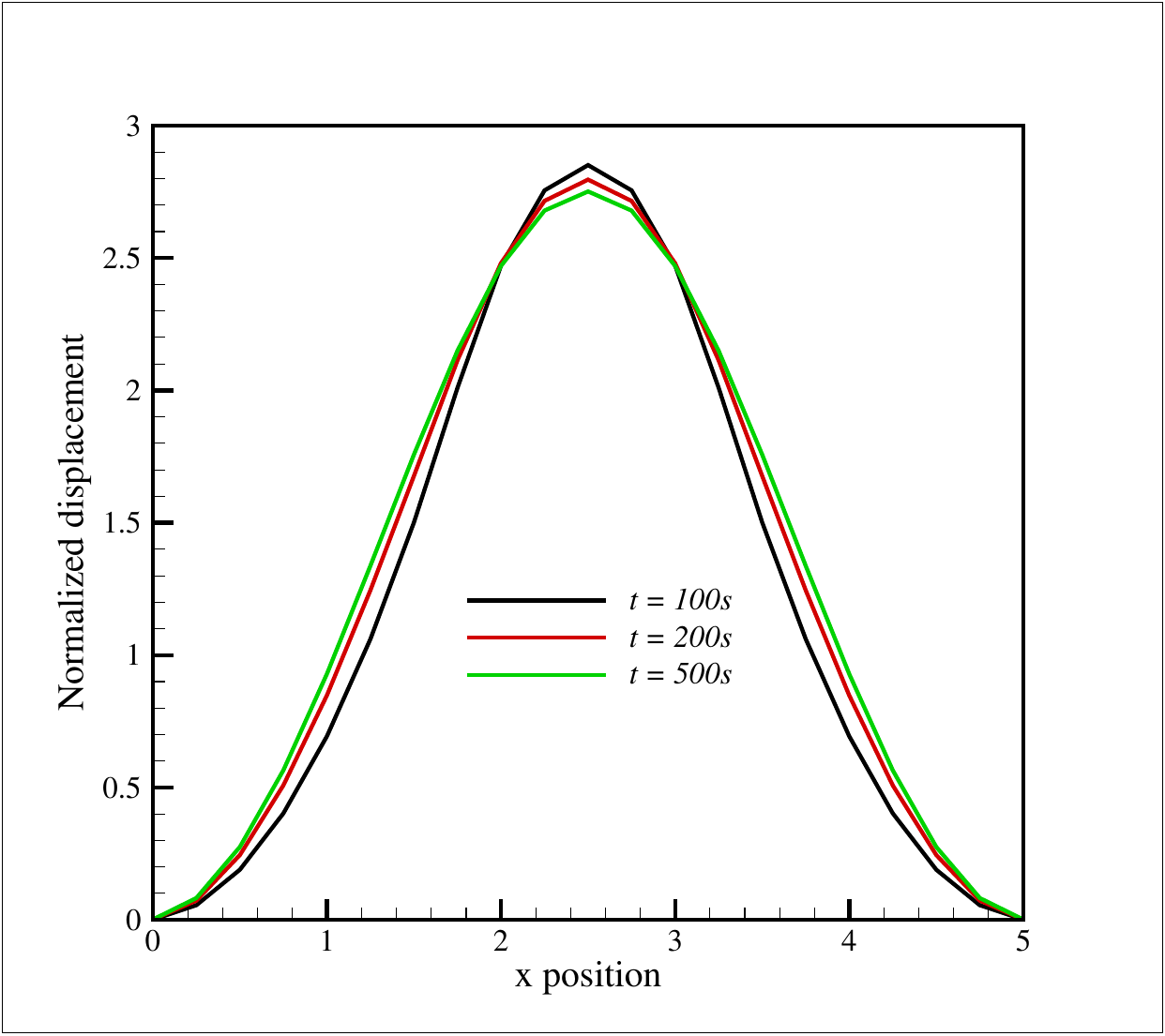}
 	\caption {2D fluid-structure interaction: the  vertical normalized displacement  $ y /PH $ versus the horizontal $ x $ position  with anisotropic ratio = 4.0 at different time instants, from up to down, $ t = 0,~100 ~200,~500$ s.}
 	\label{2dmembrane-dis-time}
 \end{figure*}
With the specified conditions, the simulation produces a deformed
configuration colored by fluid saturation, 
illustrated in  Figure. \ref{2dmembrane-diffusion-coutour}.
Initially, the presence of a water droplet in the upper 
central region induces fluid pressure, 
as explained in Eq. \ref{2dmembrane-dis-time},
leading to a localized bending in the central region. 
As time progresses, the saturation difference drives
continuous water diffusion, showing 
a smooth transition from the center to the surrounding area.  
Figure. \ref{2dmembrane-dis-time} depicts	
the vertical normalized displacement  $ y /PH $ 
versus the horizontal $ x $ position of the beam at different 
time instants. 
As a more uniform pressure distribution develops, 
a smoother flexure of the beam is obtained in the later period.
 
 \begin{figure*}[htbp]
 	\centering
  	\begin{subfigure}[b]{0.45\textwidth}
 		\includegraphics[trim =2mm 2mm 2mm 2mm, clip,width=0.95\textwidth]{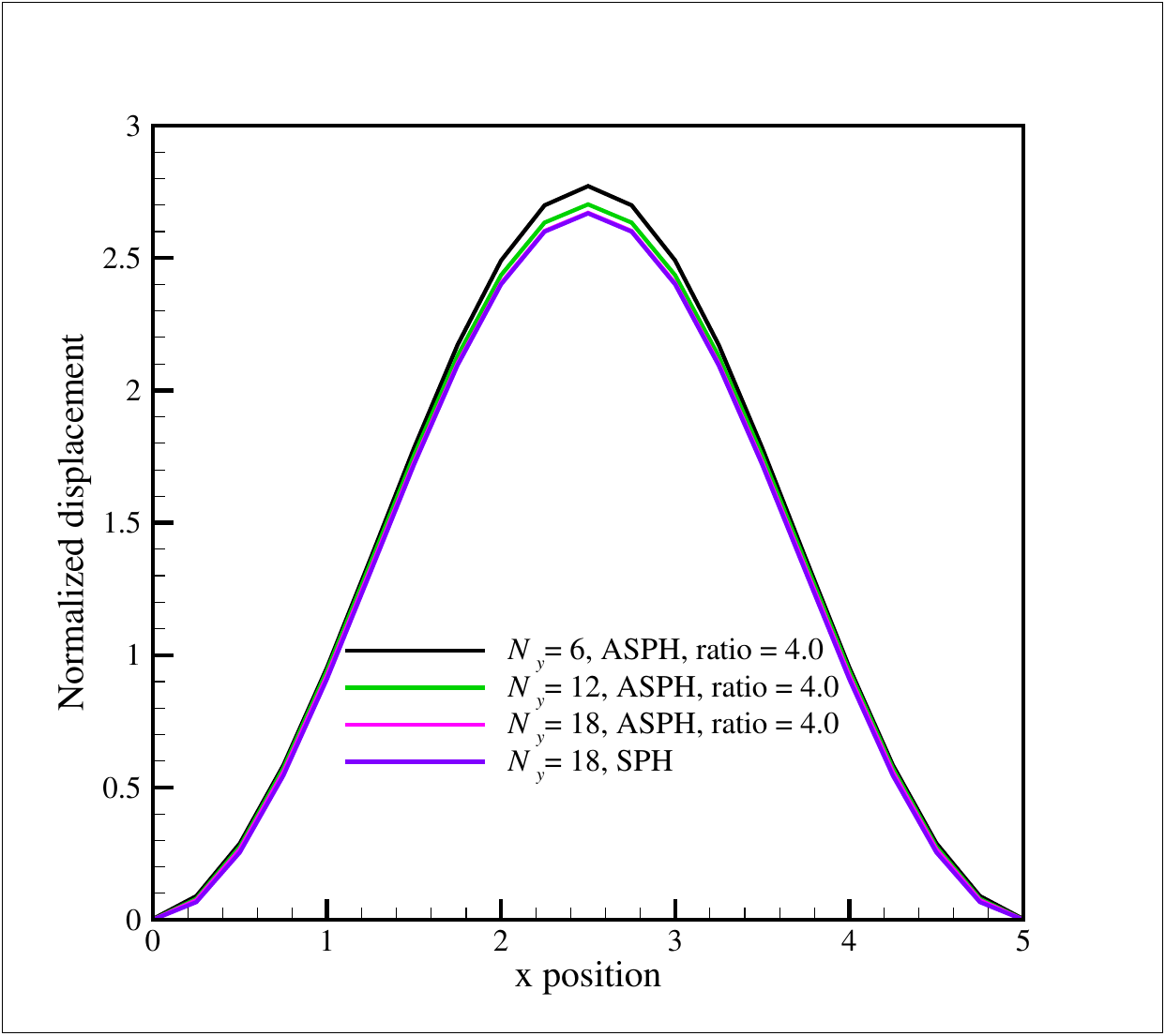}
 		\caption  {The comparison of normalized displacement.}
 		\label{membrane-dis-con}
 	\end{subfigure}
 	\begin{subfigure}[b]{0.45\textwidth}
 		\includegraphics[trim =2mm 2mm 2mm 2mm, clip,width=0.95\textwidth]{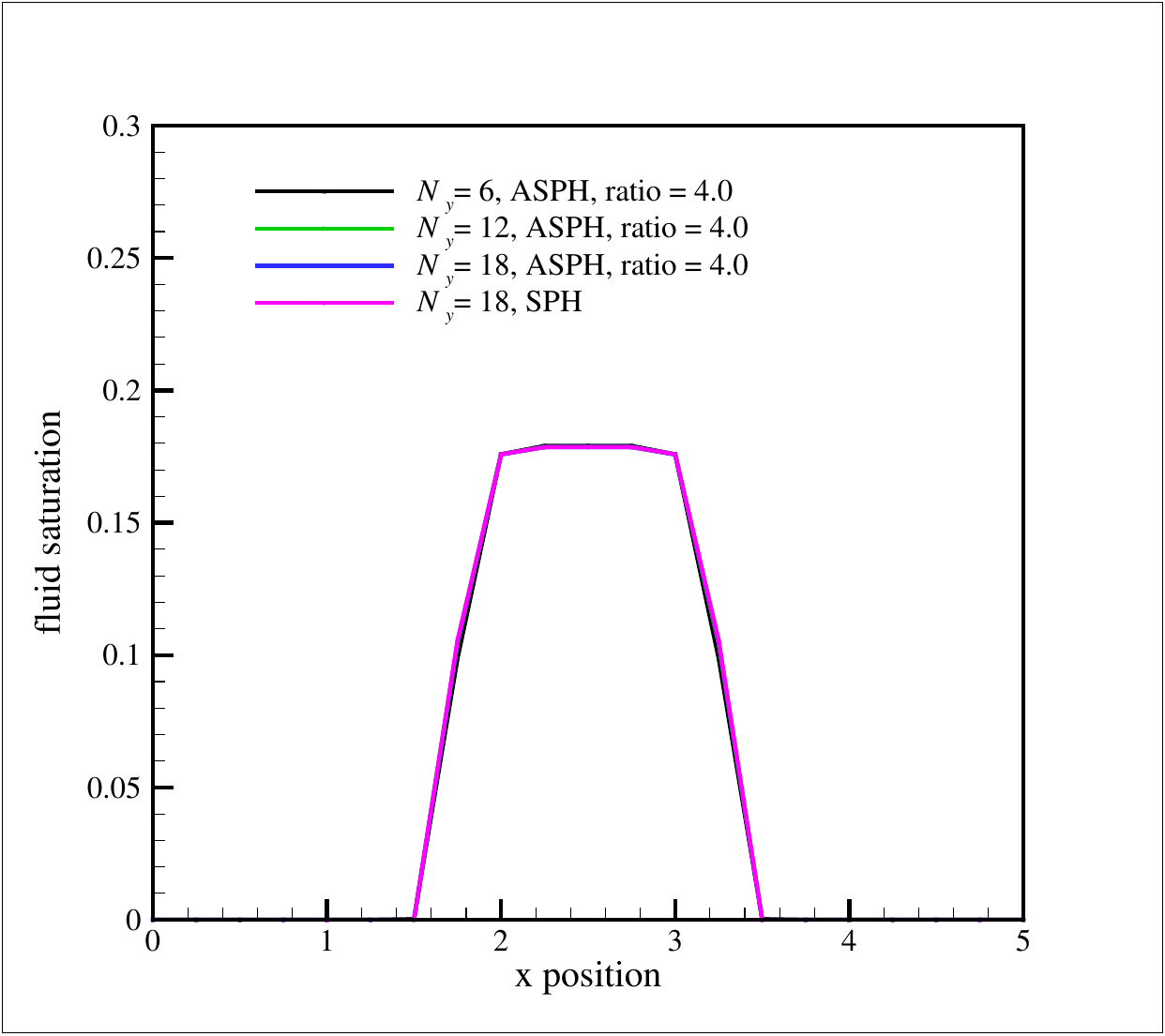}
 		\caption {The comparison of saturation $\widetilde{c}$.}
 		\label{membrane-satu-con}
 	\end{subfigure} 
 	\caption{2D fluid-structure interaction: comparisons of the  vertical normalized displacement  $ y /PH $  (a)  and the saturation  (b)  versus the horizontal $ x $ position  obtained from  the ASPH solution with anisotropic ratio = 4.0  under different resolutions to the converged results from SPH method.}
 	\label{membrane-con}
 \end{figure*}
To test the accuracy, we refine the particle resolution to perform
the convergence test.
Taking the anisotropic ratio r = 4.0 as an example, 
we maintain $ r $ constant while varying the total particle with different
particle distribution densities $ N_y $ = H/$ dp_y $ = 6,~12,~18,
with $ dp_x $ = 4$ dp_y $.
Figure. \ref{membrane-con} shows the final vertical normalized displacement $ y /PH $ and the saturation $\widetilde{c}$ distributions versus the horizontal $ x $ position  of the beam by applying the isotropic SPH method and 
the present ASPH with ratio = 4.0 under different resolutions.
With increasing particle density, the disparity in
 $ y /PH $  and  $\widetilde{c}$  between different 
resolutions using ASPH method diminishes,
indicating a convergence pattern.
Additionally, the converged outcomes derived from
ASPH with $ r $ being 4.0 and SPH   
are visually indistinguishable in both the final
vertical normalized displacement $ y /PH $ and
$\widetilde{c}$ distribution of the beam,
validating the reliability of the ASPH method in simulating 
diffusion problems in thin structures.

The efficiency of the proposed approach is demonstrated through Table \ref{2d-membrane-comparison}, 
providing a quantitative comparison of the present ASPH method 
against the SPH approach in terms of 
particle number N, computation time t as well as 
the normalized root mean square error (RMSE) of saturation $ \widetilde{c} $ 
at the final time instant. 
The root mean square error (RMSE) is defined as 
\begin{equation}
\rm{RMSE}~ ( \left<\widetilde{c}\right>^i) = \frac {\sqrt{ \sum_i (\widetilde{c}^i - \left<\widetilde{c}\right>^i )^2}}{\sqrt{ \sum_i (\widetilde{c}^i )^2}},
\label{2dmembrane-error}
\end{equation}
where $ \widetilde{c}^i$ indicates the converged solution, 
and  $ \left<\widetilde{c} \right>^i$  is a numerical value related to $ \widetilde{c} $  evaluated by SPH and ASPH methods.
From this table, it can be inferred that the present ASPH method 
reduces both the number of particles
and the computational time. 
Despite small deviations existing among different ASPH
ratios, the results from ASPH are all very close to the converged results from SPH method, with the
difference no exceeding 4\%, while achieving notable time savings.
\begin{table*}[htp!]
	\centering
	\caption{2D fluid-structure interaction: quantitative comparison of the accuracy and efficiency of this present ASPH algorithm with different anisotropic ratios.}
	\begin{tabular}{ccccc}
		\hline
		Method   & $ N $  & $ t~(s) $ &   Time saved  &  $ \rm{RMSE}~ (\left<\widetilde{c} \right>^i) $ \\ 	
		\hline
		SPH   & 1482 & $ 655.5  $& - & $ 0.00388 $  \\
		ASPH ratio = 2.0 & 761  & $ 398.1 $& 39.2\%  & $ 0.00985 $ \\
		ASPH ratio = 4.0 &  402  & $ 233.9  $&  64.2\%  & $ 0.02267 $  \\
		ASPH ratio = 6.0 &  282  & $ 147.7 $&   77.5\% & $ 0.03847$   \\
		\hline	
	\end{tabular}
	\label{2d-membrane-comparison}
\end{table*}

\subsubsection{Three dimensional fluid diffusion coupling solid deformation}
Next, we explore fluid diffusion coupling swelling in a three-dimensional film,
specifically the diffusion of water within a porous Nafion membrane.
This system has been previously investigated numerically by Zhao \cite{zhao2013modeling}
and experimentally by Goswami \cite{goswami2008wetting}.  
This reference thin porous body takes the form of a polymer film
with an x-y plane of dimensions $L_x = 10.0$ mm,  $L_y = 10.0$ mm and a height of $L_z = 0.125$ mm. 
Four boundary sides are  constrained to prevent any curling or movement.
The physical parameters are consistent with those listed in Table \ref{parameter-table}, 
and the initial conditions resemble those employed in the two-dimensional case.
The central square part of the membrane in contact with water occupies 
a region of dimensions $0.3L_x \times 0.3L_y \times 0.5 L_z$, 
and this contact lasts for 450 seconds, 
with the total physical time  set to 2500 seconds. 
No fluid is allowed to diffuse out from the membrane.
The fluid saturation $\widetilde{c} $ 
in the central square part is constrained to 
$\widetilde{c}={c} = 0.4$ for the initial 450 seconds,
while in other regions $ \widetilde{c}_0 = 0$.
In order to provide a more accurate representation of the experiment, 
the evaporation process is taken into consideration,
acknowledging the gradual loss of water over time. 
During the initial period, deformation and flexure manifest,
and subsequently, as the fluid mass diminishes from the membrane,
a gradual restoration of the original shape is observed.

 \begin{figure*}[htbp]
 	\centering
 	\begin{subfigure}[b]{0.45\textwidth}
 		\includegraphics[trim = 1mm 2mm 2mm 2mm, clip,width=0.9\textwidth]{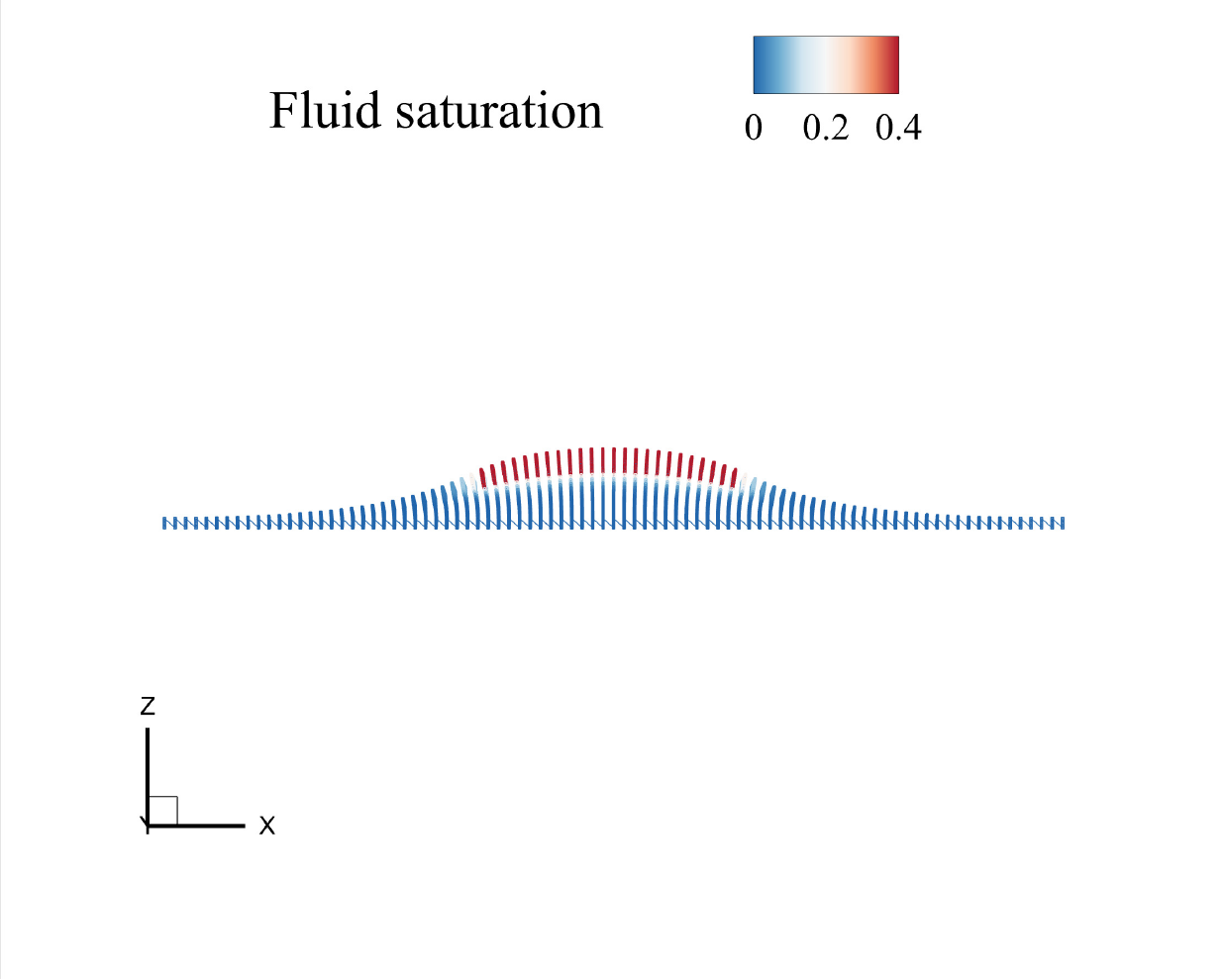}
 		\caption {t = 450s, front view}
 		\label{3d-saturaion-450-3}
 	\end{subfigure}
 	\begin{subfigure}[b]{0.45\textwidth}
 		\includegraphics[trim =  1mm 2mm 2mm 2mm, clip,width=0.9\textwidth]{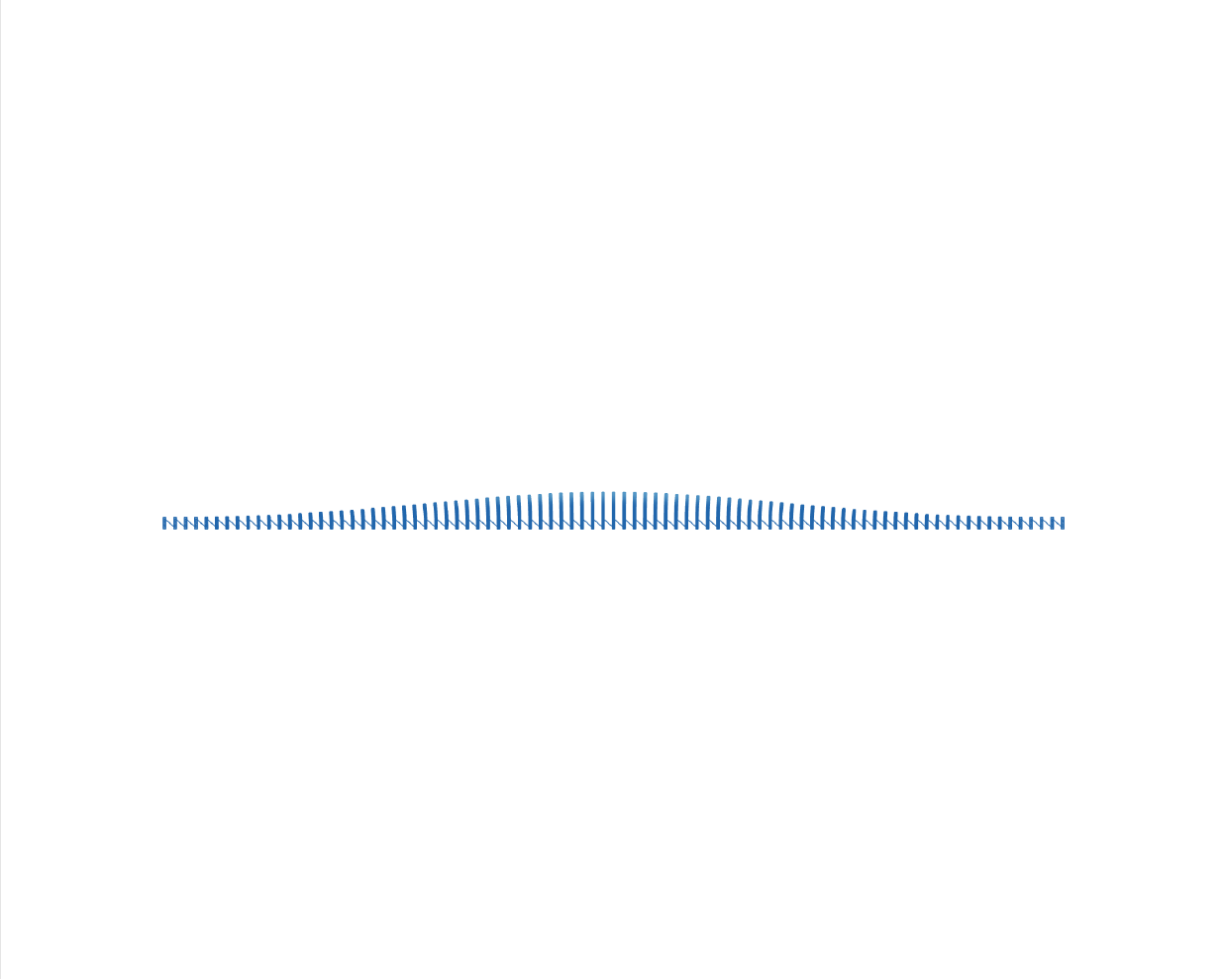}
 		\caption {t = 1500s, front view}
 		\label{3d-saturaion-1500}
 	\end{subfigure}
 	\begin{subfigure}[b]{0.45\textwidth}
 		\includegraphics[trim = 2mm 2mm 2mm 2mm, clip,width=0.9\textwidth]{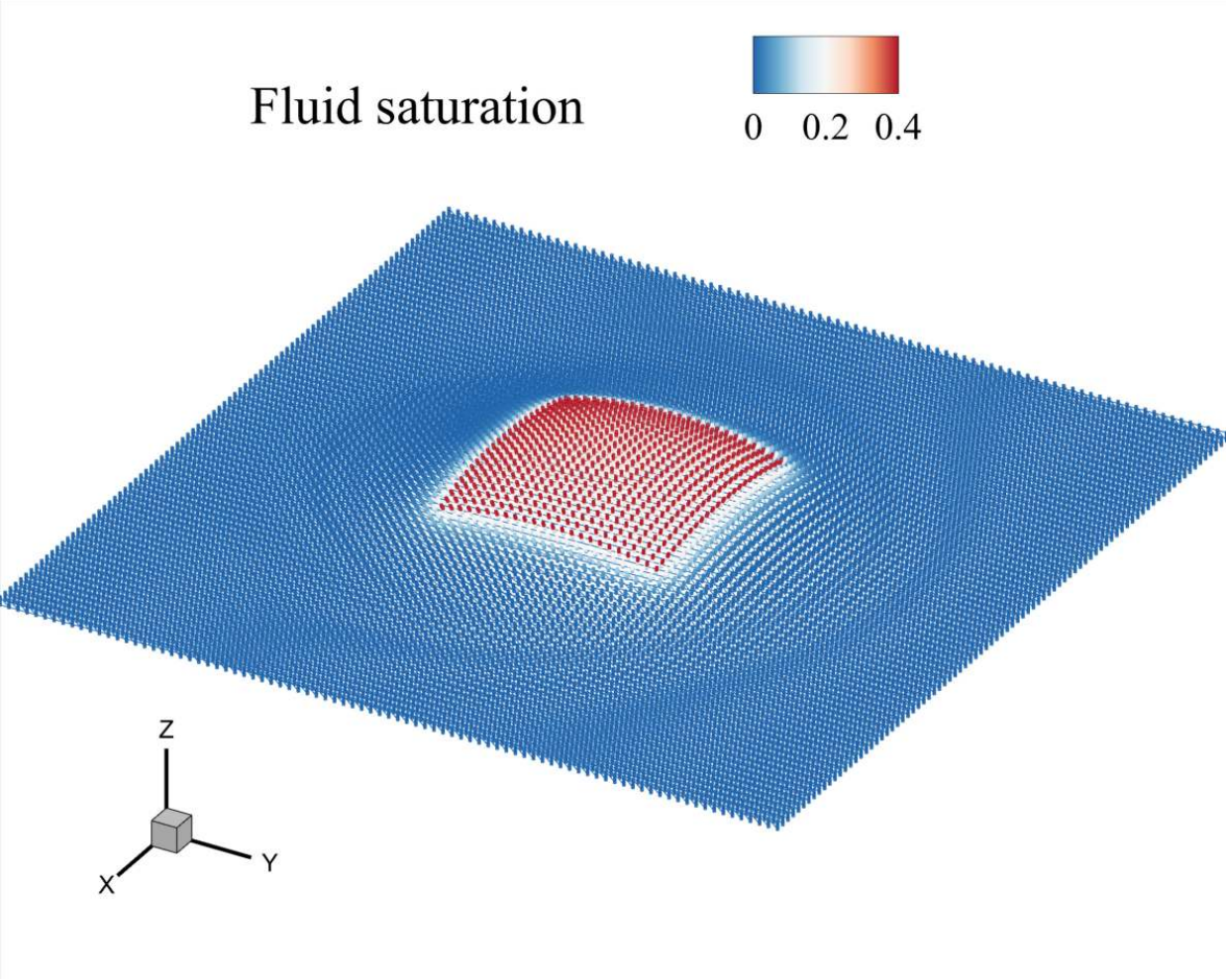}
 		\caption {t = 450s, top side view}
 		\label{3d-saturaion-450}
 	\end{subfigure}
 	\begin{subfigure}[b]{0.45\textwidth}
 		\includegraphics[trim = 2mm 2mm 2mm 2mm, clip,width=0.9\textwidth]{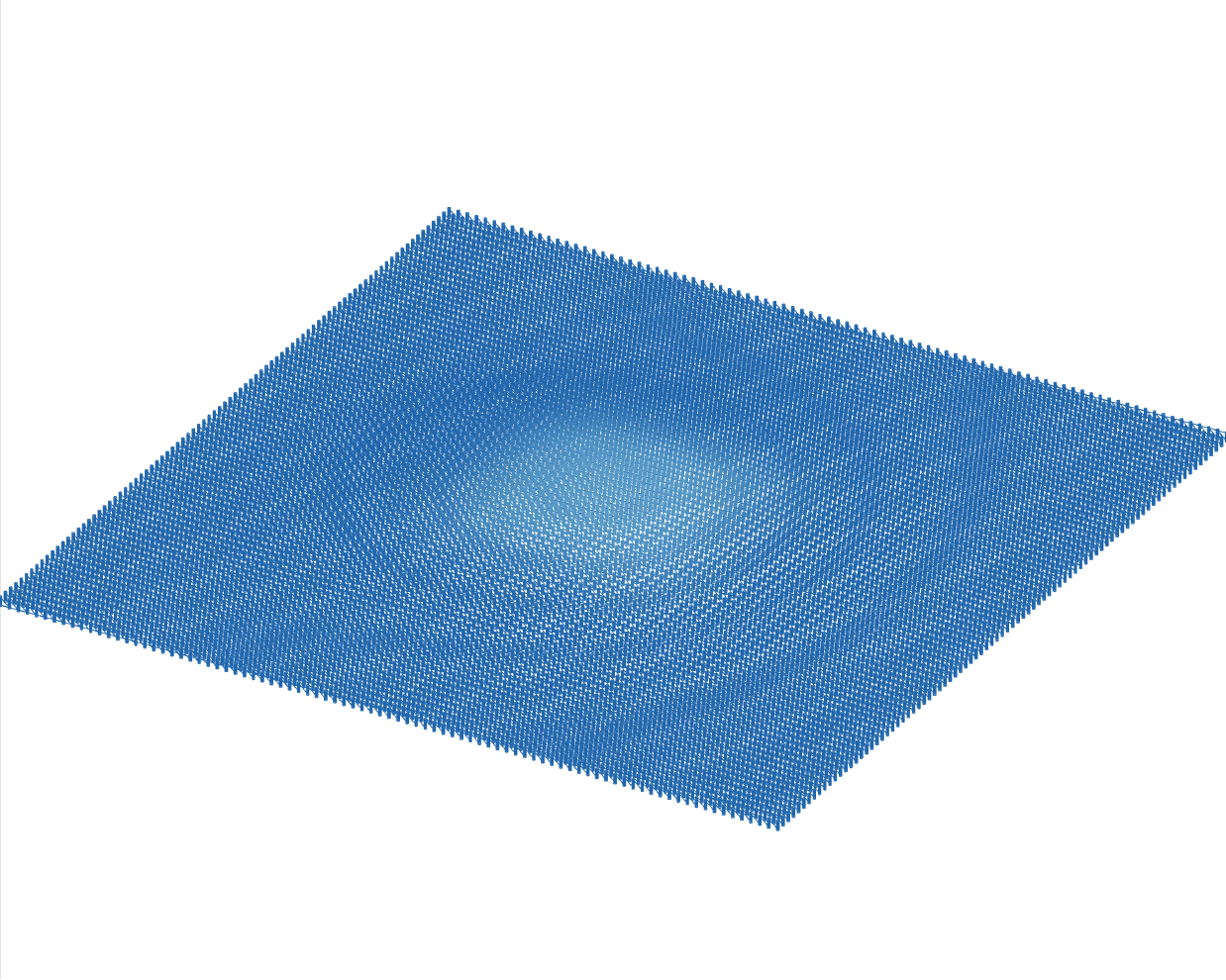}
 		\caption {t = 1500s, top side view}
 		\label{3d-saturaion-100}
 	\end{subfigure}
 	\begin{subfigure}[b]{0.45\textwidth}
 		\includegraphics[trim = 1mm 2mm 1mm 2mm, clip,width=0.9\textwidth]{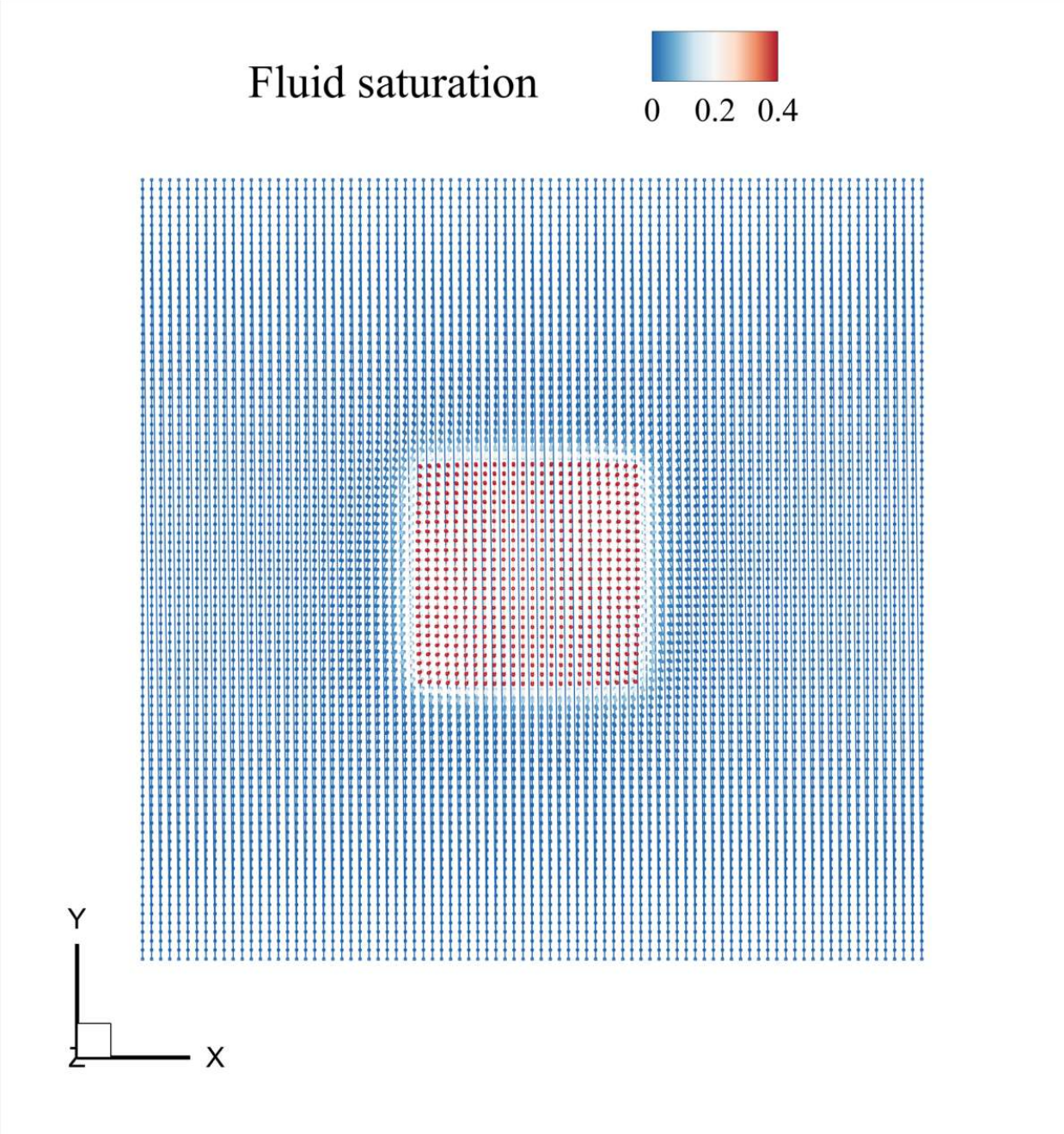}
 		\caption {t = 450s, top view}
 		\label{3d-saturaion-450-2}
 	\end{subfigure}
 	\begin{subfigure}[b]{0.45\textwidth}
 		\includegraphics[trim = 1mm 2mm 1mm 1mm, clip,width=0.9\textwidth]{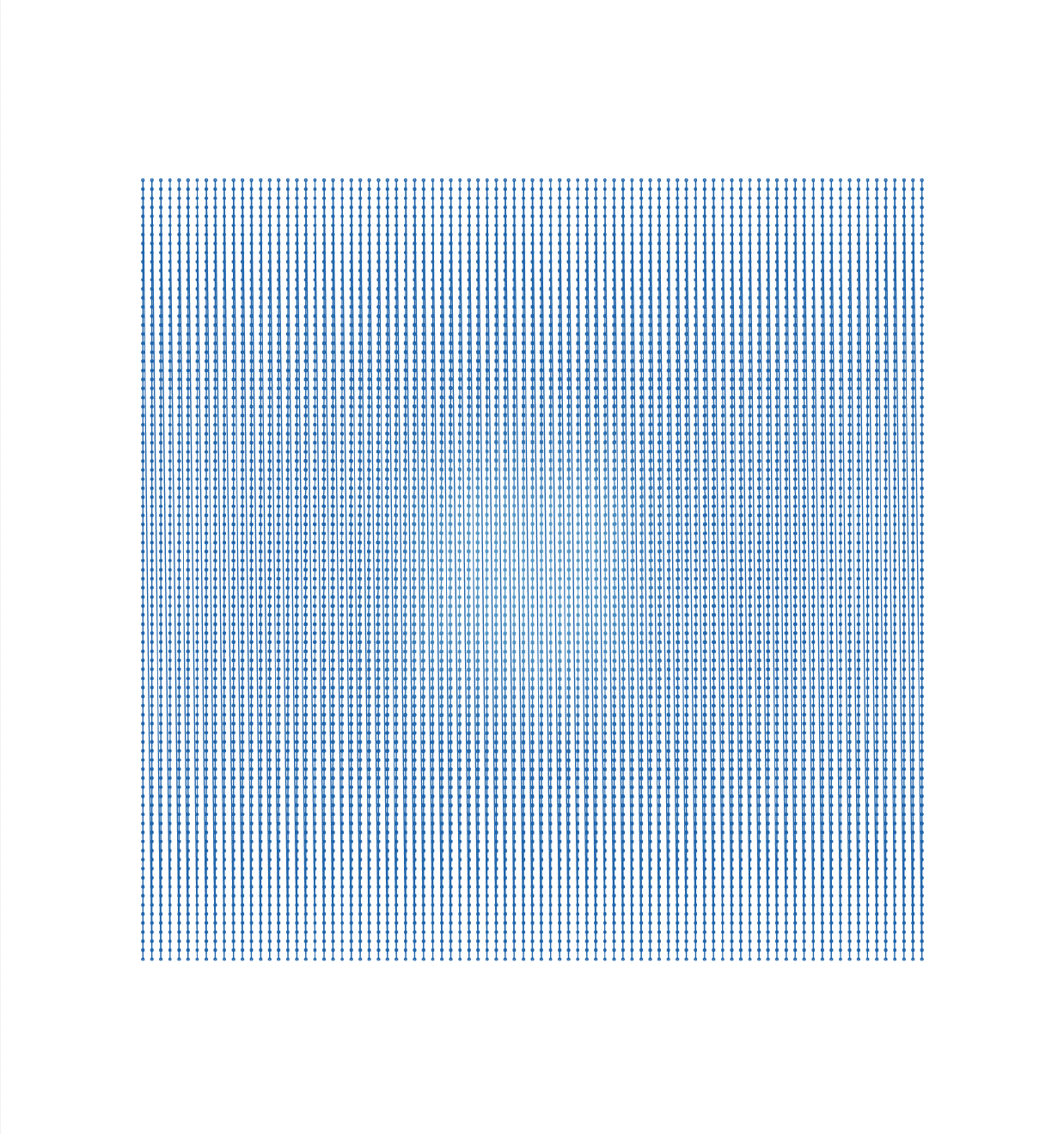}
 		\caption {t = 1500s, top view}
 		\label{3d-saturaion-1500-3}
 	\end{subfigure}
 	\caption{3D fluid-structure interaction: the deformation colored by fluid saturation at different time instants. }
 	\label{3ddiffusion}
 \end{figure*}
 
 \begin{figure*}[htbp]
 	\centering
 	\includegraphics[trim = 2mm 2mm 2mm 20mm, clip,width=0.65\textwidth]{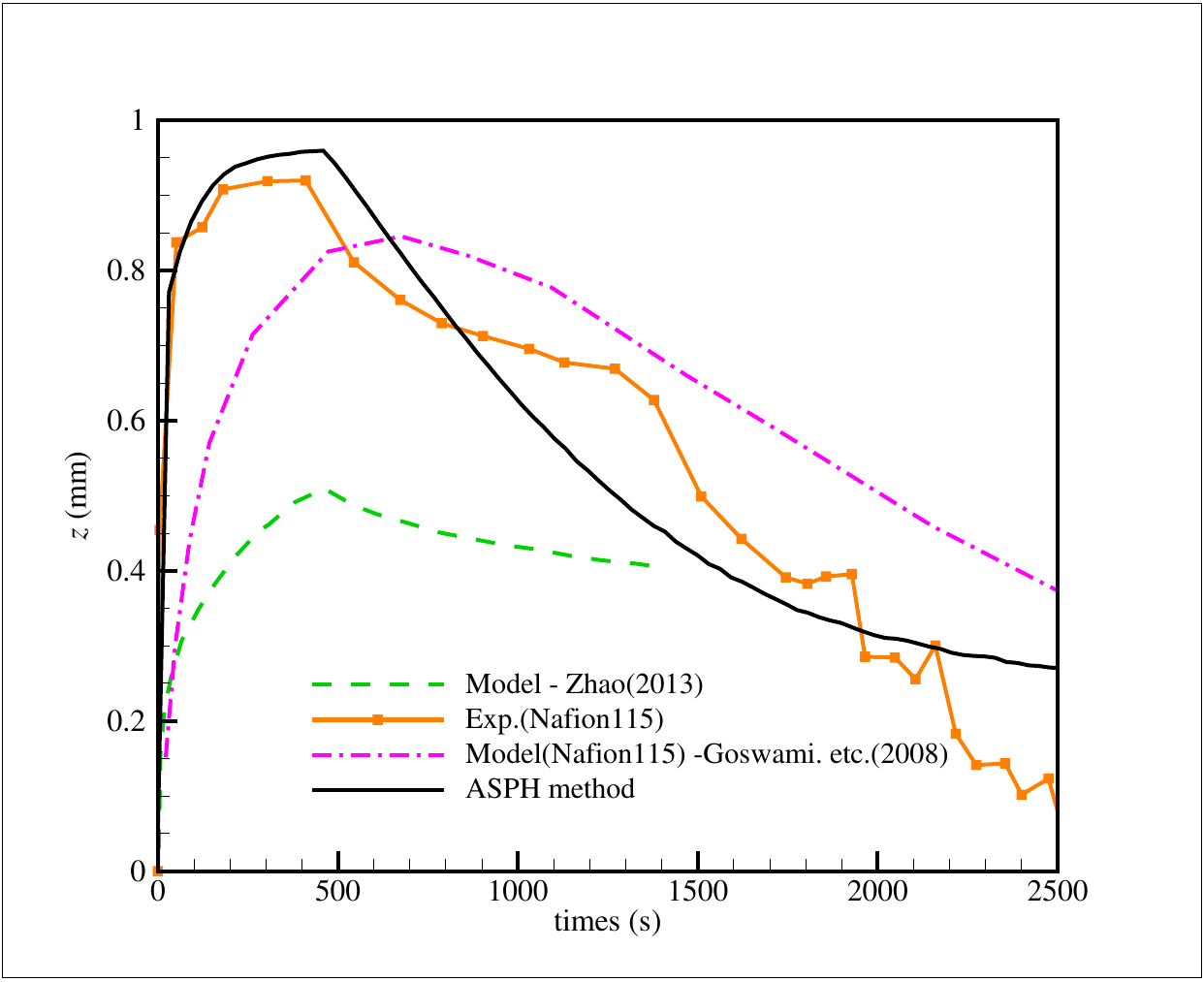}
 	\caption {3D fluid-structure interaction: bending amplitude of the center point compared with experimental data and results from other numerical models.}
 	\label{3dpoint_position}
 \end{figure*}

Figure \ref{3ddiffusion} illustrates membrane deformation 
colored by water saturation at different time instants,
indicating the diffusion evolution within the membrane.
Figure \ref{3dpoint_position}  depicts
the swelling degree  $z$  of the central 
point versus different time instants,
which are derived from the present method as well as other numerical models
and the corresponding data 
points from Goswami's experimental measurements \cite{goswami2008wetting}.
The numerical simulation results obtained from the present ASPH method
exhibit good agreement with experimental results,
capturing the deformation amplitude pattern, 
reproducing the increasing flexure  during the water contact period
and the subsequent decrease after the contact phase, consistent with saturation variations.
\newpage
\subsection{Nonisotropic diffusion with anisotropic diffusion tensor}
In many cases, diffusion processes in the real problems  
are directionally dependent or anisotropic.  
In this section, 
simulations for a contaminant source diffusing in water
are firstly carried out, 
and the results are compared with the corresponding 
analytical solutions to demonstrate the accuracy.
Then the cardiac function coupling with nonisotropic diffusion
is simulated to verify the application of this method to complex 
biological problems.

\subsubsection{2d source diffusion}
In this section, we consider a contaminant source in water,
which has been studied by Tran-Duc et al. \cite{tran2016simulation}
with numerical and analytical results available to compare with.
The contaminant source is located in the center of 
a square computation domain spanning dimensions of 200 m × 200 m.
Throughout the simulation,
the initial contaminant spatial distributions
are generated utilizing analytical solutions detailed as
\begin{equation}
C(x, t)=\frac{1}{4 \pi t \prod_{\alpha = 1}^{2} \sqrt{\mathbf{D}_{\alpha \alpha}}} \prod_{\alpha=1}^{2} \exp \left[-\frac{\left(x_{\alpha}-x_{\alpha, 0}\right)^{2}}{4 t \mathbf{D}_{\alpha \alpha}}\right],
\end{equation}
 at a temporal instance of t = 120 s (2 min).  
Subsequently, the contaminant distributions following 30 minutes
of diffusion are juxtaposed against corresponding analytical solutions.
Spatial resolution, denoted as the particle spacing, is set at  x= 0.5 m. Consequently, the respective quantities of SPH particles  are determined to be 160,000.
In the first case, the anisotropic diffusion tensor is 
\begin{equation}
	\mathbf{D} = \begin{bmatrix}
	0.1 &0.0 \\ 
	0.0 &0.01   
	\end{bmatrix} \rm{(m^2 ~ s^{-1})},
\end{equation}
in which diffusion rates in the x-direction surpasses that in the y-direction. 

\begin{figure*}[htbp]
	\centering
	\begin{subfigure}[b]{0.45\textwidth}
		\includegraphics[trim =2mm 2mm 2mm 2mm, clip,width=0.95\textwidth] {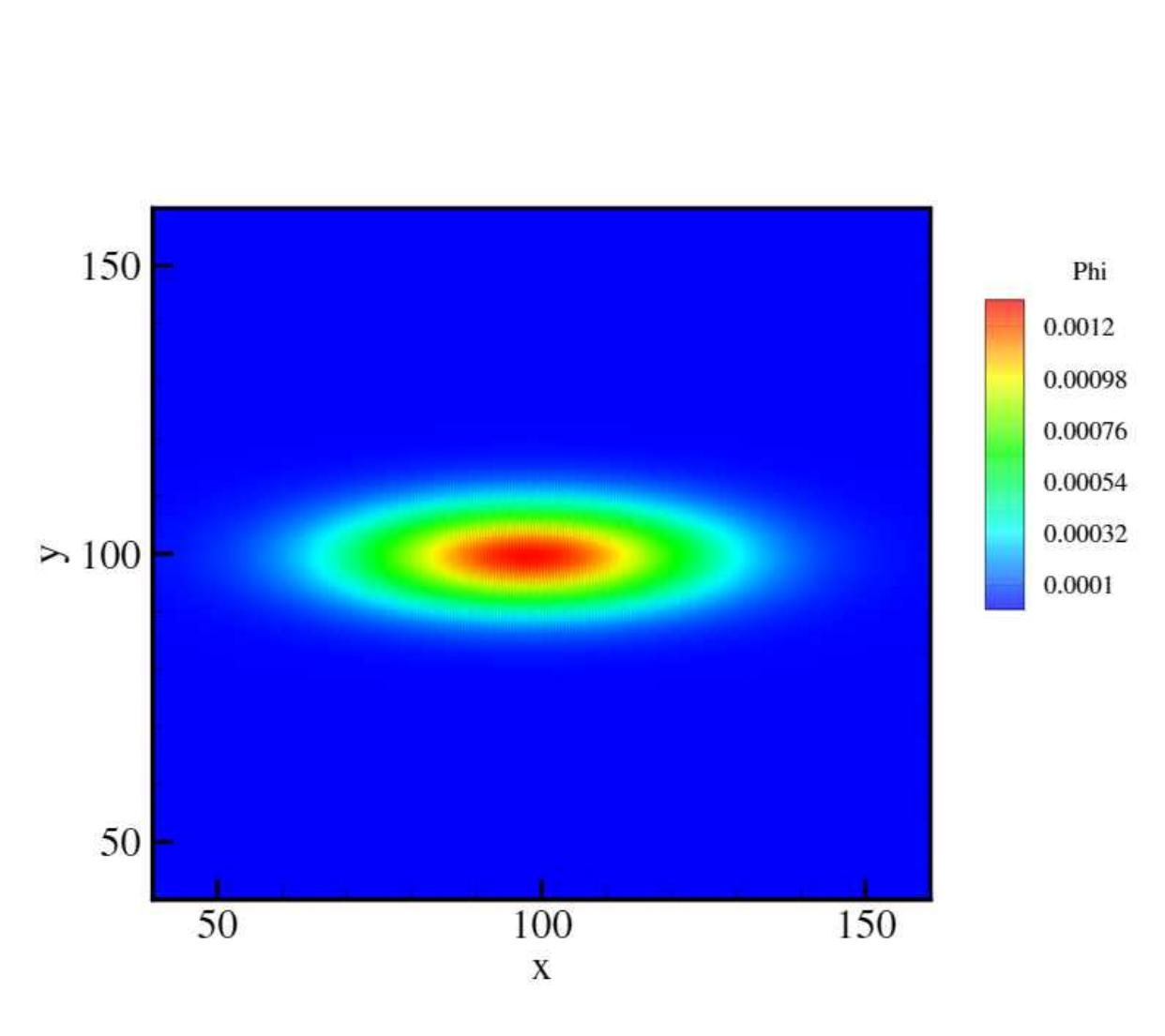}
		\caption {The present ASPH method.}
		\label{D3ASPH_solution}
	\end{subfigure}
	\begin{subfigure}[b]{0.45\textwidth}
		\includegraphics[trim =2mm 2mm 2mm 2mm, clip,width=0.95\textwidth]{R10D3_countour-eps-converted-to.pdf}
		\caption  {The analytical solution.}
		\label{D3analytical_solution}
	\end{subfigure}
	\caption{2D anisotropic diffusion: comparisons of the concentration distribution obtained from (a)the present ASPH method and (b) the analytical solution.}
	\label{D3concentration_solutions}
\end{figure*}

\begin{figure*}[htbp]
	\centering
	\begin{subfigure}[b]{0.45\textwidth}
		\includegraphics[trim =2mm 2mm 2mm 2mm, clip,width=0.95\textwidth]{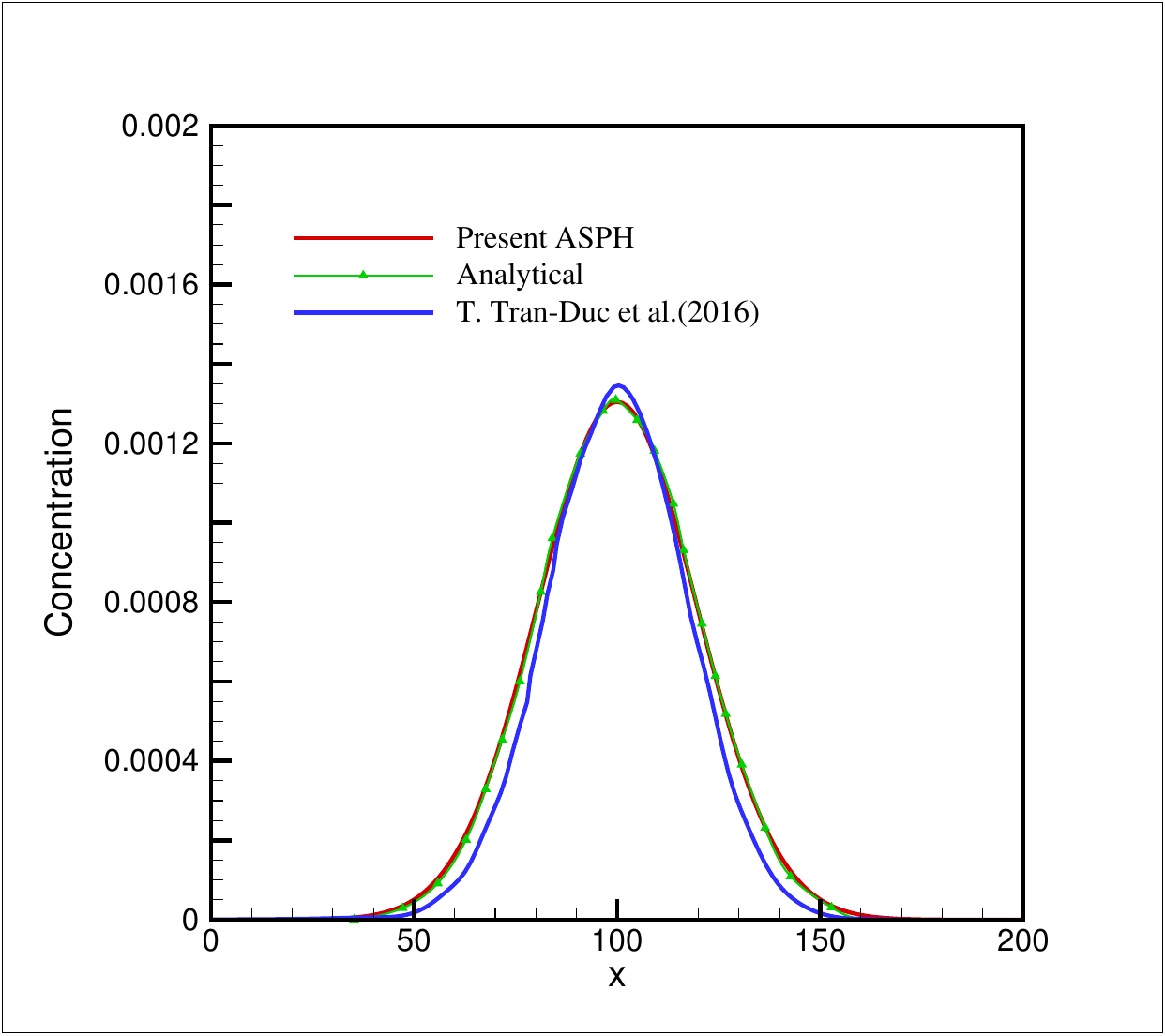}
		\caption{}
		\label{D3xcross-section}
	\end{subfigure}
	\begin{subfigure}[b]{0.45\textwidth}
		\includegraphics[trim =2mm 2mm 2mm 2mm, clip,width=0.95\textwidth]{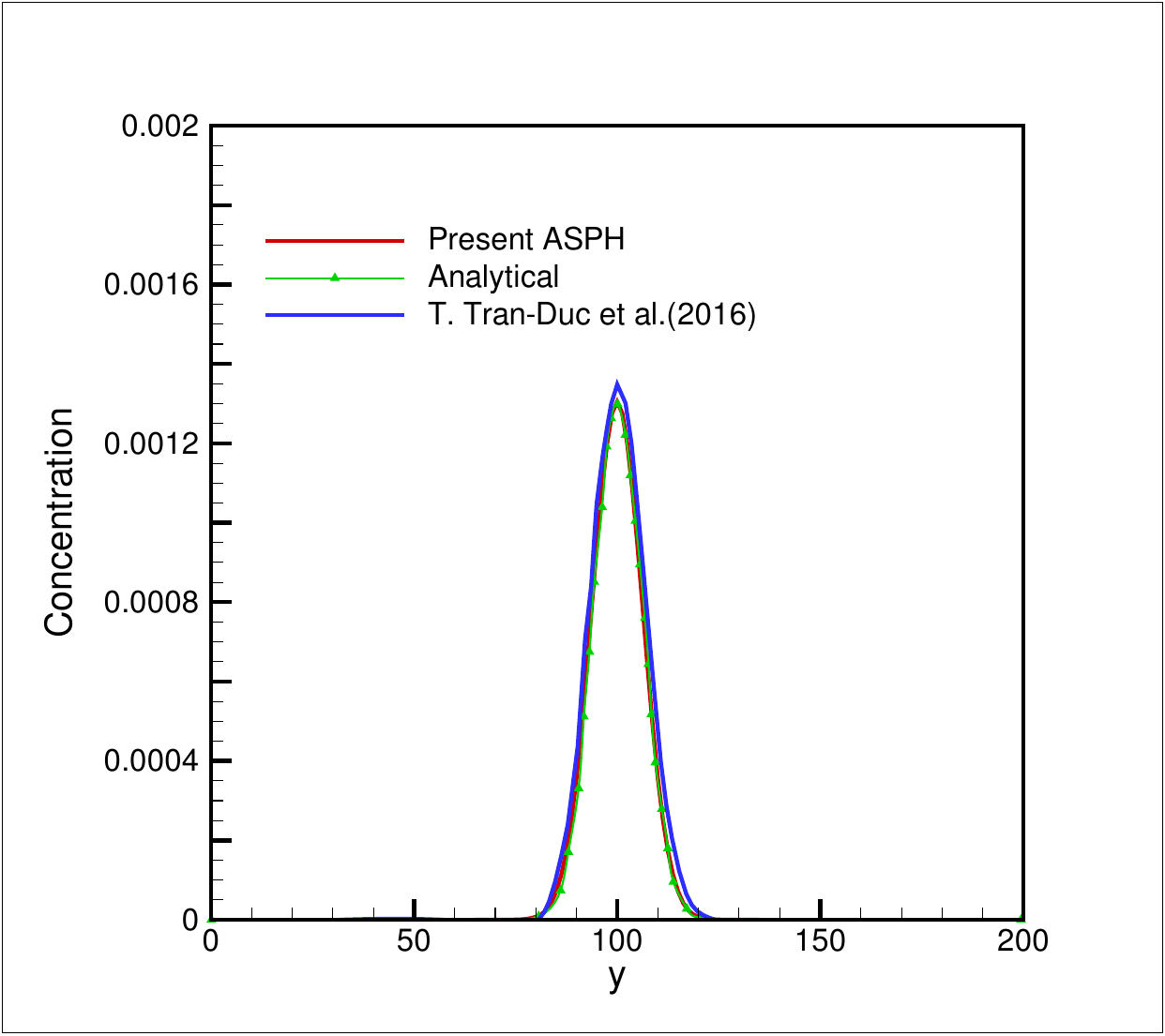}
		\caption{}
		\label{D3ycross-section}
	\end{subfigure}
	\caption{2D anisotropic diffusion: comparisons of the concentration distribution   of (a) a horizontal cross-section at y = 100 m (b) a vertical cross-section at x = 100 m.}
	\label{D3xycross-section}
\end{figure*}

Figure \ref{D3concentration_solutions} illustrates both the numerical and analytical distributions of time t = 1920 s. 
Generally, the concentration distributions exhibit 
elliptical shapes with the major axis aligned in the x-direction
and the minor axis in the y-direction,
which arises from the inherent discrepancy 
in diffusion rates between the x and y directions.
Notably, a notable concordance is observed between 
the numerical solution and the analytical counterpart 
in both their shapes and value profiles.
For a more rigorous comparison, the current numerical concentration distributions at a horizontal cross-section at y = 100 m (Figure. \ref{D3xcross-section}) and a vertical cross-section at x = 100 m (Figure. \ref{D3ycross-section}) are depicted, 
alongside a comparative analysis with analytical solutions 
and the results by Tran-Duc et al. \cite{tran2016simulation} in Figure 4. 
The  concentration profiles approximated through ASPHAD in the refer \cite{tran2016simulation} is thinner in  x-direction and more stretched in y-direction,
while the current ASPH distribution 
demonstrate a more commendable agreement with the analytical solution.
The current results mitigate the level of anisotropy inherent in the diffusion process,
displaying an agreed anisotropy in comparison to the analytical solution.  

\begin{figure*}[htbp]
 	\centering
 	\begin{subfigure}[b]{0.45\textwidth}
 		\includegraphics[trim =2mm 2mm 2mm 2mm, clip,width=0.95\textwidth]{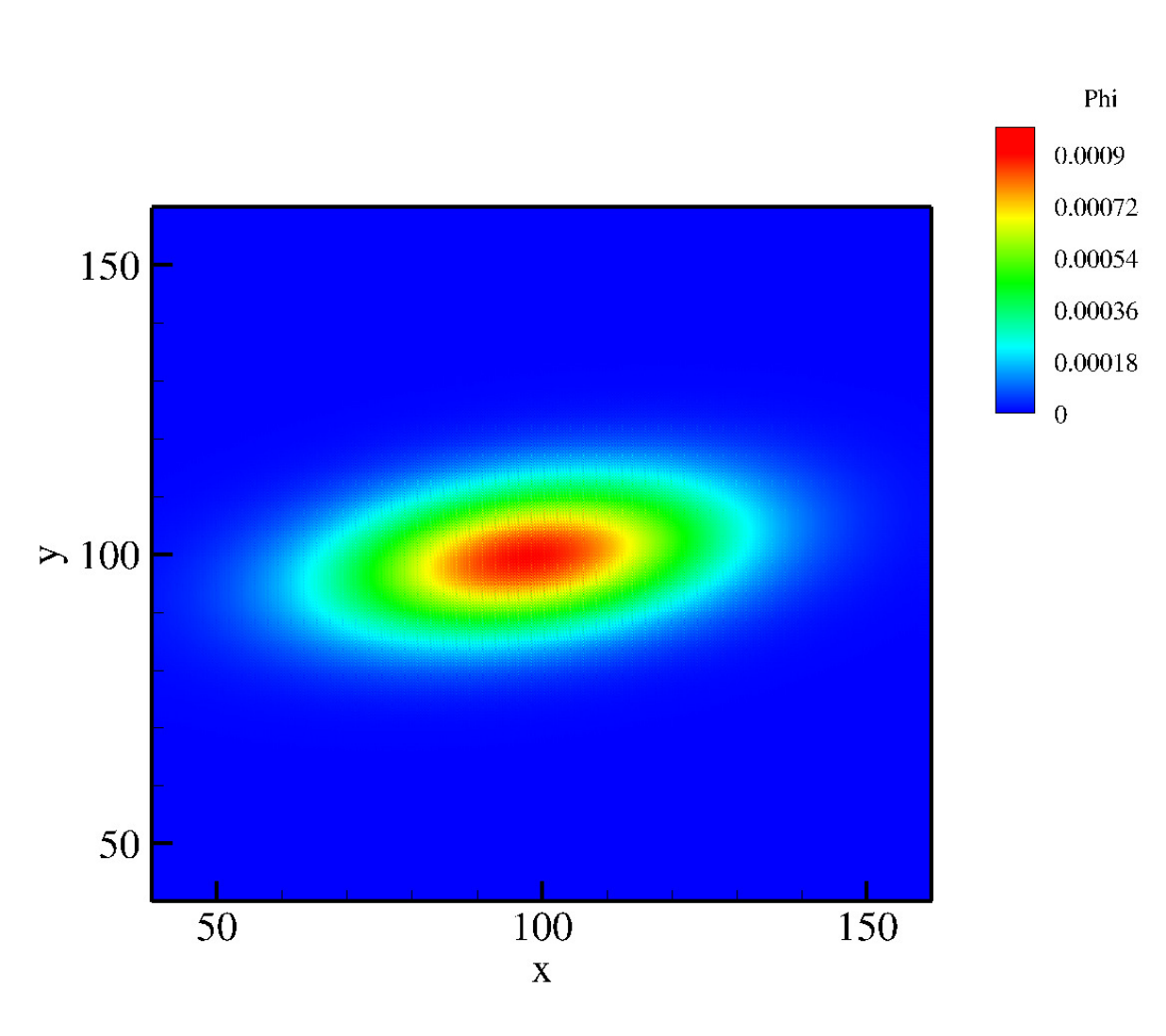}
 		\caption {The present ASPH method.}
 		\label{D4ASPH_solution}
 	\end{subfigure}
 	\begin{subfigure}[b]{0.45\textwidth}
 		\includegraphics[trim =2mm 2mm 2mm 2mm, clip,width=0.95\textwidth]{D4_countour-eps-converted-to.pdf}
 		\caption  {The analytical solution.}
 		\label{D4analytical_solution}
 	\end{subfigure}
 	\caption{2D anisotropic diffusion: comparisons of the concentration distribution obtained from (a)the present ASPH method and (b) the analytical solution.}
 	\label{D4concentration_solutions}
 \end{figure*}

\begin{figure*}[htbp]
	\centering
	\begin{subfigure}[b]{0.45\textwidth}
		\includegraphics[trim =2mm 2mm 2mm 2mm, clip,width=0.95\textwidth]{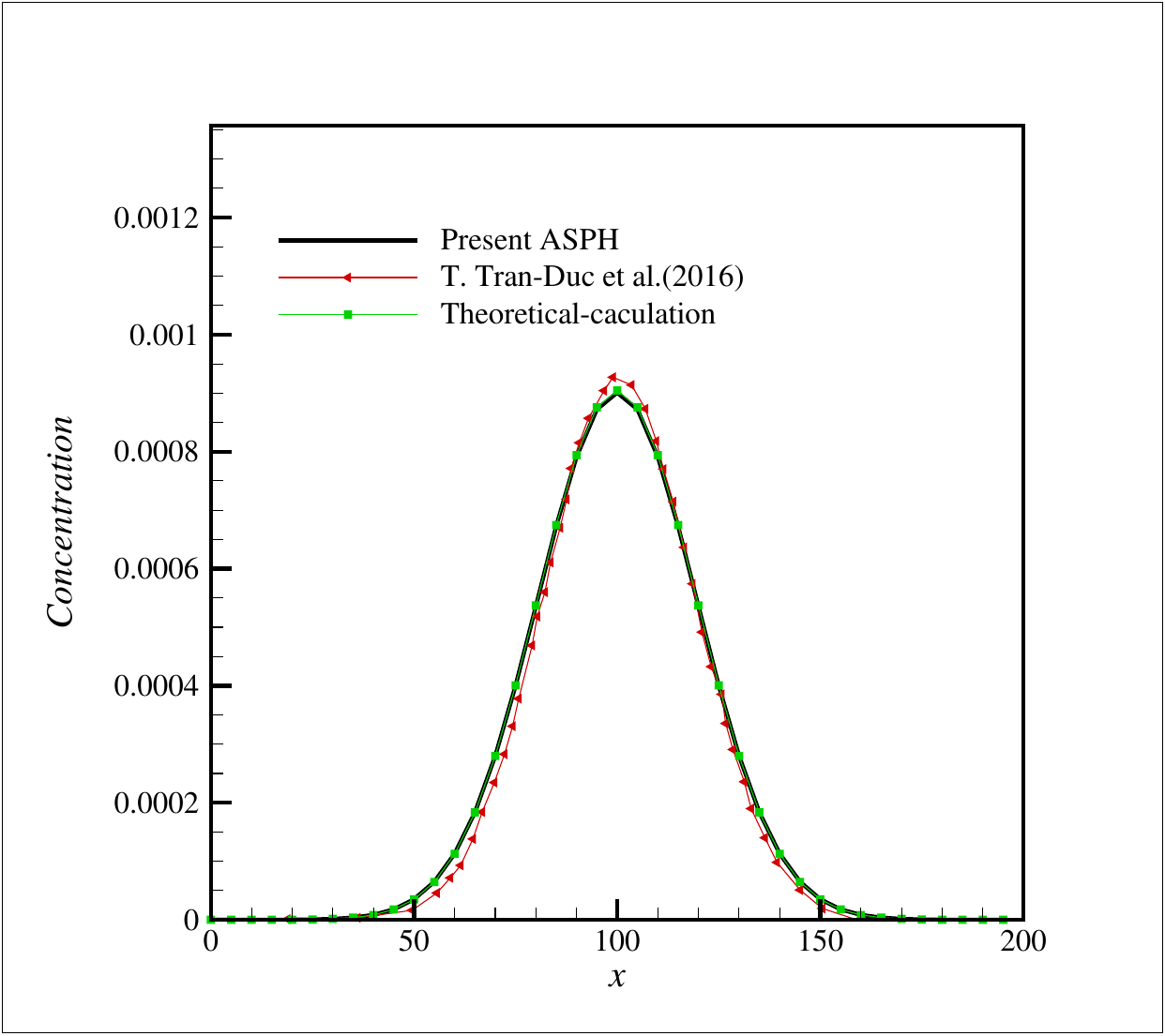}
		\caption{}
		\label{D4xcross-section}
	\end{subfigure}
	\begin{subfigure}[b]{0.45\textwidth}
		\includegraphics[trim =2mm 2mm 2mm 2mm, clip,width=0.95\textwidth]{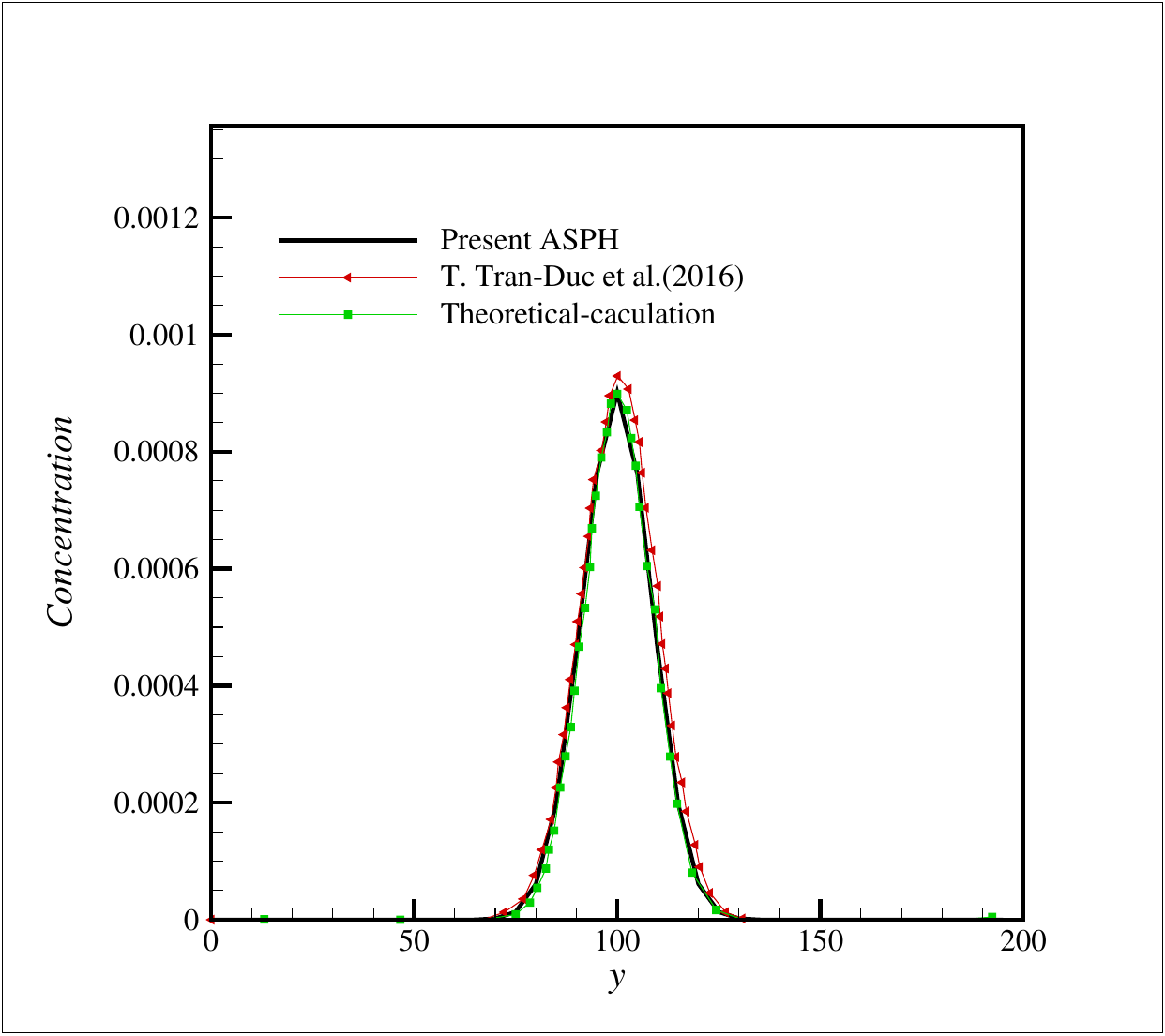}
		\caption{}
		\label{D4ycross-section}
	\end{subfigure}
	\caption{2D anisotropic diffusion: comparisons of the concentration distribution   of (a) a horizontal cross-section at y = 100 m (b) a vertical cross-section at x = 100 m.}
	\label{D4xycross-section}
\end{figure*}

In the subsequent test,  we consider a more general case  with 
a  full diffusion tensor   
\begin{equation}
	\mathbf{D} = \begin{bmatrix}
		0.1 &0.03 \\ 
		0.03 &0.03   
	\end{bmatrix} \rm{(m^2 ~ s^{-1})}.
\end{equation}  
The simulated concentration distribution portrayed 
in Figure. \ref{D4concentration_solutions} 
exhibits a notable coherence with the analytical solution, 
evident in both shape and concentration values at time t = 1920 s. 
As anticipated, the concentration distributions of present ASPH 
has orientation but maintain an elliptical form as similar as the analytical distribution, 
which means that ASPH conserves the principal diffusing
directions.
Figures. \ref{D4xcross-section} and \ref{D4ycross-section} present the current numerical concentration distributions 
as well as the analytical solution at the horizontal cross-section, 
specifically at x = 100 m, and the vertical cross-section at y = 100 m, respectively.
Again, the ASPH demonstrates enhanced computational accuracy. 
while discernible reduction in the level of anisotropy 
is observed in the ASPHAD distribution when juxtaposed with previous findings.

\subsubsection{Simulation of transmembrane potential propagation}
\begin{figure*}[htbp]
	\centering
	\begin{subfigure}[b]{0.32\textwidth}
		\includegraphics[trim =2mm 2mm 2mm 2mm, clip,width=0.99\textwidth]{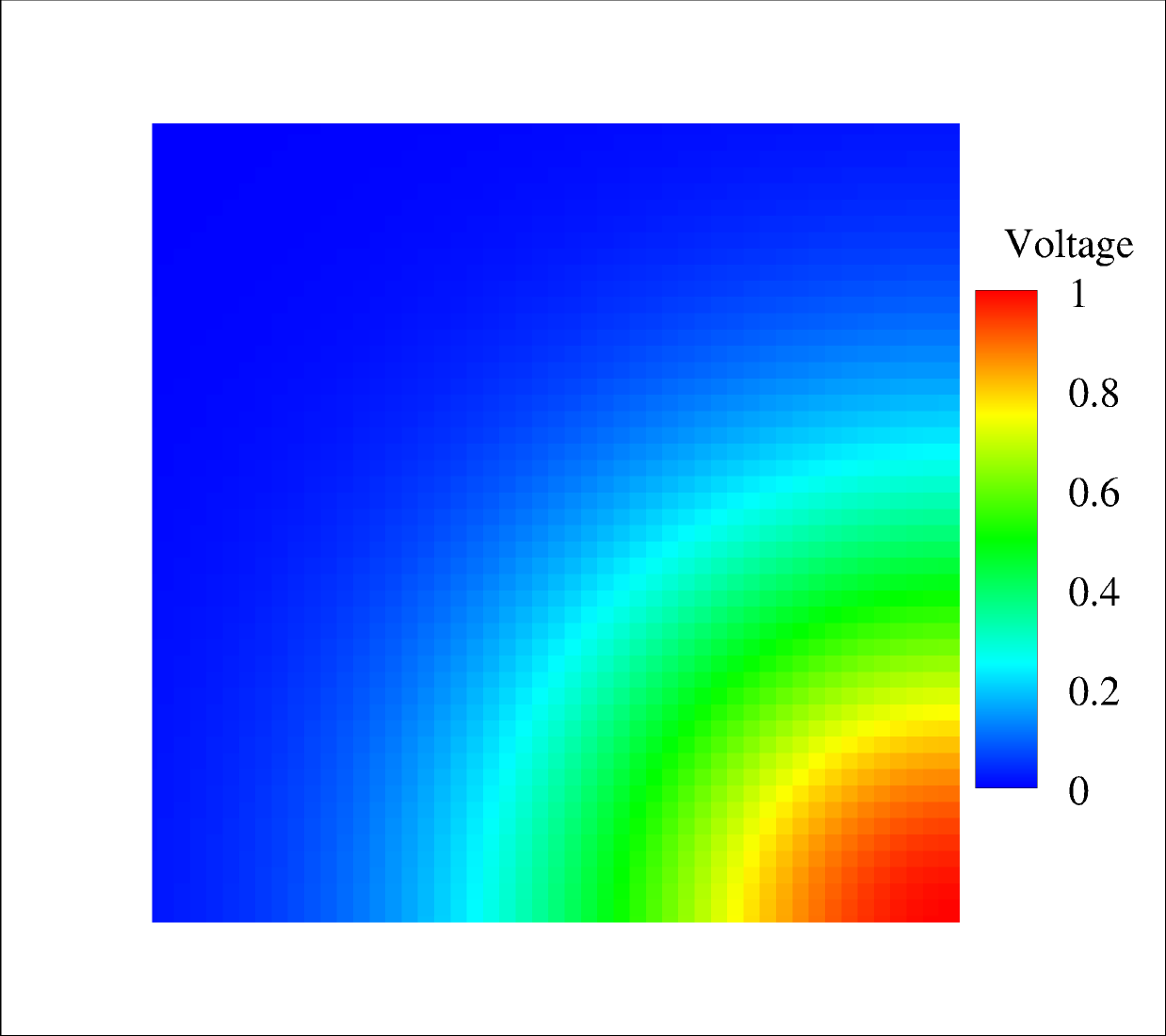}
		\caption  {t = 0.}
	\end{subfigure}
	\begin{subfigure}[b]{0.32\textwidth}
		\includegraphics[trim =2mm 2mm 2mm 2mm, clip,width=0.99\textwidth]{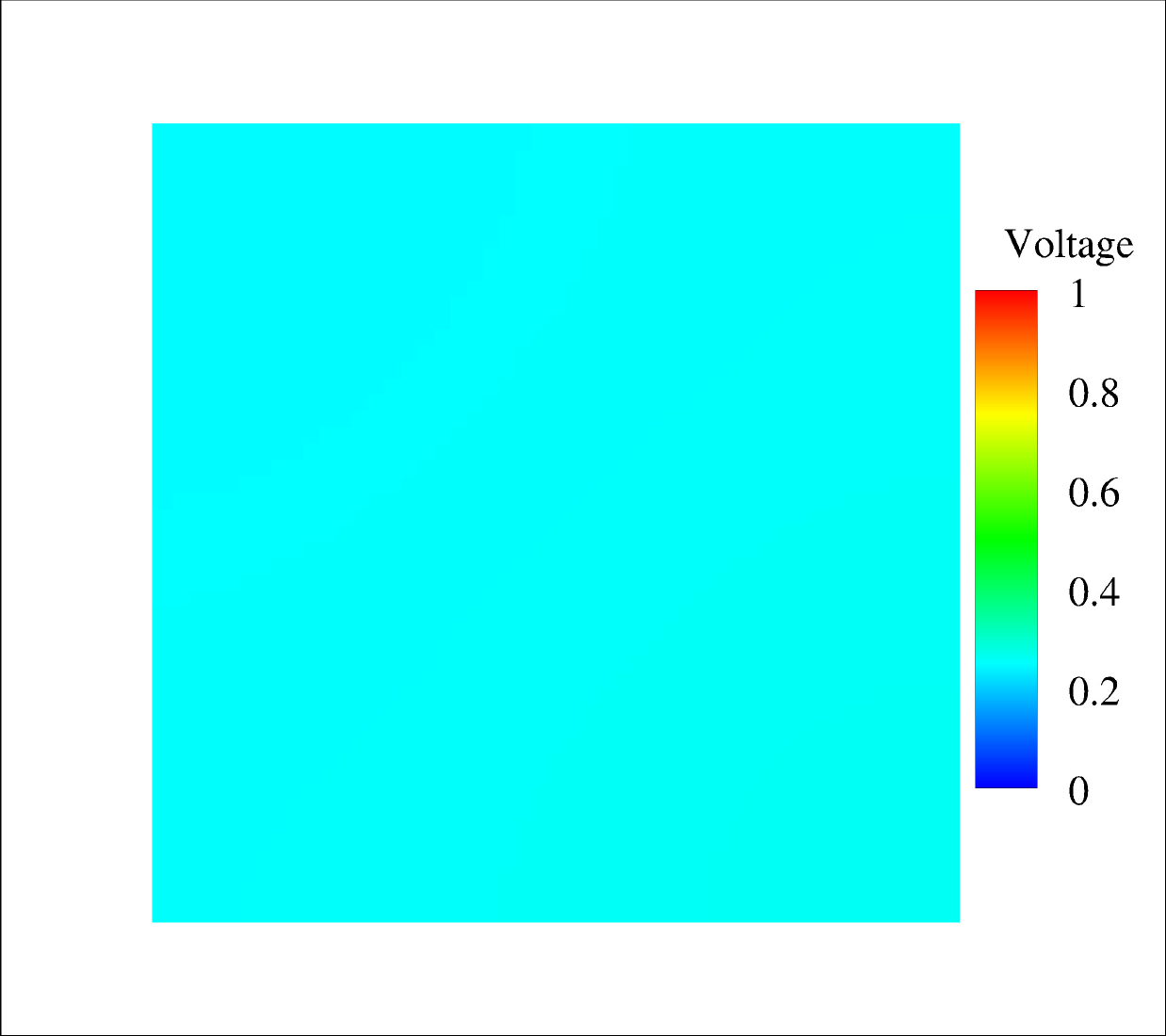}
		\caption  {t = 0.5.}
	\end{subfigure}
	\begin{subfigure}[b]{0.32\textwidth}
		\includegraphics[trim =2mm 2mm 2mm 2mm, clip,width=0.99\textwidth]{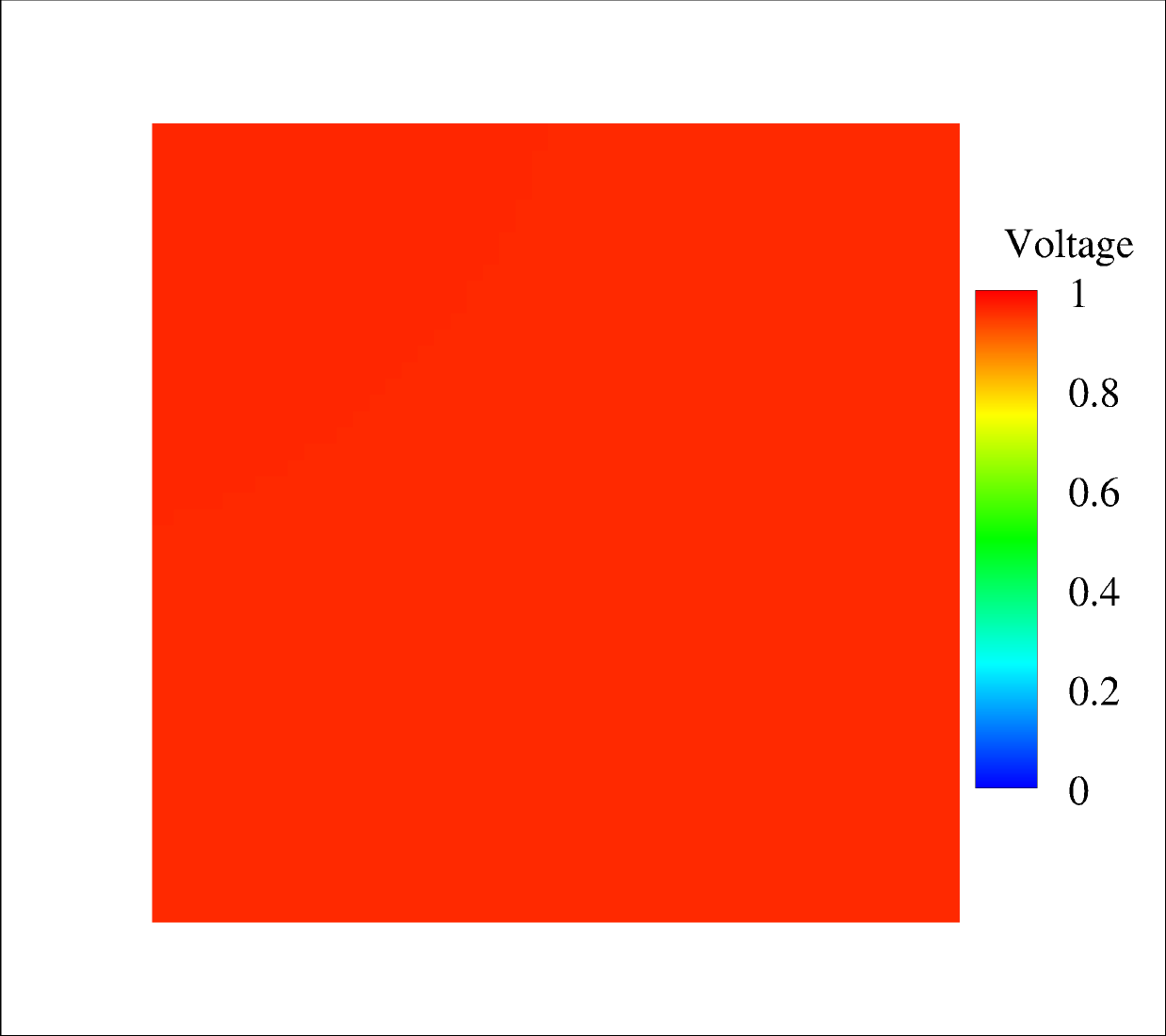}
		\caption  {t = 2.5.} 
	\end{subfigure}
	\begin{subfigure}[b]{0.32\textwidth}
		\includegraphics[trim =2mm 2mm 2mm 2mm, clip,width=0.99\textwidth]{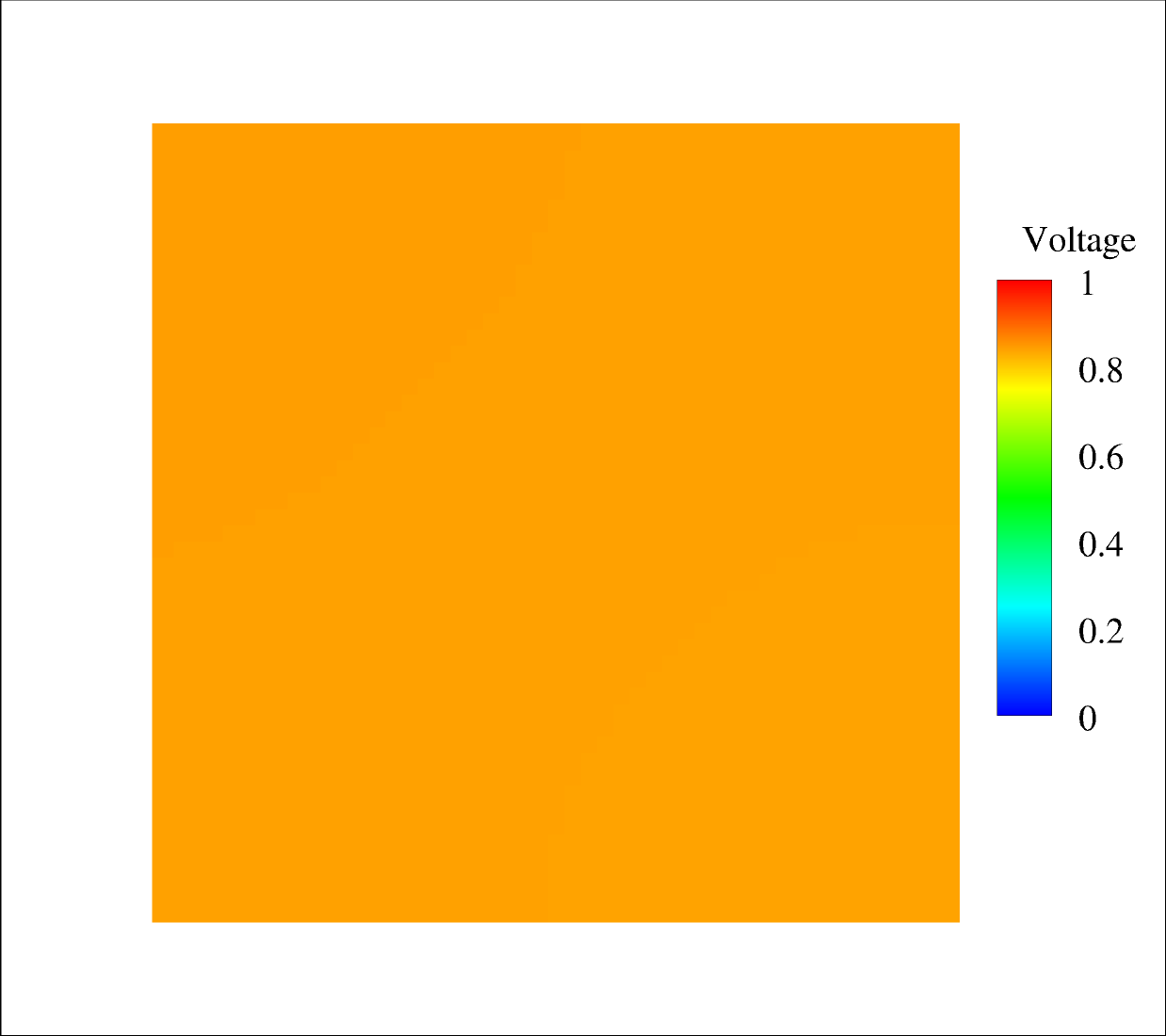}
		\caption {t = 7.}
	\end{subfigure}
	\begin{subfigure}[b]{0.31\textwidth}
		\includegraphics[trim =2mm 2mm 2mm 2mm, clip,width=0.99\textwidth]{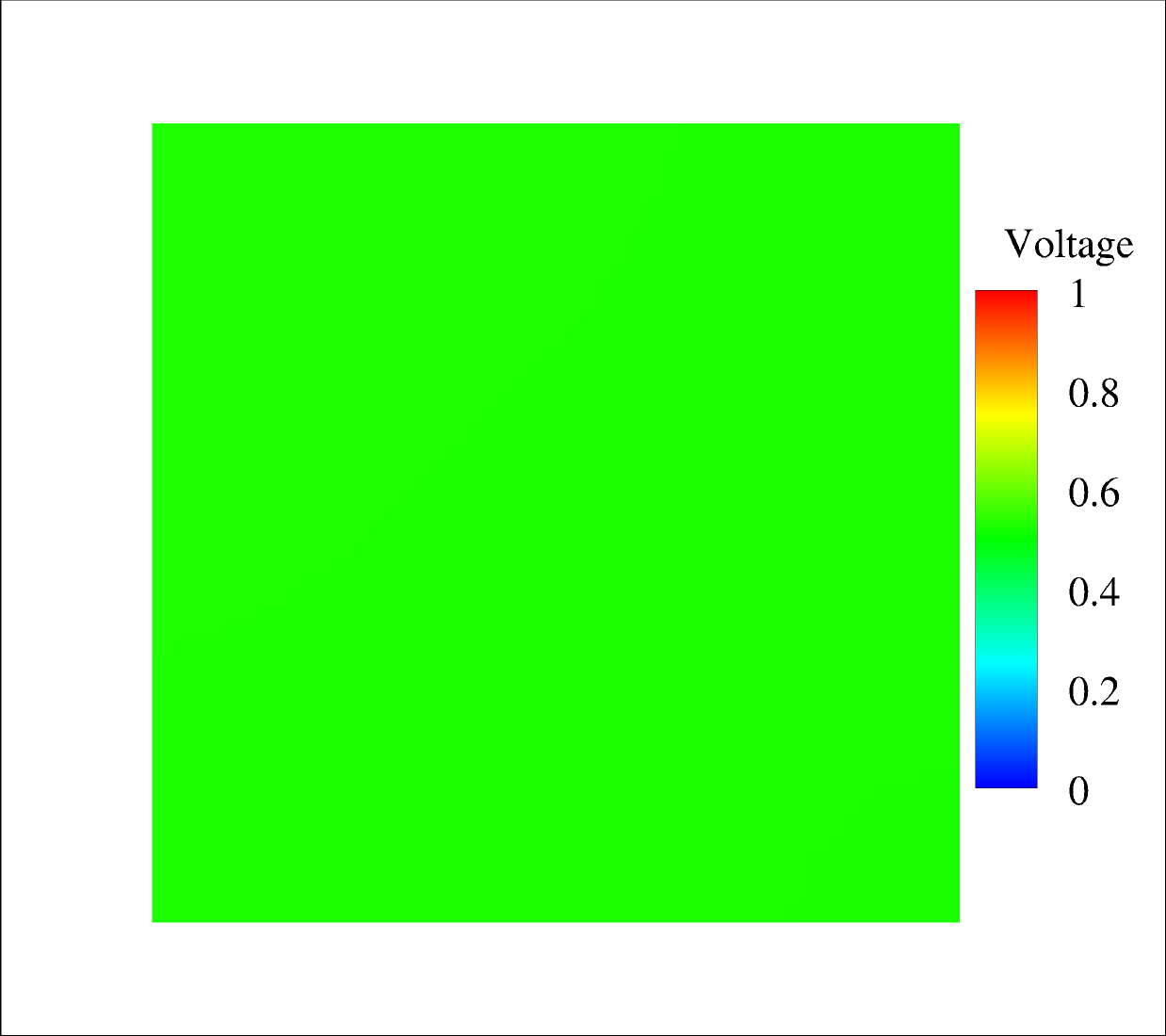}
		\caption  {t = 10.}
	\end{subfigure}
	\begin{subfigure}[b]{0.32\textwidth}
		\includegraphics[trim =2mm 2mm 2mm 2mm, clip,width=0.99\textwidth]{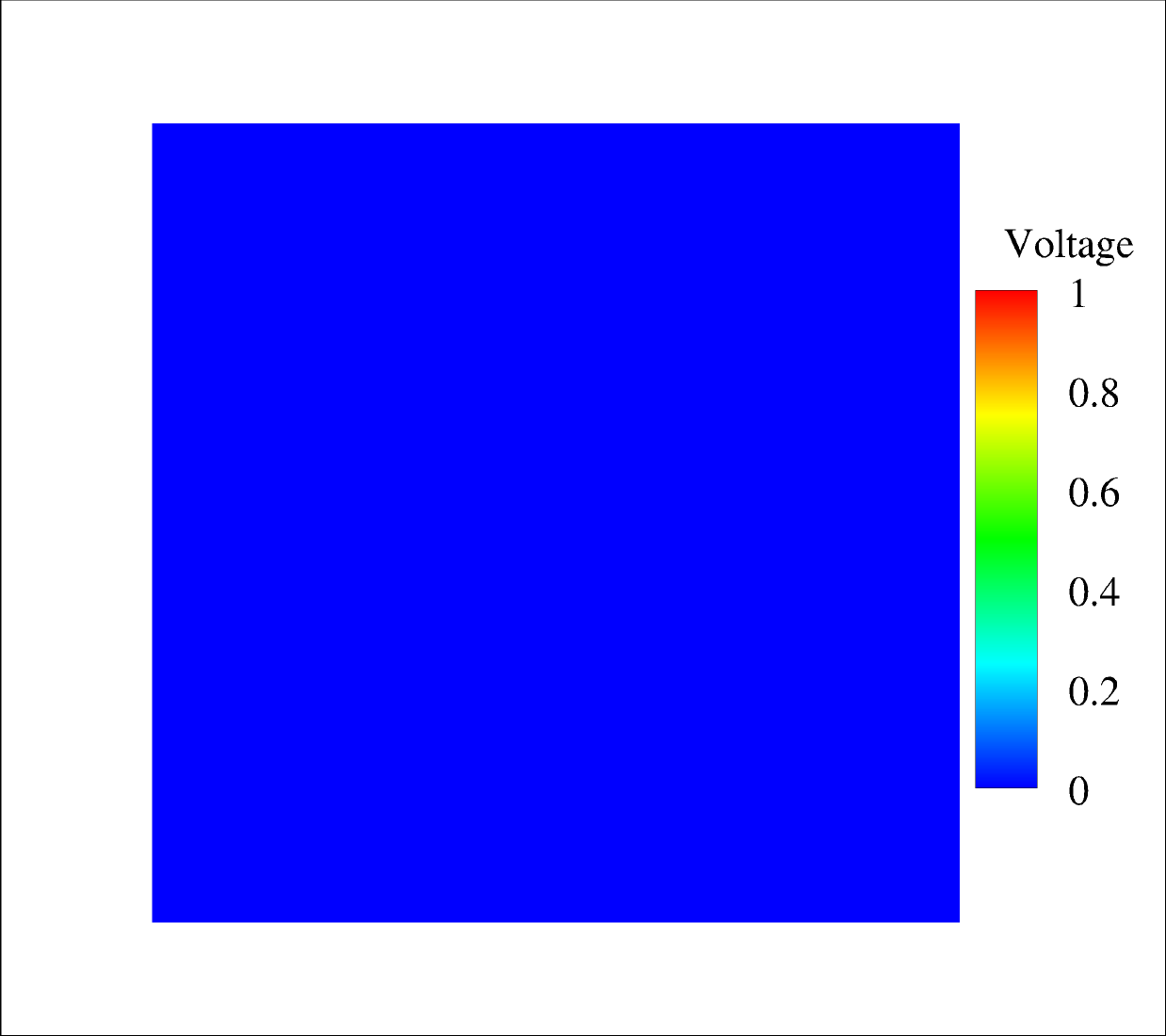}
		\caption  {t = 14.}
	\end{subfigure}
	\caption{Snapshots of the transmembrane potential evolution simulation.}
	\label{depo-coutour}
\end{figure*}

\begin{figure*}[htbp]
	\centering
	\includegraphics[trim = 2mm 2mm 2mm 2mm, clip,width=0.65\textwidth]{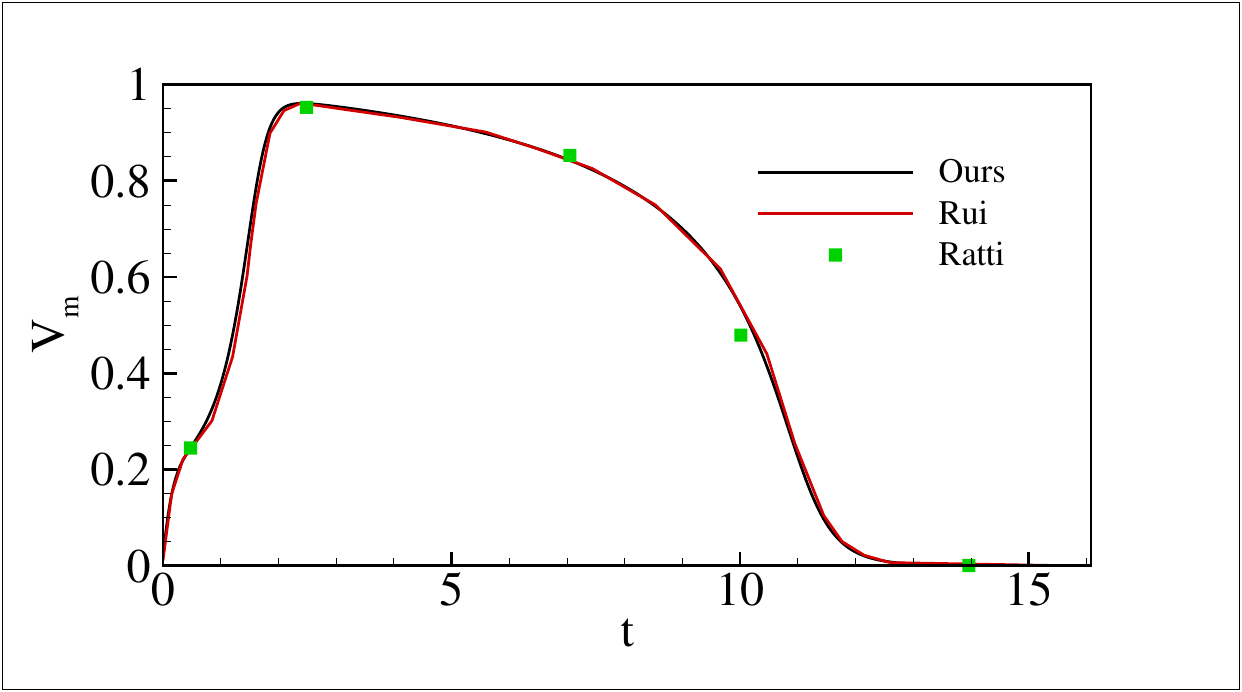}
	\caption {The time evolution of the transmembrane potential at point (0.3, 0.7), compared with the simulation results reported by Ratti and Verani \cite{ratti2019posteriori} and Rui et al. \cite{chen2024coupling}. } 
	\label{transmembrane-potential}
\end{figure*}

The application of the present method is verified by a comprehensive example of heart,
encompassing electrophysiology, passive mechanical response, and electromechanical coupling.
Initially, we conduct benchmarking on the electrical activity of the heart.
\begin{table*}[htp!]
	\centering
	\caption{Transmembrane potential propagation: parameters in Aliev-Panfilov model.}
	\begin{tabular}{cccccc}
		\hline
		 k = 8.0 &   a = 0.15 &   b = 0.15&  $ \epsilon_0 = 0.034$   & $ \mu_{1} = 0.2$  & $ \mu_{2}  =0.3$   \\ 	
		\hline	
	\end{tabular}
	\label{Aliev-Panfilov}
\end{table*}
Using the work of Ratti and Verani \cite{ratti2019posteriori} as our baseline,
we consider a problem on the propagation of transmembrane potential to
validate the accuracy of our method in solving the diffusion-reaction equations that describe electrophysiology. 
Assuming a two dimensional square domain  of  $ (0, 1)^2 $ being isotropic, 
the nondimensional time interval is set equal to (0, 16).
Being isotropic tissue, the nondimensional diffusion coefficients are $ d_{iso} $ = 1.0 and $ d_{ani} $ = 0.0.
The transmembrane potential and gating variable are initialized by Eq. \eqref{Aliev–Panfilov-potential},
where we apply the Aliev–Panfilov model with the constant parameters given in Table \ref{Aliev-Panfilov}.
\begin{equation}\label{Aliev–Panfilov-potential}
\begin{cases}
V_m = \rm exp[-\frac{(x-1)^2 + y^2}{0.25}] \\
w = 0
\end{cases}.
\end{equation} 
The sequential progression of the potential is displayed  in Figure. \ref{depo-coutour}.
Figure. \ref{transmembrane-potential} reports the predicted evolution profile of the transmembrane potential at point (0.3, 0.7),
and compares it values with the work of Ratti and Verani \cite{ratti2019posteriori} and Rui et al. \cite{chen2024coupling}.
In accordance with the previous numerical estimation  and experimental observation \cite{franzone2014mathematical},
our implementation accurately replicates the  quick propagation
of the stimulus in the tissue and the gradual decrease 
in transmembrane potential following a plateau phase.

\subsubsection{3d cardiac function coupling with nonisotropic diffusion}
To illustrate the capabilities of the current ASPH method 
in comprehensive cardiac simulations,
we examine the propagation of transmembrane potentials as free pulses,
along with the associated excitation-contraction dynamics 
within a three-dimensional left ventricle model.
Building upon the work of \cite{zhang2023multi, patelli2017isogeometric},
the ionic current is modeled using the Aliev–Panfilow model  
with constant parameters given in Table. \ref{Aliev–Panfilov-model} 
and the diffusion coefficients are set as 
$ d_{iso} $ = 0.8  mm$^2$/ms and d$ _{ani}$ = 1.2 mm$^2$/ms.  
For the momentum equation,  Holzapfel–Ogden constitution model is 
employed,  with parameters listed in Table. \ref{Holzapfel–Ogden}.
\begin{table*}[htp!]
	\centering
	\caption{3d cardiac function: parameters in Holzapfel–Ogden constitution model.}
	\begin{tabular}{cccc}
		\hline
		$ a = 0.059$ kPa  & $ a_f = 18.472$ kPa  &  $ a_s = 2.841$ kPa &  $ a_{fs} = 0.216$ kPa   \\ 	
		\hline
		$b$ = 8.023  & 	$b_f$ = 16.026 & $b_s$ = 11.12  &$b_{fs}$ = 11.436 \\
		\hline	
	\end{tabular}
	\label{Holzapfel–Ogden}
\end{table*}
\begin{table*}[htp!]
\centering
\caption{3d cardiac function: parameters in the Aliev–Panfilov model.}
\begin{tabular}{cccccc}
	\hline
$ k $  & $a$ & $ b$    &  $ \epsilon_0 $ & $\mu_{1}$ &  $\mu_{2}$ \\ 	
		\hline
		8.0  & 0.01 & $ 0.15  $&0.002 & 0.2 & 0.23\\
		\hline	
	\end{tabular}
	\label{Aliev–Panfilov-model}
\end{table*}

\begin{figure*}[htbp]
	\centering
		\begin{subfigure}[b]{0.10\textwidth}
		\includegraphics[trim = 120mm 50mm 40mm 40mm, clip,width=0.95\textwidth]{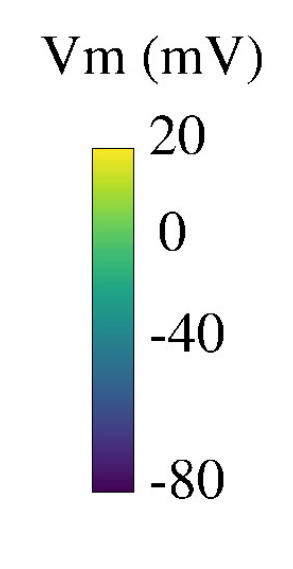}
	\end{subfigure}
	\begin{subfigure}[b]{0.17\textwidth}
		\includegraphics[trim = 18mm 60mm 100mm 20mm, clip,width=0.95\textwidth]{heart-contour-t0-eps-converted-to.pdf}
	\caption  {t = 0 ms.}
	\end{subfigure}
	\begin{subfigure}[b]{0.17\textwidth}
		\includegraphics[trim = 18mm 60mm 100mm 20mm, clip,width=0.95\textwidth]{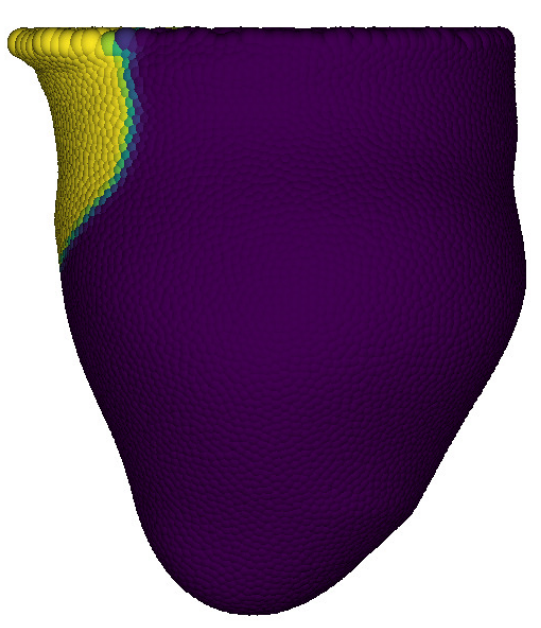}
		\caption  {t = 130 ms.}
	\end{subfigure}
	\begin{subfigure}[b]{0.17\textwidth}
	\includegraphics[trim = 18mm 60mm 100mm 20mm, clip,width=0.95\textwidth]{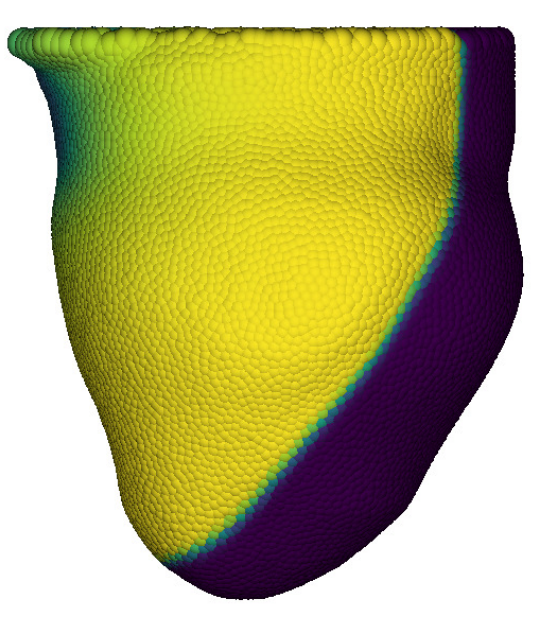}
	\caption  {t = 387 ms.} 
\end{subfigure}
\begin{subfigure}[b]{0.17\textwidth}
	\includegraphics[trim = 18mm 65mm 100mm 20mm, clip,width=0.95\textwidth]{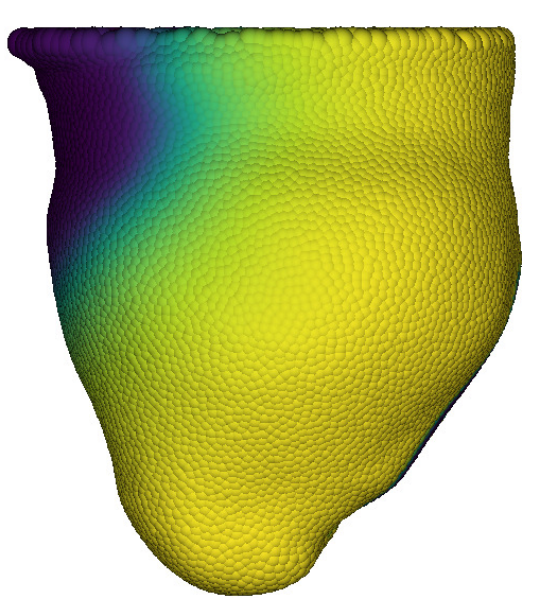}
	\caption {t = 516 ms.}
\end{subfigure}
\begin{subfigure}[b]{0.17\textwidth}
	\includegraphics[trim =18mm 65mm 100mm 20mm, clip,width=0.95\textwidth]{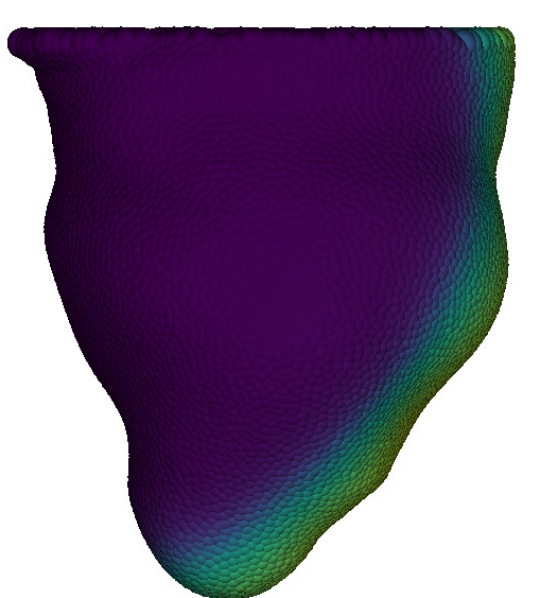}
	\caption {t = 774 ms.}
\end{subfigure}
	\begin{subfigure}[b]{0.10\textwidth}
	\includegraphics[trim = 120mm 50mm 40mm 40mm, clip,width=0.95\textwidth]{heart-contour-t0-eps-converted-to.pdf}
\end{subfigure}
	\begin{subfigure}[b]{0.17\textwidth}
	\includegraphics[trim =18mm 60mm 100mm 20mm, clip,width=0.95\textwidth]{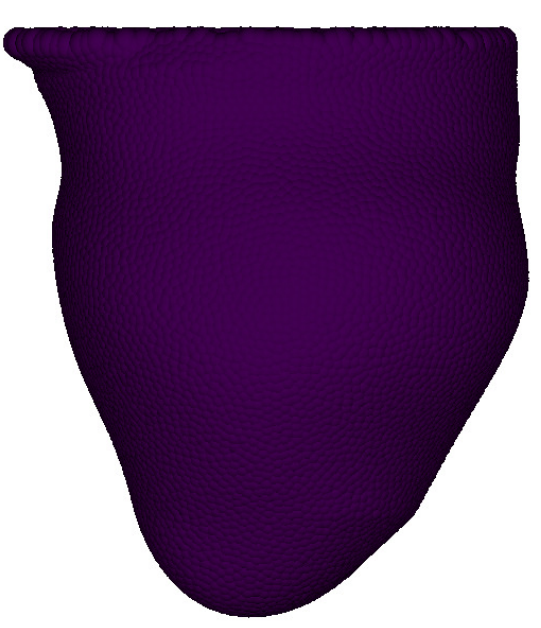}
	\caption  {t = 0 ms.}
\end{subfigure}
\begin{subfigure}[b]{0.17\textwidth}
	\includegraphics[trim =18mm 60mm 100mm 20mm, clip,width=0.95\textwidth]{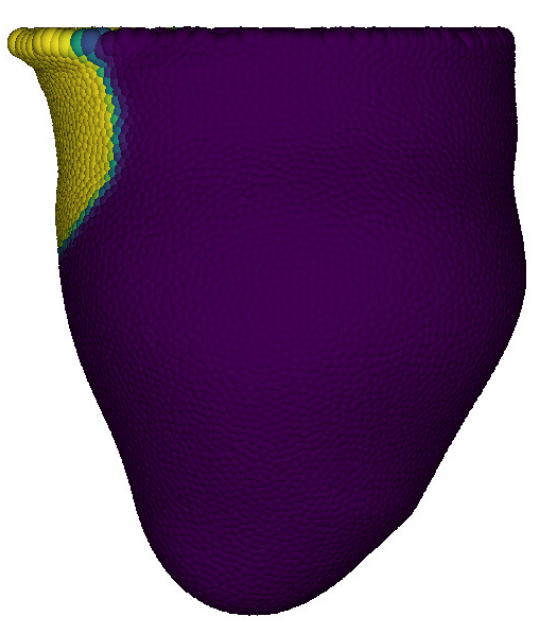}
	\caption  {t = 130 ms.}
\end{subfigure}
\begin{subfigure}[b]{0.17\textwidth}
	\includegraphics[trim =18mm 60mm 100mm 20mm, clip,width=0.95\textwidth]{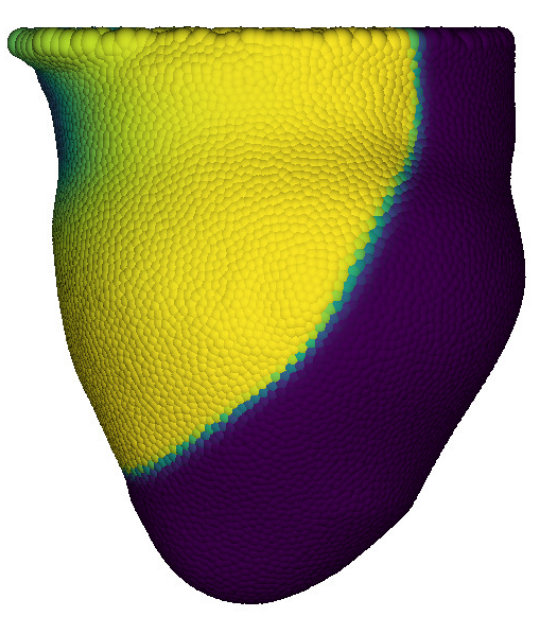}
	\caption  {t = 387 ms.} 
\end{subfigure}
\begin{subfigure}[b]{0.17\textwidth}
	\includegraphics[trim =18mm 65mm 100mm 20mm, clip,width=0.95\textwidth]{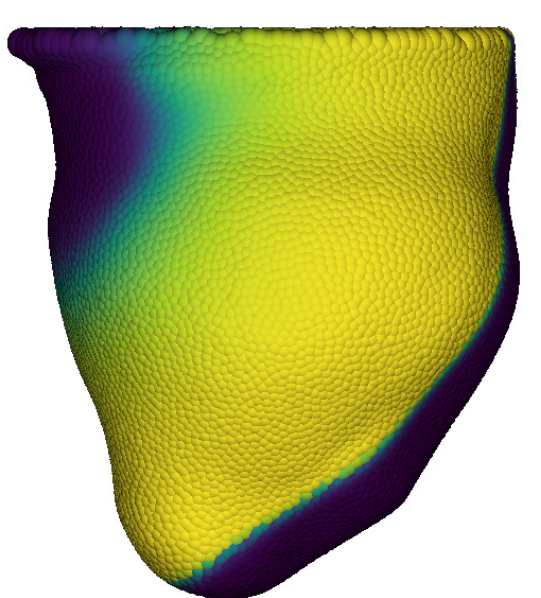}
	\caption {t = 516 ms.}
\end{subfigure}
\begin{subfigure}[b]{0.17\textwidth}
	\includegraphics[trim =18mm 65mm 100mm 20mm, clip,width=0.95\textwidth]{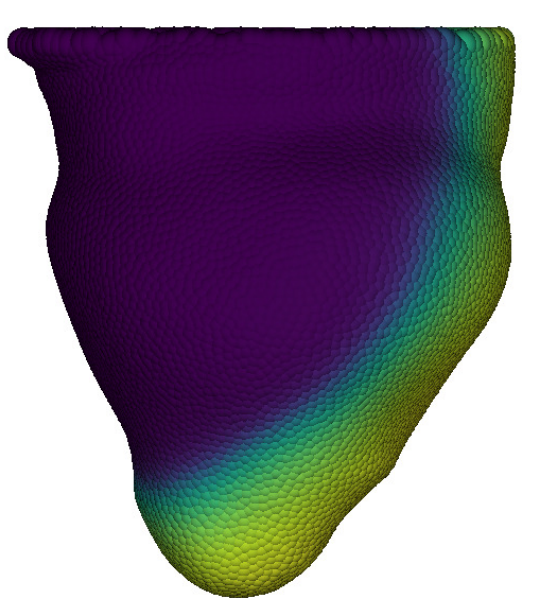}
	\caption {t = 774 ms.}
\end{subfigure}
	\caption{Left ventricle model: The propagation of the transmembrane potential in the heart triggering by a free-pulse pattern. 
	The snapshots depict contours of the transmembrane potential $ V_m $.}
	\label{heart-coutour}
\end{figure*}

\begin{figure*}[htbp]
	\centering
	\includegraphics[trim = 2mm 2mm 2mm 2mm, clip,width=0.65\textwidth]{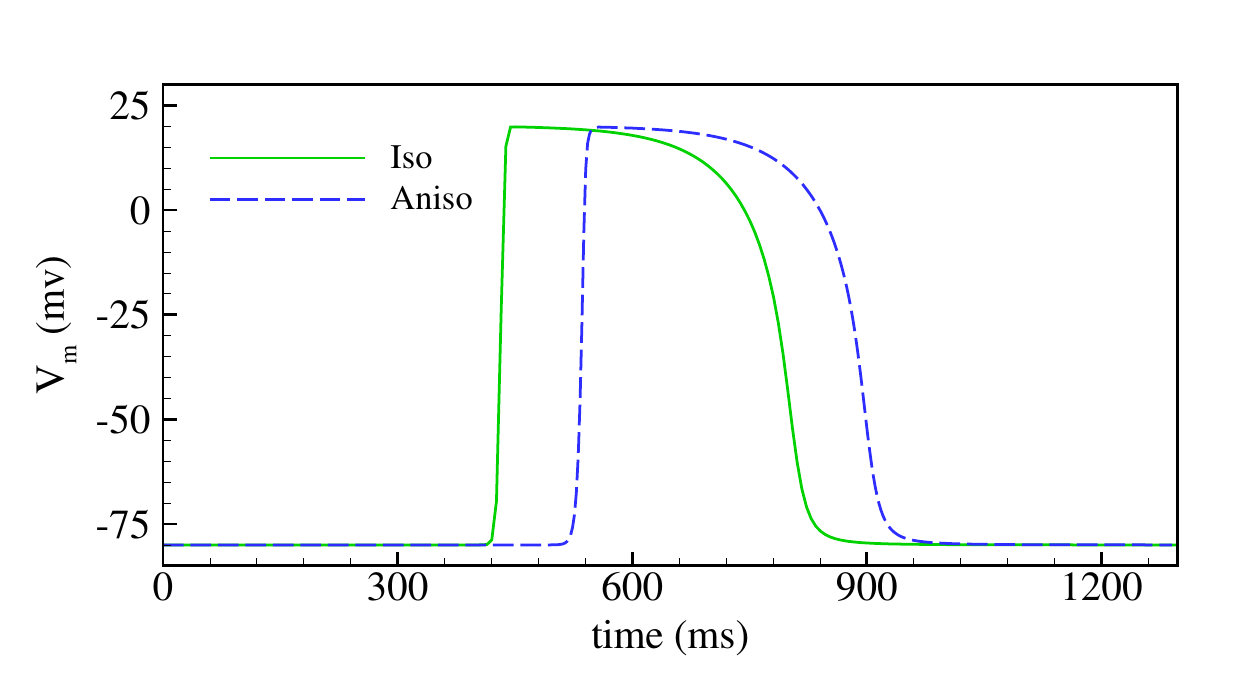}
	\caption {Left ventricle model: the time history of the transmembrane potential recorded on the apex.}
	\label{heart-line}
\end{figure*}

\begin{figure*}[htbp]
	\centering
	\includegraphics[trim = 2mm 2mm 2mm 2mm, clip,width=0.65\textwidth]{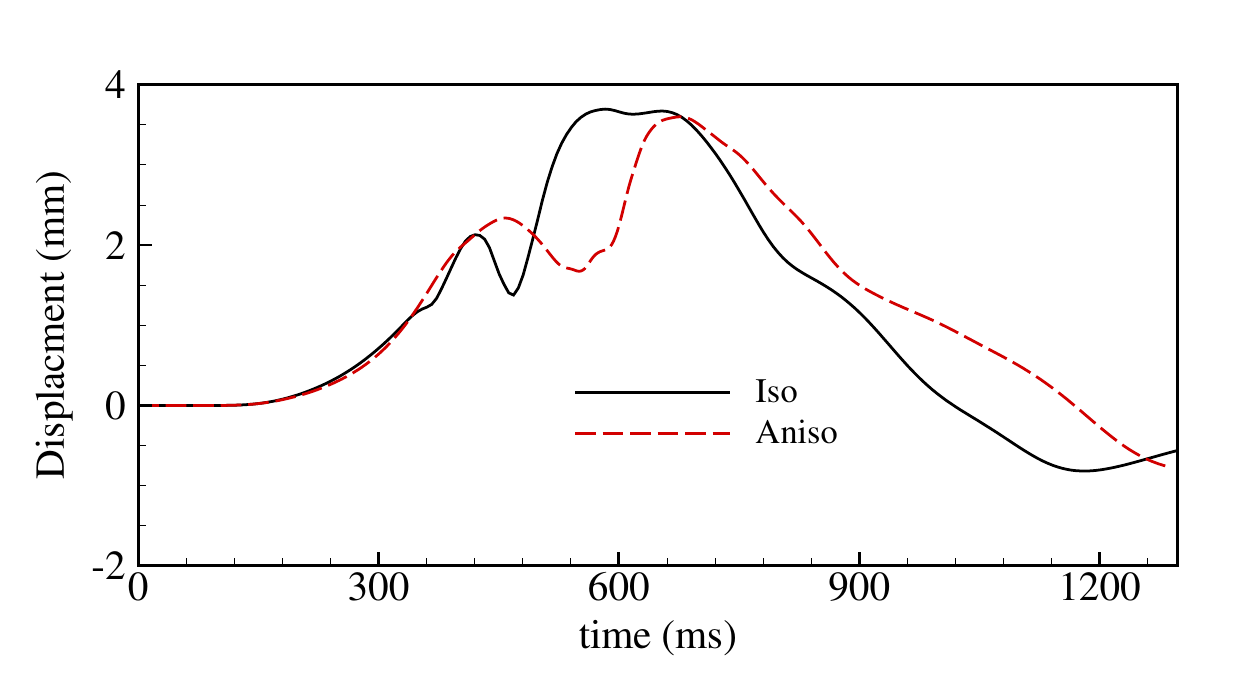}
	\caption {Generic biventricluar heart: the time history of displacement recorded on the apex.}
	\label{heart-dis-line}
\end{figure*}

The transmembrane potential travels in the left ventricle
from the base to the apex,
with a free-pulse stimulus applied at the muscular source
for time interval $t \in [0, 0.5]$ ms with a stimulation
$ V_m $ = 20 mV 
and the potential evolution is depicted in Figure. \ref{heart-coutour}.
It can be observed that the transmembrane potential propagates 
in the similar pattern for both iso- and anisotropic material model.
Figure. \ref{heart-line} and Figure. \ref{heart-dis-line} 
 report the time evolution of 
the transmembrane potentials and the $ z $
components of the displacements 
at the apex separately for both isotropic and anisotropic material models.
Compared with the isotropic condition, 
the anisotropic one induces slightly slower propagation of the
transmembrane potential in the ventricle and  corresponding different mechanical response.
For both isotropic and anisotropic materials,
the transmembrane potential profiles show similar pattern 
to the results reported by Zhang et al. \cite{zhang2023multi}.


\section{Conclusion}  
 
In this study,  
we incorporate a second derivative model based on smooth particle hydrodynamics (SPH) to solve the diffusion process with anisotropic characteristics.
A full-version formulation  with second order accuracy that contains the elements of the hessian matrix is applied to deduce the diffusion operator.
With the Hessian matrix in hand,
a coordinate transformation tensor is applied to 
obtain the  anisotropic diffusion operator.
For diffusion in thin structures, 
anisotropic resolution coupling anisotropic kernel 
are employed to enhance the computation efficiency.
Anisotropic kernel and gradient functions are considered when applying the Taylor series expansion to the second derivative model, producing the Laplacian operators.

The behavior of the present scheme was analyzed by applying it to typical diffusion problems with analytical solution,
including the diffusion of scalar with pre-function within a thin structure and anisotropic diffusion
of a contaminant contaminant in fluid.
In both cases,  
the comparison with the theoretical solution demonstrating the present scheme can achieve the second order accuracy.
Even with large ansiotropic characteristic,  
the present method perfectly reproduces the analytical result. 
Convergence tests in terms of spatial resolution are also examined.
Furthermore,  the application of the proposed SPH second derivative model 
to the diffusion in thin membrane and the anisotropic transport of 
transmembrane potential within the left ventricle have been conducted with promising results.
Generally, the introduced ASPH formulations of the anisotropic
Laplacian show improved performance when 
compared to classical SPH formulations, 
in terms of better approximation accuracy and higher stability.
The non-physical values of the scalar quantities 
which are usually occur in other methods are effectively avoided, 
and the oscillations are eliminated.
More comprehensive studies and detailed comparisons 
with traditional SPH method are needed to fully 
explore the advantages of present ASPH algorithm.
\newpage
\appendix
\section{Fluid-structure interaction model}\label{appendixB}
In this appendix, 
we reference Zhao's algorithm \cite{zhao2013modeling} 
to provide a concise overview of the porosity assumption 
for porous media model
and its associated relations, 
including porosity and fluid saturation (\ref{appendixB1}), 
as well as stress relations (\ref{appendixB3}).
In this mixture model, 
the definition of state variables including solid density $\rho^s$,
locally fluid density  $\rho^l$,
solid velocity $\mathbf{v}^s$, 
and fluid saturation $\widetilde{c}$ 
enables the  fluid velocity  to be 
calculated with reference to the solid velocity.
This simplified approach is practically significant, 
notably reducing the system complexity
by eliminating the necessity
for two separate sets of equations to describe the fluid and solid separately.

\subsection{Porosity and fluid saturation}\label{appendixB1}
Considering  a representative volume element $dV$,
the macroscopic porosity $c$ is defined as 
the ratio of the total volume of the 
pores $dV^p$ to $dV$, yielding $c = \frac{dV^p}{dV}$. 
Note that $ 0 < {c} < 1 $ holds for all cases.
When the porous solid is partially saturated by fluid,  
the fluid saturation level $ \widetilde{c} $ can be defined as
\begin{equation}
	\label{s_defination}
	\widetilde{c} =\frac{dV^l}{dV},
\end{equation}
where $ dV^l $ denotes the fluid volume 
in the representative element $ dV $. 
Clearly, $\widetilde{c}$ is always less than or equal to the maximum possible saturation ${c}$, i.e., $\widetilde{c} \leq {c} $.
The locally effective fluid density $ \rho^l $, 
defined as the mass of the fluid per unit volume, 
varies depending on the extent of fluid saturation 
and can be expressed as 
\begin{equation}
	\label{fluid_density-appendix}
	\rho^l =  \frac{dm^{l}} {dV } = \frac{dm^{l}} {dV^l }\frac{dV^{l}} {dV } =\rho^L \widetilde{c} ,
\end{equation} 
where $dm^l$ represents
the mass of the fluid within $dV$, 
$ {\rho}^L$ the 
fluid density 
which is assumed to be a constant for 
incompressible fluids. 

\subsection{Effective  stress on solid}\label{appendixB3}
Following \cite{gawin1995coupled,korsawe2006finite,ghaboussi1973flow, atkin1976continuum}, 
the total stress acting on the solid 
is the sum of Cauchy stress $ \boldsymbol{\sigma}^s $ 
due to deformation 
and the pressure stress $\boldsymbol{\sigma}^l $
due to the presence of the fluid phase, 
written as:
\begin{equation}
	\boldsymbol{\sigma} =\boldsymbol{\sigma}^s +\boldsymbol{\sigma}^l = \boldsymbol{\sigma}^s - p^l \mathbf{I}.
\end{equation}
where $ p^l $ is fluid pressure. 
For a hyper-elastic material, 
the constitutive equation for the 
solid component is given by
\begin{equation}
	\label{cauchy-stress}
	\boldsymbol{\sigma}^s = 2\mu \mathbf{e} + \lambda \text{tr}(\mathbf{e})\mathbf{I},
\end{equation}
where  the Eulerian-Almansi finite strain tensor  $\mathbf{e}$  
can be evaluated  by
\begin{equation}
	\mathbf{e} = \frac{1}{2} ( \mathbf{I} - \mathbf{F}^T \mathbf{F}).
\end{equation}
The Lam$\acute{e}$ parameter $ \lambda $ can be calculated via  
shear modulus $\mu$  and bulk modulus $K$ as 
$ \lambda =K - \frac{2\mu}{3}$.  

The fluid pressure solely depends on 
the fluid saturation level within the porous solid element,
represented by a function $ p^l = p^l (\widetilde{c}) $. 
In the present model, 
this behavior is described mathematically by applying a linear relation, 
taking the form
\begin{equation}
	\label{fluid_pressure}
	p^l = C (\widetilde{c} -   \widetilde{c}_0),
\end{equation} 
where $ C $ is a material constant,  
$\widetilde{c}_0$ the  initial  saturation. 
Details can be referred to \cite{atkin1976continuum}.

\section[\appendixname~\thesection]{Transformation tensor $\mathbf{G}$}
\label{appendix-G}
For a two-dimensional case, 
with $ h_1 $ denoting the length in semimajor axis direction
and $ h_2 $ in the semiminor axis in an ellipse, 
$ \theta $  being the rotation angle of the semimajor axis compared to the 
	real frame, $\mathbf{G}$ is given by,
\begin{equation}	 
	\mathbf{G}=\begin{bmatrix}  
		h_{1}^{-1} \cos^2\theta + h_{2}^{-1} \sin^2\theta   &   (h_{1}^{-1} - h_{2}^{-1}) \cos\theta \sin\theta  \\  
		(h_{1}^{-1} - h_{2}^{-1}) \cos\theta \sin\theta & h_{1}^{-1} \sin^2\theta + h_{2}^{-1} \cos^2\theta \\   
	\end{bmatrix}.  
\end{equation} 
If a kernel frame is consistent with the real frame,
in other words, rotation angle $\theta = 0 $, 
$ \mathbf{G} $ can be simplified into 

\begin{equation}	 
	\mathbf{G}=\begin{pmatrix}  
		h_{1}^{-1}    &   0  \\  
		0 &   h_{2}^{-1}\\   
	\end{pmatrix}.  
\end{equation} 
While for a more complex  three-dimensional case,
with a vector $ (h_1, h_2, h_3) $ representing the smoothing lengths along different axes in the kernel frame, 
and the rotation angle $ (\omega,\psi, \chi ) $ between the kernel frame and the real $ (x, y, z) $ frames, 
$\mathbf{G}$ can be expressed as

\begin{equation}
	\mathbf{G}=\begin{pmatrix}  
		G_{11} & G_{21} & G_{31} \\  
		G_{21} & G_{22} & G_{32} \\  
		G_{31} & G_{32} & G_{33}  
	\end{pmatrix} ,
\end{equation}
where the six elements are defined as
\begin{equation}
	\begin{array}{l}
		G_{11}=h_{1}^{-1} \omega_{1}^{2} \psi_{1}^{2}+h_{2}^{-1}\left(\omega_{1} \psi_{2} \chi_{2}-\omega_{2} \chi_{1}\right)^{2}+h_{3}^{-1}\left(\omega_{1} \psi_{2} \chi_{1}+\omega_{2} \chi_{2}\right)^{2}, \\
		\begin{split}
			G_{21}&=h_{1}^{-1} \omega_{1} \omega_{2} \psi_{1}^{2}+h_{2}^{-1}\left(\omega_{1} \psi_{2} \chi_{2}-\omega_{2} \chi_{1}\right)\left(\omega_{2} \psi_{2} \chi_{2}+\omega_{1} \chi_{1}\right)\\
			&+h_{3}^{-1}\left(\omega_{2} \psi_{2} \chi_{1}-\omega_{1} \chi_{2}\right)\left(\omega_{1} \psi_{2} \chi_{1}+\omega_{2} \chi_{2}\right), \\
		\end{split}\\
		G_{31}=-h_{1}^{-1} \omega_{1} \psi_{1} \psi_{2}+h_{2}^{-1} \psi_{1} \chi_{2}\left(\omega_{1} \psi_{2} \chi_{2}-\omega_{2} \chi_{1}\right)+h_{3}^{-1} \psi_{1} \chi_{1}\left(\omega_{1} \psi_{2} \chi_{1}+\omega_{2} \chi_{2}\right), \\
		G_{22}=h_{1}^{-1} \omega_{2}^{2} \psi_{1}^{2}+h_{2}^{-1}\left(\omega_{2} \psi_{2} \chi_{2}+\omega_{1} \chi_{1}\right)^{2}+h_{3}^{-1}\left(\omega_{2} \psi_{2} \chi_{1}-\omega_{1} \chi_{2}\right)^{2}, \\
		G_{32}=-h_{1}^{-1} \omega_{2} \psi_{1} \psi_{2}+h_{2}^{-1} \psi_{1} \chi_{2}\left(\omega_{2} \psi_{2} \chi_{2}+\omega_{1} \chi_{1}\right)+h_{3}^{-1} \psi_{1} \chi_{1}\left(\omega_{2} \psi_{2} \chi_{1}-\omega_{1} \chi_{2}\right), \\
		G_{33}=h_{1}^{-1} \psi_{2}^{2}+h_{2}^{-1} \psi_{1}^{2} \chi_{2}^{2}+h_{3}^{-1} \psi_{1}^{2} \chi_{1}^{2}.
	\end{array}
\end{equation}
Here, for a angle $a$ ($a = \omega, \psi~\text{or}~\chi $), $ a_{1} = \cos a$ and $ a_{2} = \sin a $. 
When $\omega=  \psi= \chi =0 $,
\begin{equation}	 
	\mathbf{G}=\begin{pmatrix}  
		h_{1}^{-1} & 0 & 0\\  
		0 & h_{2}^{-1} & 0\\   
		0 & 0 & h_{3}^{-1}\\   
	\end{pmatrix}.  
\end{equation} 
Clearly, SPH can be considered as a special case
of ASPH with $ h_1=h_2=h_3=h $ and $\omega=  \psi= \chi =0 $.
Tensor $\mathbf{G}$ is only needed to be calculated once. After $\mathbf{G}$  is initialized, the kernel expressions are accordingly determined.

\section[\appendixname~\thesection]{The second derivative model}
\label{full-expression}
{\linespread{1.5} \selectfont%
Following Ref .\cite{asai2023class}, 
the full expression of the second derivative model  with the second order accuracy in 2D	case is written as
\begin{equation}
\begin{split}
\begin{bmatrix}
\vspace{1.3ex}	({r^{1}_{ij}})^2 \\ \vspace{1.3ex}	({r^{2}_{ij}})^2
\\ \vspace{1.3ex}	 r^{1}_{ij}  r^{2}_{ij}  
\end{bmatrix}
& \left[\begin{array}{c}
\vspace{1.1ex}	
\left(r^{1}_{i j}\right)^{2}-\boldsymbol{r}_{i j} \cdot \bigcup_{k \in \mathbb{S}^{i}} \frac{m_{k}}{\rho_{k}}\left(r^{1}_{i k}\right)^{2} \widetilde{\nabla}^{0}  W_{i k} \\
\vspace{1.1ex}	
\left(r^{2}_{i j}\right)^{2}-\boldsymbol{r}_{i j} \cdot \bigcup_{k \in \mathbb{S}^{i}} \frac{m_{k}}{\rho_{k}}\left(r^{2}_{i k}\right)^{2} \widetilde{\nabla}^{0}  W_{i k}\\	\vspace{1.1ex}	
r^{1}_{i j} r^{2}_{i j}-\boldsymbol{r}_{i j} \cdot \bigcup_{k \in \mathbb{S}^{i}} \frac{m_{k}}{\rho_{k}} r^{1}_{i k} r^{2}_{i k} \widetilde{\nabla}^{0}  W_{i k}
\end{array}\right]^{T}\left[\begin{array}{c}
\vspace{1.1ex}	
\frac{\partial^{2} \phi_{i}}{\left(\partial x^{1}\right)^{2}} \\
\vspace{1.1ex}	
\frac{\partial^{2} \phi_{i}}{\left(\partial x^{2}\right)^{2}}  \\
\vspace{1.1ex}	
2 	\frac{\partial^{2} \phi_{i}}{\partial x^{1}\partial x^{2}} 
\end{array}\right]_{2 \mathrm{D}} \\
& = 2 \begin{bmatrix}
\vspace{1.3ex}	({r^{1}_{ij}})^2 \\ \vspace{1.3ex}	({r^{2}_{ij}})^2
\\ \vspace{1.3ex}	 r^{1}_{ij}  r^{2}_{ij}  
\end{bmatrix} 
\left\{\phi_{i j}-\boldsymbol{r}_{i j} \cdot\left(\bigcup_{k \in \mathbb{S}^{i}} \frac{m_{k}}{\rho_{k}} \phi_{i k} \widetilde{\nabla}^{0}  W_{i k}\right)\right\},
\end{split}
\label{laplacian_2d}
\end{equation}
while in 3D cases,
the equation is expressed as
\begin{equation}
\begin{split}
\begin{bmatrix}
\vspace{1.3ex}	({r^{1}_{ij}})^2 \\ \vspace{1.3ex}	({r^{2}_{ij}})^2\\
\vspace{1.3ex}	({r^{3}_{ij}})^2 \\
\vspace{1.3ex} r^{1}_{ij}  r^{2}_{ij}   \\
\vspace{1.3ex} r^{2}_{ij}  r^{3}_{ij}   \\
\vspace{1.3ex} r^{3}_{ij}  r^{1}_{ij}  
\end{bmatrix}
&\left[\begin{array}{c}
\vspace{1.1ex}	
\left(r^{1}_{i j}\right)^{2}-\boldsymbol{r}_{i j} \cdot \bigcup_{k \in \mathbb{S}^{i}} \frac{m_{k}}{\rho_{k}}\left(r^{1}_{i k}\right)^{2} \widetilde{\nabla}^{0}  W_{i k} \\ 	\vspace{1.1ex}	
\left(r^{2}_{i j}\right)^{2}-\boldsymbol{r}_{i j} \cdot \bigcup_{k \in \mathbb{S}^{i}} \frac{m_{k}}{\rho_{k}}\left(r^{2}_{i k}\right)^{2} \widetilde{\nabla}^{0}  W_{i k}\\ 	\vspace{1.1ex}	
\left(r^{3}_{i j}\right)^{2}-\boldsymbol{r}_{i j} \cdot \bigcup_{k \in \mathbb{S}^{i}} \frac{m_{k}}{\rho_{k}}\left(r^{3}_{i k}\right)^{2} \widetilde{\nabla}^{0}  W_{i k} \\	\vspace{1.1ex}	
r^{1}_{i j} r^{2}_{i j}-\boldsymbol{r}_{i j} \cdot \bigcup_{k \in \mathbb{S}^{i}} \frac{m_{k}}{\rho_{k}} r^{1}_{i k} r^{2}_{i k} \widetilde{\nabla}^{0}  W_{i k}
\\	\vspace{1.1ex}	
r^{2}_{i j} r^{3}_{i j}-\boldsymbol{r}_{i j} \cdot \bigcup_{k \in \mathbb{S}^{i}} \frac{m_{k}}{\rho_{k}} r^{2}_{i k} r^{3}_{i k} \widetilde{\nabla}^{0}  W_{i k}
\\	\vspace{1.1ex}	
r^{3}_{i j} r^{1}_{i j}-\boldsymbol{r}_{i j} \cdot \bigcup_{k \in \mathbb{S}^{i}} \frac{m_{k}}{\rho_{k}} r^{3}_{i k} r^{1}_{i k} \widetilde{\nabla}^{0}  W_{i k}
\end{array}\right]^{T}
\begin{bmatrix} 
\vspace{1.1ex}
\frac{\partial^{2} \phi_{i}}{\left(\partial x^{1}\right)^{2}} \\
\vspace{1.1ex}
\frac{\partial^{2} \phi_{i}}{\left(\partial x^{2}\right)^{2}}  \\
\vspace{1.1ex}	
\frac{\partial^{2} \phi_{i}}{\left(\partial x^{3}\right)^{2}} \\ 	\vspace{1.1ex}	
2 	\frac{\partial^{2} \phi_{i}}{\partial x^{1}\partial x^{2}} \\ 	\vspace{1.1ex}	
2 	\frac{\partial^{2} \phi_{i}}{\partial x^{2}\partial x^{3}} \\ 	\vspace{1.1ex}	
2 	\frac{\partial^{2} \phi_{i}}{\partial x^{3}\partial x^{1}} 
\end{bmatrix}_{3 \mathrm{D}}\\
& = 2 \begin{bmatrix}
\vspace{1.3ex}	({r^{1}_{ij}})^2 \\ \vspace{1.3ex}	({r^{2}_{ij}})^2\\
\vspace{1.3ex}	({r^{3}_{ij}})^2 \\
\vspace{1.3ex} r^{1}_{ij}  r^{2}_{ij}   \\
\vspace{1.3ex} r^{2}_{ij}  r^{3}_{ij}   \\
\vspace{1.3ex} r^{3}_{ij}  r^{1}_{ij}  
\end{bmatrix}
\left\{\phi_{i j}-\boldsymbol{r}_{i j} \cdot\left(\bigcup_{k \in \mathbb{S}^{i}} \frac{m_{k}}{\rho_{k}} \phi_{i k} \widetilde{\nabla}^{0}  W_{i k}\right)\right\},
\end{split}
\label{laplacian_3d}
\end{equation}
where $ \tilde{\nabla}^{0} W_{i k}  $ can be obtained by applying matrix
$  \mathbf{B}^{0^T}$ as  $ \tilde{\nabla}^{0} W_{i k} =   \mathbf{B}^{0^T} {\nabla}^{0} W_{i k} $ .
Note that a scalar function $ F_{ij} = (: \boldsymbol{r}_{ij} ) \cdot \tilde{\nabla} W_{ij} / |r_{ij}|^4 $ is multiplied  both sides  in Eq. \eqref{laplacian_2d} and Eq. \eqref{laplacian_3d}
and the volume integral using the particle summation is token.
Subsequently, the elements ${\frac{\partial^2 \phi_i} {(\partial x^1)^2 }} ,
 ~{\frac{\partial^2 \phi_i} {(\partial x^2)^2 }},
 ~2{\frac{\partial^2 \phi_i} {\partial x^1\partial x^2}} $ in Hessian matrix $ \mathbf{H}_x $ are obtained.

 \bibliographystyle{IEEEtran}
\bibliography{mybio}
\end{document}